\RequirePackage[pagewise,mathlines]{lineno}
\documentclass[twocolumn,showpacs,amsmath,amssymb,superscriptaddress,nofootinbib]{revtex4-1}

\usepackage{graphicx}
\usepackage{dcolumn}
\usepackage{bm}
\usepackage{color}
\usepackage{ulem}
\usepackage{url}
\usepackage{multirow}
\usepackage{enumitem}

\newcommand{\bfg }{\begin{figure}[htpb]}
\newcommand{\efg }{\end{figure}}
\newcommand{\bmn }{\begin{minipage}}
\newcommand{\emn }{\end{minipage}}
\newcommand{\bt }{\begin{table}[htpb]}
\newcommand{\et }{\end{table}}

\newcommand{\sNN}{$\sqrt{s_{\rm NN}}$ }

\newcommand{\GeVc}{GeV/$c$ }
\newcommand{\GeVcsq}{GeV/$c^2$ }
\newcommand{\phiV}{$\phi_{\rm V}$}

\newcommand{ \be }{\begin{equation}}
\newcommand{ \ee }{\end{equation}}
\newcommand{ \bea }{\begin{eqnarray}}
\newcommand{ \eea }{\end{eqnarray}}
\newcommand{ \la }{\langle}
\newcommand{ \ra }{\rangle}

\newcommand{ \pT}{$p_{\rm T}$}
\newcommand{ \Nbin}{$N_{\rm bin}$}
\newcommand{ \Npart}{$N_{\rm part}$}

\begin{document}
\def\Journal#1#2#3#4{{#1} {\bf #2}, #3 (#4)}

\def\NCA{Nuovo Cimento}
\def\NIM{Nucl. Instr. Meth.}
\def\NIMA{{Nucl. Instr. Meth.} A}
\def\NPB{{Nucl. Phys.} B}
\def\NPA{{Nucl. Phys.} A}
\def\PLB{{Phys. Lett.}  B}
\def\PRL{{Phys. Rev. Lett.}}
\def\PRC{{Phys. Rev.} C}
\def\PRD{{Phys. Rev.} D}
\def\ZPC{{Z. Phys.} C}
\def\JPG{{J. Phys.} G}
\def\EPJ{{Eur. Phys. J.} C}
\def\EPJA{{Eur. Phys. J.} A}
\def\EPJST{{Eur. Phys. J.} - Special Topics}
\def\RPP{{Rep. Prog. Phys.}}
\def\PR{{Phys. Rep.}}
\def\ANP{{Adv. Nucl. Phys.}}

\preprint{}

{\it version 0.28}

\title{Measurements of Dielectron Production in Au$+$Au Collisions at \sNN = 200\,GeV from the STAR Experiment}

\affiliation{AGH University of Science and Technology, Cracow 30-059, Poland}
\affiliation{Argonne National Laboratory, Argonne, Illinois 60439, USA}
\affiliation{Brookhaven National Laboratory, Upton, New York 11973, USA}
\affiliation{University of California, Berkeley, California 94720, USA}
\affiliation{University of California, Davis, California 95616, USA}
\affiliation{University of California, Los Angeles, California 90095, USA}
\affiliation{Central China Normal University (HZNU), Wuhan 430079, China}
\affiliation{University of Illinois at Chicago, Chicago, Illinois 60607, USA}
\affiliation{Creighton University, Omaha, Nebraska 68178, USA}
\affiliation{Czech Technical University in Prague, FNSPE, Prague, 115 19, Czech Republic}
\affiliation{Nuclear Physics Institute AS CR, 250 68 \v{R}e\v{z}/Prague, Czech Republic}
\affiliation{Frankfurt Institute for Advanced Studies FIAS, Frankfurt 60438, Germany}
\affiliation{Institute of Physics, Bhubaneswar 751005, India}
\affiliation{Indian Institute of Technology, Mumbai 400076, India}
\affiliation{Indiana University, Bloomington, Indiana 47408, USA}
\affiliation{Alikhanov Institute for Theoretical and Experimental Physics, Moscow 117218, Russia}
\affiliation{University of Jammu, Jammu 180001, India}
\affiliation{Joint Institute for Nuclear Research, Dubna, 141 980, Russia}
\affiliation{Kent State University, Kent, Ohio 44242, USA}
\affiliation{University of Kentucky, Lexington, Kentucky, 40506-0055, USA}
\affiliation{Korea Institute of Science and Technology Information, Daejeon 305-701, Korea}
\affiliation{Institute of Modern Physics, Lanzhou 730000, China}
\affiliation{Lawrence Berkeley National Laboratory, Berkeley, California 94720, USA}
\affiliation{Max-Planck-Institut fur Physik, Munich 80805, Germany}
\affiliation{Michigan State University, East Lansing, Michigan 48824, USA}
\affiliation{Moscow Engineering Physics Institute, Moscow 115409, Russia}
\affiliation{National Institute of Science Education and Research, Bhubaneswar 751005, India}
\affiliation{Ohio State University, Columbus, Ohio 43210, USA}
\affiliation{Institute of Nuclear Physics PAN, Cracow 31-342, Poland}
\affiliation{Panjab University, Chandigarh 160014, India}
\affiliation{Pennsylvania State University, University Park, Pennsylvania 16802, USA}
\affiliation{Institute of High Energy Physics, Protvino 142281, Russia}
\affiliation{Purdue University, West Lafayette, Indiana 47907, USA}
\affiliation{Pusan National University, Pusan 609735, Republic of Korea}
\affiliation{University of Rajasthan, Jaipur 302004, India}
\affiliation{Rice University, Houston, Texas 77251, USA}
\affiliation{University of Science and Technology of China, Hefei 230026, China}
\affiliation{Shandong University, Jinan, Shandong 250100, China}
\affiliation{Shanghai Institute of Applied Physics, Shanghai 201800, China}
\affiliation{Temple University, Philadelphia, Pennsylvania 19122, USA}
\affiliation{Texas A\&M University, College Station, Texas 77843, USA}
\affiliation{University of Texas, Austin, Texas 78712, USA}
\affiliation{University of Houston, Houston, Texas 77204, USA}
\affiliation{Tsinghua University, Beijing 100084, China}
\affiliation{United States Naval Academy, Annapolis, Maryland, 21402, USA}
\affiliation{Valparaiso University, Valparaiso, Indiana 46383, USA}
\affiliation{Variable Energy Cyclotron Centre, Kolkata 700064, India}
\affiliation{Warsaw University of Technology, Warsaw 00-661, Poland}
\affiliation{Wayne State University, Detroit, Michigan 48201, USA}
\affiliation{World Laboratory for Cosmology and Particle Physics (WLCAPP), Cairo 11571, Egypt}
\affiliation{Yale University, New Haven, Connecticut 06520, USA}
\affiliation{University of Zagreb, Zagreb, HR-10002, Croatia}

\author{L.~Adamczyk}\affiliation{AGH University of Science and Technology, Cracow 30-059, Poland}
\author{J.~K.~Adkins}\affiliation{University of Kentucky, Lexington, Kentucky, 40506-0055, USA}
\author{G.~Agakishiev}\affiliation{Joint Institute for Nuclear Research, Dubna, 141 980, Russia}
\author{M.~M.~Aggarwal}\affiliation{Panjab University, Chandigarh 160014, India}
\author{Z.~Ahammed}\affiliation{Variable Energy Cyclotron Centre, Kolkata 700064, India}
\author{I.~Alekseev}\affiliation{Alikhanov Institute for Theoretical and Experimental Physics, Moscow 117218, Russia}
\author{J.~Alford}\affiliation{Kent State University, Kent, Ohio 44242, USA}
\author{A.~Aparin}\affiliation{Joint Institute for Nuclear Research, Dubna, 141 980, Russia}
\author{D.~Arkhipkin}\affiliation{Brookhaven National Laboratory, Upton, New York 11973, USA}
\author{E.~C.~Aschenauer}\affiliation{Brookhaven National Laboratory, Upton, New York 11973, USA}
% ...

\collaboration{STAR Collaboration}\noaffiliation

\date{\today}

\begin{abstract}
We report on measurements of dielectron ($e^+e^-$) production in Au$+$Au
collisions at a center-of-mass energy of 200 GeV per nucleon-nucleon pair using the STAR detector 
at RHIC. Systematic measurements of the dielectron yield as a function of transverse momentum (\pT) and 
collision centrality show an enhancement compared to a cocktail simulation of hadronic sources in the low 
invariant-mass region ($M_{ee}<$ 1\,GeV/$c^2$).
This enhancement cannot be reproduced by the $\rho$-meson vacuum spectral function.
In minimum-bias collisions, in the invariant-mass range of 0.30 $-$ 0.76\,GeV/$c^2$, 
integrated over the full \pT\ acceptance, 
the enhancement factor is 1.76\,$\pm$\,0.06\,(stat.)\,$\pm$\,0.26\,(sys.)\,$\pm$\,0.29\,(cocktail).
The enhancement factor exhibits weak centrality and \pT\ dependence in STAR's accessible kinematic regions, 
while the excess yield in this invariant-mass region as a function of 
the number of participating nucleons follows a power-law shape with a power of 1.44 $\pm$ 0.10.  
Models that assume an in-medium broadening of the $\rho$ meson spectral function consistently describe the
observed excess in these measurements. 
Additionally, we report on measurements of $\omega$ and $\phi$-meson production 
through their $e^+e^-$ decay channel. 
These measurements show good agreement with Tsallis Blast-Wave model predictions as well as, 
in the case of the $\phi$-meson, results through its $K^+K^-$ decay channel.
In the intermediate invariant-mass region (1.1$<M_{ee}<$ 3\,GeV/$c^2$), 
we investigate the spectral shapes from different collision centralities. 
Physics implications for possible in-medium modification of charmed hadron production and other physics sources are discussed.
\end{abstract}

\pacs{25.75.Cj, 25.75.Dw}
\maketitle


\section{Introduction}
A major scientific goal of the ultra-relativistic heavy-ion program is 
to study Quantum ChromoDynamics (QCD) matter at high temperature and density in the laboratory. 
Previous measurements at Relativistic Heavy Ion Collider (RHIC) have established the formation of a
strongly-coupled Quark Gluon Plasma (sQGP) in high-energy heavy-ion
collisions~\cite{STARwhitepaper}. 
Throughout the evolution of the hot, dense, and strongly interacting system, electromagnetic probes are
produced and escape with little interaction. 
Thus, they provide direct information about the various stages of the system's evolution.

Following convention, the dilepton invariant-mass spectrum is typically divided into three ranges:
the low mass region - LMR ($M_{ll}<M_{\phi}$), the intermediate mass region - IMR ($M_{\phi}<M_{ll}<M_{J/\psi}$) and 
the high mass region - HMR ($M_{ll}>M_{J/\psi}$). 
As will be described next, distinctively different physical processes contribute or even dominate within these particular ranges.

The initial hard perturbative QCD process, Drell-Yan production, ($q\bar{q}\rightarrow l^+l^-$) can 
produce high-mass dileptons and is expected to be an important mechanism in the HMR~\cite{RappWambach}.   
Moreover, initial hard scattering processes can allow for
bremsstrahlung emission of virtual photons which convert into low invariant mass, 
high transverse momentum ($p_{\rm T}$) dielectrons (``internal conversion'').
These dileptons, in principle, are calculable within the perturbative QCD framework.

The colliding participant system is expected to quickly reach the partonic sQGP phase where dileptons
can be produced through electromagnetic radiation via parton-parton scatterings.
Theoretical calculations indicate that at top RHIC energy, QGP thermal
dilepton production will become a dominant source in the IMR while thermal dileptons
with higher masses originate from earlier stages~\cite{RappThermalEE}.  This
suggests that investigating the thermal dilepton production as a function of $M_{ll}$ \&
\pT\ allows for probing the medium properties at different stages of the
system's space-time evolution. Measuring thermal dilepton collective flow and
polarization can reveal information about the relevant degrees of freedom which may relate to
deconfinement and the equilibrium of the strongly interacting matter created in
heavy-ion collisions~\cite{DengDileptonFlow,ChatterjeeDileptonV2,ShuryakDileptonPol,STARv2,NA60polarization}.
Thermal radiation can produce real photons as well as virtual photons that decay to dileptons.
Comparative analysis of distributions for these dileptons with respect to those produced in initial hard scattering, can shed light on direct real photon production from the QGP medium.

When the system expands, cools down, and enters into the hadronic phase, dileptons
are produced via multiple hadron-hadron scattering by coupling to vector mesons
($\rho$, $\omega$, $\phi$, {\it etc.}). They are expected to dominate the LMR and 
their mass spectra may be related to
the chiral symmetry restoration in the medium~\cite{RappWambach,RappPLB14}.
Theoretical calculations suggest that the vector meson spectral functions will undergo
modifications in a hot and dense hadronic medium, which may be connected to the
restoration of chiral symmetry. Two scenarios have been proposed for
the change of vector meson spectral functions when chiral symmetry is restored:
a shift of the pole mass~\cite{BrownRho} and/or a broadening of the mass
spectrum~\cite{RappWambach2}. Measurements of the dielectron continuum in the low
mass region will help expose the vector meson production mechanisms, and hence the
chiral properties of the medium in heavy-ion collisions.

Finally, when all particles decouple from the system, long-lived hadrons ($\pi^0$,
$\eta$, $D\overline{D}$, {\it etc.}) can decay into lepton pairs and are measured by the
detector system. Their contributions can be calculated based on the measured or predicted invariant yields
of the respective parent particles, and incorporated in the so-called hadron cocktail. 

Dilepton measurements in heavy-ion collisions have been pursued for decades
from relatively low energies to relativistic and ultrarelativistic
energies~\cite{DLS,HADES,HELIOS3,CERES,NA60,PHENIX}. The CERES measurements
of $e^+e^-$ spectra in Pb+Au collisions at the Super Proton Synchrotron
(SPS) showed an enhancement in the mass region below $\sim$700 MeV/$c^2$ with 
respect to the hadron cocktail that included  the vacuum line shape for the $\rho$ meson~\cite{CERES}.
High-statistics measurements from the NA60 experiment at \sNN = 17.2\,GeV suggested that this enhancement is
consistent with in-medium broadening of the $\rho$-meson spectral function rather than
a drop of its pole mass~\cite{NA60,Rapp4SPS,Renk4SPS,Dusling4SPS,PHSD4SPS}. 
Strikingly, after removal of correlated charm contributions, the NA60 collaboration also 
observed that the slope parameters of the dimuon transverse mass ($m_{\rm T}$) spectrum showed
a roughly linear increase with dimuon invariant mass below the $\phi$-meson mass, followed by a
sudden decline at higher masses. This observation provided a first indication of thermal leptons from a partonic source~\cite{NA60}. 

Thermal radiation of dileptons is expected to be significantly enhanced due to a well-developed
partonic phase in the heavy-ion collision systems created at RHIC. 
The PHENIX collaboration has measured dielectrons at mid-rapidity in Au+Au collisions within its detector acceptance. 
For minimum-bias collisions, in the mass region between 150 and 700~MeV/$c^2$ 
an enhancement of 4.7\,$\pm$\,0.4\,(stat.)\,$\pm$\,1.5\,(syst.)\,$\pm$\,0.9\,(model) 
has been reported by the PHENIX collaboration~\cite{PHENIX}. 
Several theoretical calculations, which have successfully explained the SPS
data~\cite{Rapp4SPS,Renk4SPS,Dusling4SPS,PHSD4SPS} and the STAR data~\cite{STARPRL} previously, 
were unable to reproduce the magnitude of the low mass dielectron enhancement observed by PHENIX 
through expected vector meson contributions in the hadronic medium. 
The PHENIX measured IMR yields are consistent with the charm contribution from $p+p$ scaled with the number of binary collisions. 
However, within the limits of the data precision and our present understanding of the modification of 
charmed hadron production in Au + Au collisions, no conclusive evidence for thermal radiation can be inferred from this measurement. 
A detailed dilepton program to investigate the in-medium chiral and thermal properties is one of the
main focuses of future heavy-ion projects from Schwerionen Synchrotron (SIS) energies up to LHC energies.


In this paper, we report on detailed measurements of dielectron production in Au +
Au collisions at \sNN = 200\,GeV with the Solenoidal Tracker At RHIC (STAR) experiment. The data used in
this analysis were recorded during the RHIC runs in 2010 and 2011. 
The barrel Time-Of-Flight (TOF) detector system was completed before these runs 
thus significantly improving the electron identification over a wide momentum range in
STAR's Time Projection Chamber (TPC).

The paper is organized as follows: Section II describes the experimental setup and
the data sets used in this analysis. Section III explains in detail the analysis
techniques including electron identification, dielectron invariant mass reconstruction, background
subtraction, detector acceptance efficiency correction, and systematic
uncertainties.  Section IV presents our results on dielectron production
yields within the STAR detector acceptance and a comparison to the hadron cocktails.
Results are compared with theoretical calculations of in-medium
modified vector meson line shapes as well as QGP thermal radiation
contributions. Systematic studies on the centrality and \pT\ dependence of
the dielectron yields are presented. 
Our results and conclusions are summarized in Section V. 

\section{Experimental Setup}
The data used in this analysis were collected by the STAR detector~\cite{starnim}. The major detector subsystems 
used in this analysis are the TPC, the TOF, and two trigger subsystems: the Vertex Position Detectors
(VPDs) and the Zero Degree Calorimeters (ZDCs).

\subsection{Time Projection Chamber}

The TPC~\cite{tpcnim} is the main tracking detector and consists of a 4.2 m long solenoidal cylinder
concentric with the beam pipe. 
It is operated in a uniform 0.5 Tesla magnetic field parallel to the beam direction (defined as $z$ direction in STAR).
The inner and outer radii of the active volume
are 0.5 and 2.0 m, respectively. It covers the full azimuth and a
pseudorapidity range of $|\eta|<2$ for the inner radius and $|\eta|<1$ for the outer
radius. The TPC has 45 readout layers allowing measurements of charged particle momenta with
a resolution of $\sim$ 1\% at \pT$\sim$ 1\,GeV/$c$ for tracks originating from
the collision vertices. It is also used for particle identification (PID) via
the ionization energy loss ($dE/dx$) in the TPC gas with a mean $dE/dx$
resolution of about 7\% .

\subsection{Time Of Flight System}

The TOF system consists of the Barrel TOF (BTOF) detector covering the TPC cylinder
and the VPDs located in the forward pseudorapidity regions. The latter provide the
common start time. BTOF detector utilizes the multi-gap resistive
plate chamber technology~\cite{tofproposal}. It covers the full azimuth and
$|\eta|<$ 0.9. The VPD detector has two parts, sitting along the beam pipe on both sides of the STAR
detector at $\pm$5.7~m from the center. The detectors cover a pseudorapidity range of 4.4$<|\eta|<$5.1~\cite{vpdnim}. The time stamps recorded by the VPD and the
BTOF detectors are used to calculate the particle time-of-flight (tof).
The tof is further combined with the track length and momentum, both measured by the TPC, to identify
charged particles. The timing resolution of the TOF system, including the start
timing resolution in Au+Au 200\,GeV collisions, is less than 100~ps.

\subsection{Trigger Definitions}

The minimum bias trigger in Au+Au 200\,GeV collisions for the 2010 and 2011 runs was defined as
a coincidence between the two VPDs and an online collision-vertex cut in order to
select collision events that took place near the center of the detector. The 
central trigger in the 2010 Au$+$Au collisions includes the ZDC detectors, 
located on both sides of the STAR detector at approximately $\pm$18~m. 
This trigger requires a small signal in the ZDC detectors in combination with a large hit multiplicity
in the BTOF and corresponds to the top 10\% of the total hadronic cross section.

\section{Analysis Technique}

\subsection{Event Selection and Centrality Definition}

Events used in this analysis were required to have a reconstructed
collision vertex (primary vertex) within 30 cm of the TPC center along the beam
direction in order to ensure uniform TPC acceptance. To suppress the chance of selecting
the wrong vertex from different bunch-crossing collisions and to ensure that
the selected event indeed fired the trigger, the difference between event vertex
$z$-coordinate $V^{\rm TPC}_z$ and the $V^{\rm VPD}_z$ calculated from the VPD timing was required
to be within 3~cm. These selection criteria yield 240M (year 2010) and 490M (year 2011) 0-80\% minimum-bias triggered events and 220M (year 2010) central triggered (0-10\%) Au+Au events at \sNN = 200\,GeV. 
The results reported in this paper are from the combined year 2010 and year 2011 data.

\begin{figure}
\centering{
\includegraphics[width=0.5\textwidth] {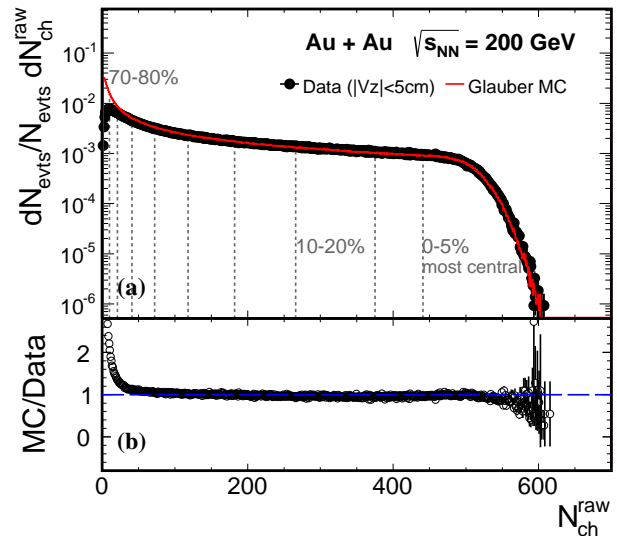}}
\caption[]{Upper panel: Uncorrected multiplicity $N_{\rm ch}^{\rm raw}$
  distribution measured within $|\eta| < 0.5$ and $|V_z|<$ 5 cm. The solid curve
  depicts the multiplicity distribution from a MC Glauber simulation. Bottom
  panel: Ratio between MC and data.}
 \label{refMult}
\end{figure}

Centrality in Au$+$Au 200\,GeV collisions was defined using the
uncorrected charged particle multiplicity $dN/d\eta$ within $|\eta|<0.5$. The
$dN/d\eta$ distribution was then compared to a Monte Carlo (MC) Glauber
calculation in order to delineate the centrality bins. 
Furthermore, the dependence of $dN/d\eta$ on the collision vertex position $V_z$ and the beam 
luminosity has been included in order to take acceptance and efficiency changes on the measured $dN/d\eta$ into account.
The measured uncorrected $dN/d\eta$ distribution from Au$+$Au 200~GeV minimum-bias events 
collected in year 2010 is shown in Fig.~\ref{refMult}. 
The $dN/d\eta$ distributions are from the $V_z$ region of $-5<V_z<5$~cm and extrapolated to a zero ZDC-coincidence rate, 
so as to correct for the detector acceptance and efficiency dependence on the $V_z$ and luminosity. 
The measured distribution matches the MC Glauber calculation well for $dN/d\eta>100$. 
In the lower multiplicity region, the VPD trigger becomes less efficient.
The bottom panel shows the ratio between MC and measured data. 
The centrality bins are defined according to the MC Glauber distribution in order to
determine the centrality cut on the measured $dN/d\eta$. 
To obtain the real minimum-bias sample, events in the low multiplicity region have been weighted with the ratio
shown in Fig.~\ref{refMult} (bottom panel) to account for the VPD inefficiency.


The average number of participants
$\la$\Npart$\ra$, and number of binary collisions $\la$\Nbin$\ra$ from MC Glauber
simulations of Au$+$Au at \sNN\ = 200\,GeV are listed in Table~\ref{CentralityiDf}.

\bt
\caption{Summary of centrality bins, average number of participants
  $\la$\Npart$\ra$, and number of binary collisions $\la$\Nbin$\ra$ from MC
  Glauber simulation of Au$+$Au at \sNN\ = 200\,GeV. The errors indicate
  uncertainties from the MC Glauber calculations.}
\centering
\begin{tabular}{c|c|c}
\hline
	\hline
	Centrality  & $\la$\Npart$\ra$ & $\la$\Nbin$\ra$  \\ 
	\hline                                                  
	0-10\%            & \hspace{0.05in}  $325.5\pm3.7$  \hspace{0.05in}   & \hspace{0.05in}   $941.2\pm26.3$   \hspace{0.05in}       \\ 
	\hline
	10-40\%                  &   $174.1\pm10.0$     &  $391.4\pm30.3$          \\ 
	\hline
	40-80\%                  &   $41.8\pm7.9$     &   $56.6\pm13.7$           \\ 
	\hline
	\hline                                                  
	0-80\%                  &   $126.7\pm7.7$     &   $291.9\pm20.5$          \\ 
	\hline                                                  
	\hline
\end{tabular}
\label{CentralityiDf}
\et

\subsection{Track Selection}

Electron candidate tracks used in this analysis were required to satisfy
the following selection criteria:

\begin{itemize}
\item the number of fit points in the TPC (nHitsFits) should be greater than 20 (out of a maximum of 45) to ensure good momentum resolution;
\item the ratio of the number of fit points over the number of possible points
  should be greater than 0.52 in order to avoid track splitting in the TPC;
\item the distance of closest approach (DCA) to the primary vertex should be less than 1~cm
  in order to reduce contributions from secondary decays;
\item the number of points used for calculating $\langle
  dE/dx\rangle$ (nHitsdEdx) should be greater than 15 to ensure good $dE/dx$ resolution;
\item the track should match to a valid TOF hit with the projected position within TOF's sensitive readout volume.
\end{itemize}

\subsection{Electron identification}

\begin{figure}
\centering{
\includegraphics[width=0.5\textwidth] {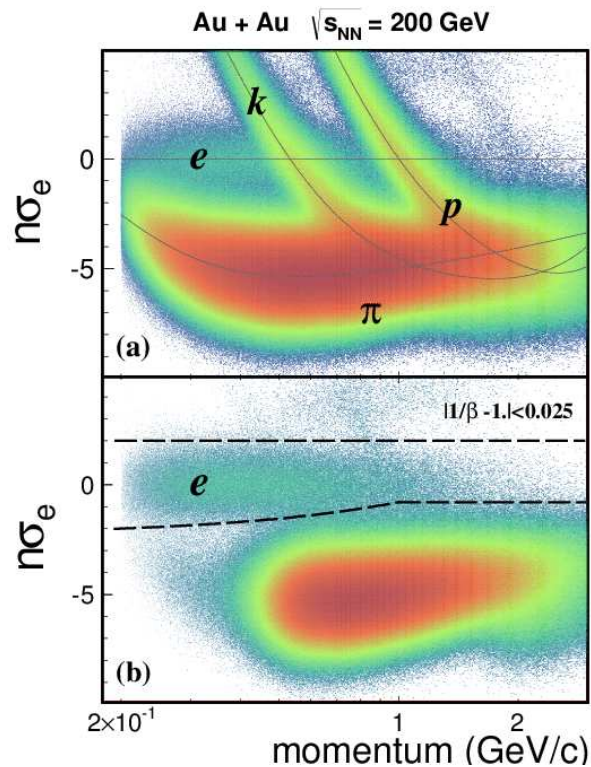}}
\caption[]{(Color online) Upper panel: normalized $dE/dx$ ($n\sigma_e$)  vs. momentum ($p$) distributions for all charged particles. Bottom panel: $n\sigma_e$ vs.
  $p$ distributions after applying the TOF velocity cut $|1/\beta-1|<0.025$.}
 \label{fig1dEdx}
\end{figure}

Electrons (including positrons if not specified) were identified based on a
combination of the TPC and TOF detectors. The electron identification procedure
has been described in~\cite{PIDNIM}. 
In low multiplicity collisions, electrons can be cleanly separated from hadrons by requiring a TOF velocity cut and using the TPC truncated mean ionization energy loss $dE/dx$ dependence on particle momentum.
However, the situation becomes more complicated in high multiplicity Au$+$Au collisions. 
The normalized $dE/dx$ is defined as follows:
\begin{eqnarray}
n\sigma_{e}& = & \frac{ \ln\big( \langle dE/dx \rangle^{\text{m}}  / \langle dE/dx \rangle^{\text{th}}_e\big)}  {R_{dE/dx}}
\label{EQdEdx}
\end{eqnarray}
where $\langle\rangle^\text{m}$ and $\langle\rangle^\text{th}$ represent measured and
theoretical values, respectively, and $R_{dE/dx}$ is the experimental $dE/dx$
resolution.
The $n\sigma_e$ {\it vs.}\ $p$ distributions for the 2010 data are shown in Fig.~\ref{fig1dEdx}. 
The upper panel shows the distribution for all charged particles, 
the bottom panel shows the distribution after applying the TOF velocity selection $|1/\beta-1|<0.025$, 
which accepts about 95\% of the electrons based on the TOF timing resolution.
Despite the TOF velocity selection, there are still some
slow hadrons that contribute to the electron band in this distribution. The source
of these remaining slow hadrons is described in the following paragraphs.

The $n\sigma_e$ distribution for TPC tracks with matched TOF hits are shown in Fig.~\ref{fig1dEdx}. 
For most cases where TOF hits are correctly associated with the
charged particle tracks, one would have a meaningful particle velocity
measurement that can then be used for particle identification. There are
also many TOF hits from electrons that originate from photon conversions in
the material between the TPC and TOF sensitive detector volumes. Since photons
do not leave a trace in the TPC, these TOF hits can be randomly associated with TPC tracks especially in high-multiplicity events.

\begin{figure}
\centering{
\includegraphics[width=0.5\textwidth] {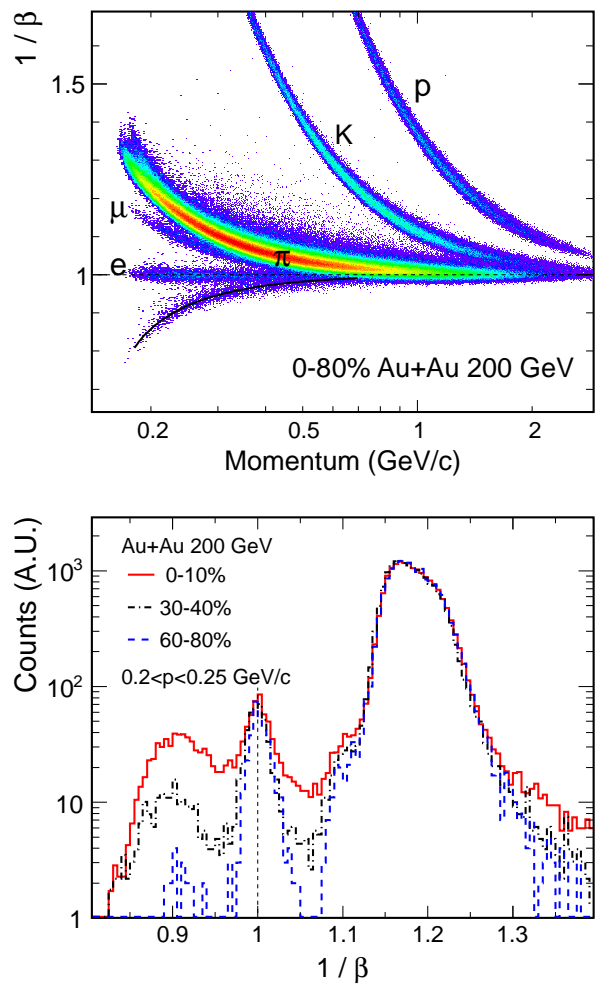}}
 \caption[]{(Color online) Upper panel: $1/\beta$ {\it vs.} particle momentum. Solid
   line depicts a prediction for those associations where TOF hits were triggered
   by conversion electrons while matched randomly with TPC charged tracks.
   Bottom panel: $1/\beta$ projection in the momentum bin $0.2<p<0.25$\,GeV/$c$
   for three centrality bins, normalized to the pion peak region.}
 \label{fig2beta}
\end{figure}

The inverted particle velocity ($1/\beta$) measured by the TOF (time) and the TPC (path length) 
versus the particle momentum ($p$) measured by the TPC is shown in the upper panel of Fig.~\ref{fig2beta} for 
all TPC-TOF associations in Au$+$Au collisions at \sNN = 200\,GeV. 
The band below $1/\beta=1$ depicts the associations between conversion electron TOF hits
and random TPC charged tracks. The bottom panel of Fig.~\ref{fig2beta} shows the
$1/\beta$ distributions in the momentum range $0.2<p<0.25$\,GeV/$c$ for three
centrality classes. The three distributions are normalized to the pion peak. One
can see that with increasing multiplicity that the fake association fraction increases substantially. 
These random associations were further confirmed using Monte Carlo (MC) {\sc Geant} \cite{geant:321} simulations.

As mentioned before, the TOF-based velocity of particles depends on 
the time-of-flight measurements from the TOF detector and the track length determined by the TPC. 
For particles from secondary vertex decays ({\it e.g.}, $\pi, K,$ and $p$ from $K^{0}_{S}, \Lambda,$ 
and $\Omega$ decays), the track length and time-of-flight measurements have some offset, 
which leads to uncertainties when calculating the velocity. 

\begin{figure}
\centering{
\includegraphics[width=0.5\textwidth]{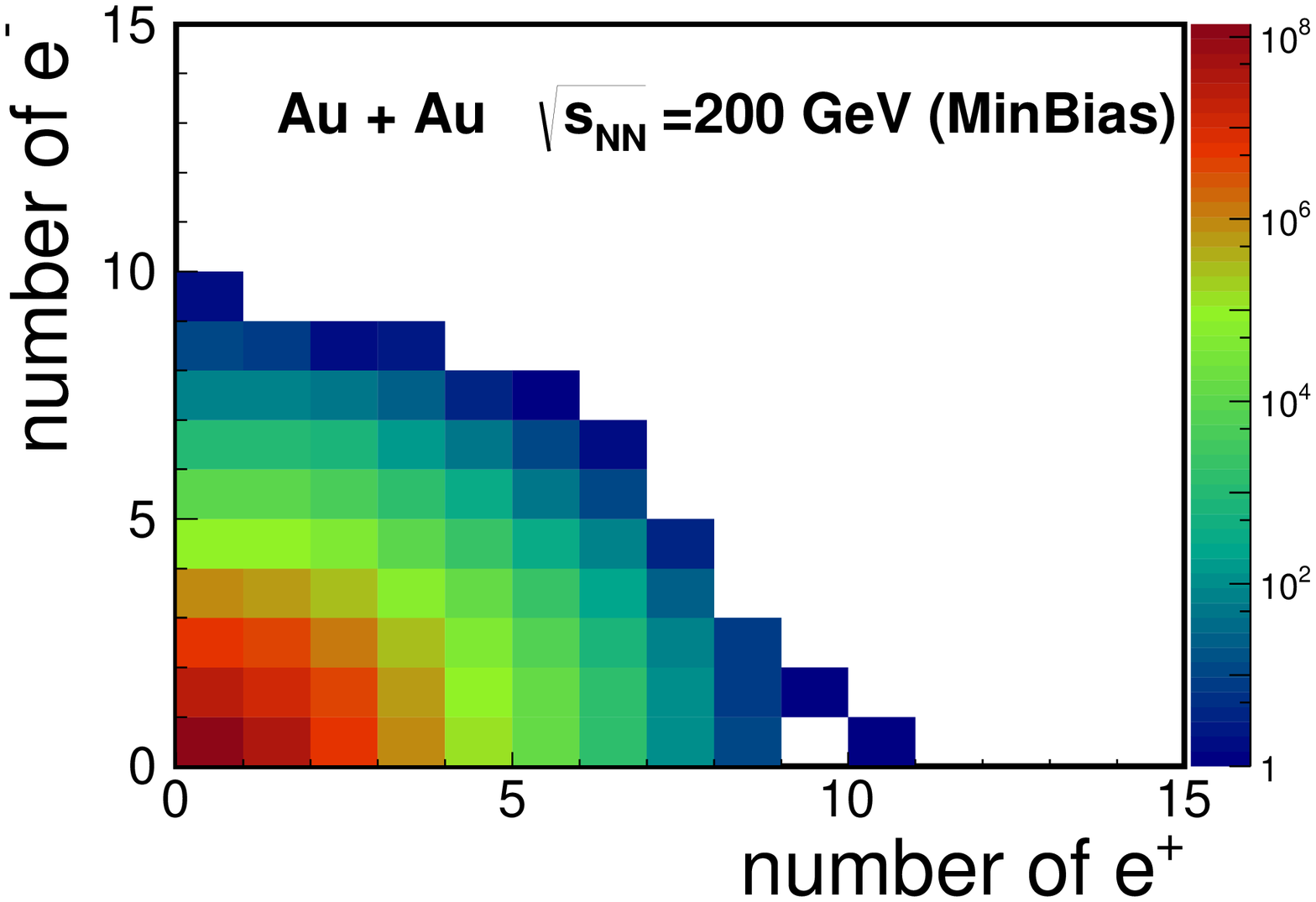}}
\centering{
\includegraphics[width=0.5\textwidth]{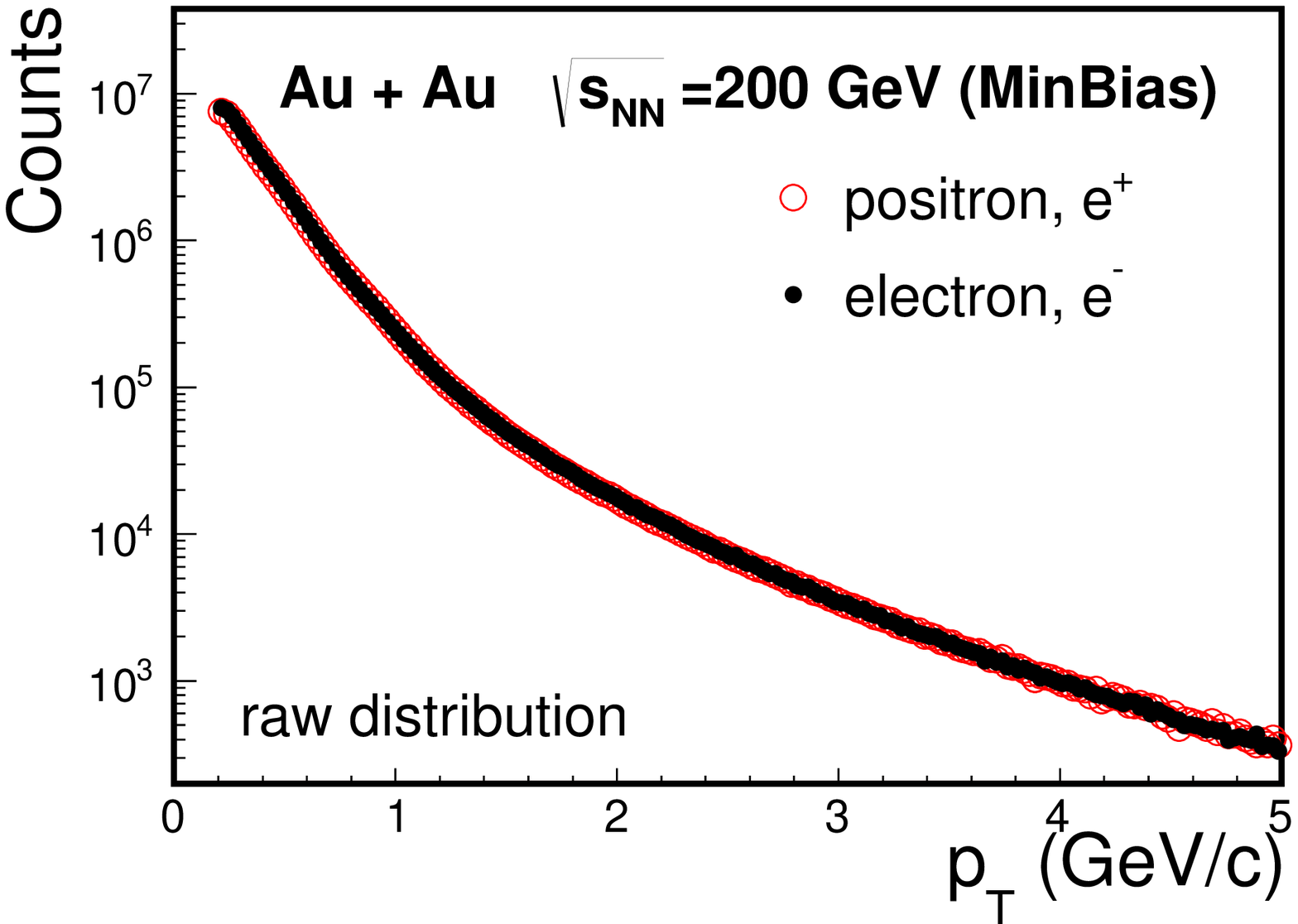}}
\caption[]{Upper panel: The distributions of number of electron candidates. 
  Bottom panel: Raw $p_{T}$ spectra of the electron candidates. 
}
\label{rawelectron}
\end{figure}

Consequently, the applied particle velocity cut cannot remove 
the random association of charged hadron tracks with TOF signals
and the particles from secondary vertex decays.
Such hadrons are mostly at momentum of 400 MeV/$c$ or above where the hadron $dE/dx$ bands cross the electron band. 
These hadrons remain in the $dE/dx$ {\it vs.} $p$ distribution in the lower panel of Fig.~\ref{fig1dEdx} and 
introduce an additional hadron background in the sample of selected electron candidates in the region 
where the electron $dE/dx$ band crosses with hadrons (mostly kaons and protons). 
The dashed black lines in the lower panel of Fig.~\ref{fig1dEdx} depict the $dE/dx$ cuts 
to select the single electron candidates in this analysis. 
Finally, distributions of the number of selected electron candidates are shown in upper panel of Fig.~\ref{rawelectron},
their raw $p_{T}$ spectra are also shown in the lower panel of Fig.~\ref{rawelectron}.

\subsection{Electron Purity and hadron contamination}

\begin{figure}
\centering{
\includegraphics[width=0.5\textwidth] {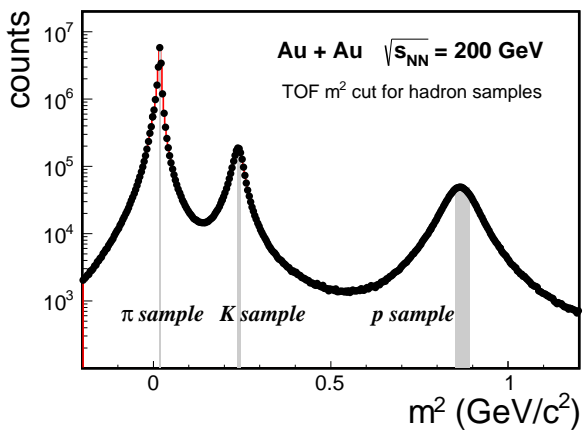}}
\caption[]{Charged particle $m^{2}$ distribution from TOF measurements in 200~GeV Au$+$Au minimum bias
  collisions. The shaded areas are the respective $m^{2}$ thresholds used for selecting high purity $\pi, K, p$ samples.}
 \label{TOFm2}
\end{figure}

\begin{figure}
\centering{
\includegraphics[width=0.5\textwidth] {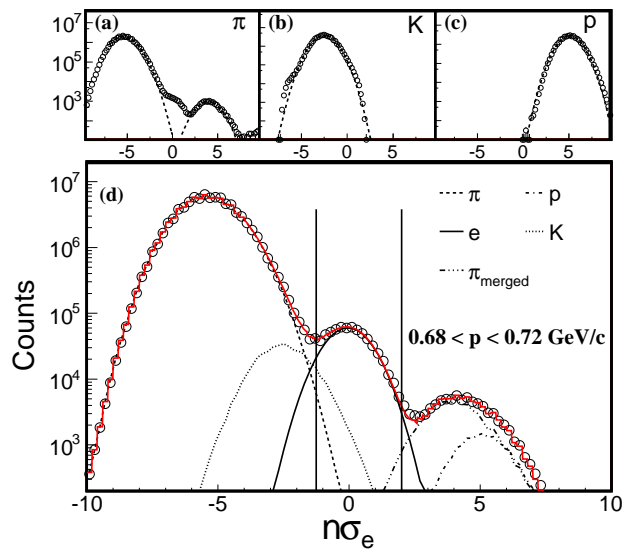}}
\caption[]{Upper panel: $n\sigma_{e}$ distribution of clean $\pi, K,
  p$ samples selected using TOF masses. Dashed lines are Gaussian fits to these distributions allowing the extrapolation of the tail regions where contamination becomes apparent. Bottom
  panel: An example of multi-component fit to the $n\sigma_{e}$ distribution
  for the momentum bin $0.68<p<0.72$\,GeV/$c$ in 200~GeV Au$+$Au minimum bias
  collisions. The two vertical lines indicate
  the selected electron range.}
 \label{fig3efit}
\end{figure}

The $n\sigma_e$ {\it vs.} $p$ distribution after the TOF velocity selection has been shown
in Fig.~\ref{fig1dEdx}. We have performed a multi-component fit to the $n\sigma_e$
distribution for individual momentum slices in order to decompose the yields of each particle
species, and thus derive the electron purity and hadron contamination for
a certain $n\sigma_e$ cut. The $n\sigma_e$ distribution for electrons is
assumed to be Gaussian, with its position and shape determined by selecting
conversion electrons using an invariant-mass reconstruction. The positions and
shapes of the $n\sigma_e$ distributions for pions, kaons, and protons were
determined by selecting pure samples of these particles with particle masses
calculated from the TOF detector. 
Figure~\ref{TOFm2} shows the respective $m^2$ thresholds for the pure hadron samples.
The positions and shapes of all components are kept fixed during the fits, 
leaving only the individual yields as free parameters to fit the $n\sigma_e$
distribution slices in Fig.~\ref{fig1dEdx}. 
An example of the fit result for the momentum bin of $0.68<p<0.72$\,GeV/$c$ is shown in Fig.~\ref{fig3efit}.
The black dotted curve at high $n\sigma_e$ region depicts a small contribution of tracks
from merged pions in the TPC. The $\langle dE/dx \rangle$ value of these tracks
are twice that of normal pion tracks, thus its position and shape is
predictable from the pion $n\sigma_e$ distribution. For completeness, we have included this
contribution in the fit although it is well separated from the electron peak. 

\begin{figure}
\centering{
\includegraphics[width=0.5\textwidth] {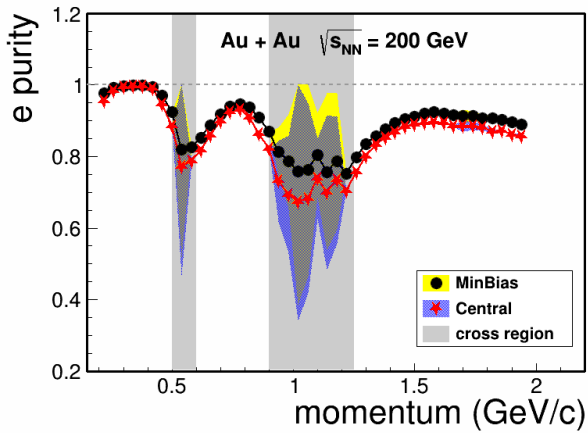}}
\caption[]{(Color online) Estimated electron purity {\it vs.} momentum in 200\,GeV Au
  + Au collisions. Gray areas indicate the momentum regions where $n\sigma_{e}$
  of kaons and protons cross with that of electrons resulting in large uncertainties in those ranges.}
 \label{fig4purity}
\end{figure}

The multi-component fits describe the full distributions  well in the regions 
where the slow hadron peaks can be separated from the electron peaks.
In the region where kaons and protons start to overlap with electrons, 
we use the hadron yields from neighboring momentum bins with clean particle identification 
to interpolate the expected hadron yield.
The systematic uncertainties on the electron purity in these overlapping bins were estimated by 
comparing the yields to the results from the free paramter fit, which take the hadron yield as free parameter. 
Figure~\ref{fig4purity} shows the electron purity for the
candidate samples used in the minimum-bias and central collisions. 
As expected, hadron contamination increases from peripheral to central collisions.
The electron purity integrated over the region of 0.2$<$\pT$<$2.0~GeV/$c$ is (94.6$\pm$1.9)\% and (92.1$\pm$2.0)\% for 0-80\% minimum bias and 0-10\% central Au+Au collisions, respectively. The impact of hadron contamination on the dielectron spectra will be further discussed in Section \ref{Sect:SystematicUncertainties}.

\subsection{Electron pairing and background subtraction}

\begin{figure}
\centering{
\includegraphics[width=0.5\textwidth]{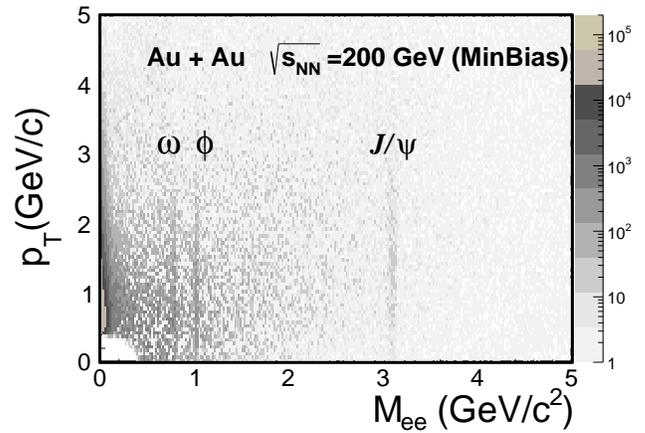}}
\caption[]{Two-dimensional ($M_{ee}$, $p_{\rm T}$) distribution of 
  unlike-sign $e^+e^-$ pairs with background subtraction from 200\,GeV Au + Au minimum bias 
  collisions in the STAR acceptance (\pT$>$0.2\,GeV/$c$, $|\eta|$$<$1, and $|y_{ee}|$$<$1).}
\label{bg2}
\end{figure}

For each individual event, all electron and positron candidates within the STAR acceptance of \pT$>$ 0.2\,GeV/$c$ 
and $|\eta|<$1 are combined to generate the inclusive unlike-sign pair ($N_{+-}$) invariant-mass distribution. 
Despite slight acceptance differences between the TPC and the TOF, 
the collision vertex distribution along beam direction ($z$) will provide finite acceptance 
and efficiency for charged tracks up to $|\eta|<$1. Therefore we used $|\eta|<$1 in this analysis and 
the dependence of efficiency and acceptance along $\eta$ has been corrected in the final spectra.
In Fig.~\ref{bg2}, a two-dimensional distribution in invariant-mass and 
transverse-momentum ($M_{ee}$, $p_{T}$) of $N_{+-}$ pairs is shown in the STAR acceptance 
with $|y_{ee}|<$1 (electron pair rapidity) and the background subtracted. 
Vector meson signals ($\omega$, $\phi$, and $J/\psi$) are fairly easy to recognize after the background subtraction.
All distributions shown in this paper are calculated within the same STAR acceptance 
including $|y_{ee}|<$1 unless specified otherwise.

In this analysis the signal ($S$) is defined as the $e^+e^-$ pairs that originate from pair production sources such as $\pi^{0}$, $\eta$, $\eta'$, $\rho$, $\omega$, $\phi$, J/$\psi$, $\gamma*$ decays, as well as correlated charm hadron decay. 
Background sources that contribute to the inclusive unlike-sign pair
distributions include:

\begin{itemize}

\item Combinatorial background pairs from two uncorrelated
  electrons.

\item Correlated background pairs. For instance, in the case of Dalitz decays
  followed by a conversion of the decay photon({\it e.g.}, $\pi^{0} \rightarrow e^{+}
  e^{-}\gamma$, then $\gamma Z \rightarrow e^{+}e^{-}Z^{*}$), the electron
  from the Dalitz decay and the positron from the conversion are not completely
  uncorrelated as they originate from the same source. Another significant
  contribution is the electron pairs from same-jet fragmentation or
  back-to-back di-jet fragmentation. This source may become more significant at
  high mass or \pT.

\end{itemize}

Contributions from uncorrelated and correlated background pairs are thoroughly studied and evaluated using like-sign pairs, $N_{++}$ and $N_{--}$, constructed from the same event.
It has been demonstrated that when the $e^+$ and $e^-$ are produced in statistically independent
pairs, the geometric mean of the like-sign pairs $2\sqrt{N_{++}\times N_{--}}$
fully describes the background in the inclusive unlike-sign pair distribution
$N_{+-}$~\cite{PHENIX}. In this analysis, we consistently used the like-sign
distribution $2\sqrt{N_{++}\times N_{--}}$ to estimate or normalize the background distribution. 
The mixed-event unlike-sign distribution $B_{+-}$ was constructed 
to estimate the combinatorial background and was used for a better statistical 
background estimation wherever the correlated background is negligible or the mixed-event unlike-sign distribution 
agrees with the same-event like-sign distribution $2\sqrt{N_{++}\times N_{--}}$. 
Mixed-event like-sign pair distributions $B_{++}$ , $B_{--}$ were also constructed to
verify the applicable kinematic region for the mixed-event technique as well as to define the normalization 
factor for the mixed-event unlike-sign distribution.

A sizeable component of the correlated electron pairs, that is not considered as part of the final signal distribution, originates from photon conversions in the detector material.
Details of the conversion electron removal will be discussed in subsection III-E.1. 

Hadron contamination in the selected electron/positron sample due to particle
misidentification may result in some residual contributions to the final signal
distribution. Most of these are from resonance decays. 
The high purity of the electron sample in this analysis allows us to demonstrate that the residual 
contribution due to hadron contamination in the final distribution is negligible. 
Such details will be discussed in the Section III-H. 


\subsubsection{Photon conversion removal}
\label{sect:photonconversionremoval}
Background pairs from photon conversion were removed from the sample using
the \phiV\ angle selection method. This method is
similar to that used by the PHENIX collaboration~\cite{PHENIX} and relies on the kinematics of the pair production process. The opening angle between the two conversion electrons should be zero, and the
electron tracks are bent only in the plane perpendicular to the magnetic field
direction, which for the STAR experiment is parallel to the beam direction ($z$).
Unit-vector definitions used for the construction of the \phiV\ angle were taken from~\cite{PHENIX} as:

\begin{equation} 
\begin{split}
&\hat{u} = \frac{{\vec{p}}_{+}+{\vec{p}}_{-}}{|{\vec{p}}_{+}+{\vec{p}}_{-}|} ,
\hat{v}={\vec{p}}_{+} \times {\vec{p}}_{-}  \\ &\hat{w} = \hat{u} \times
\hat{v} , {\hat{w}}_{c} = \hat{u} \times \hat{z} \\ &\cos{\phi}_{\rm V} =
\hat{w}\cdot{\hat{w}}_{c} \\ 
\end{split}
\label{EQphoton}
\end{equation}
where $\vec{p}_{\pm}$ are momentum vectors of $e^\pm$ tracks, and
$\hat{z}$ is the magnetic field direction.

\begin{figure}
\centering{
\includegraphics[width=0.5\textwidth]{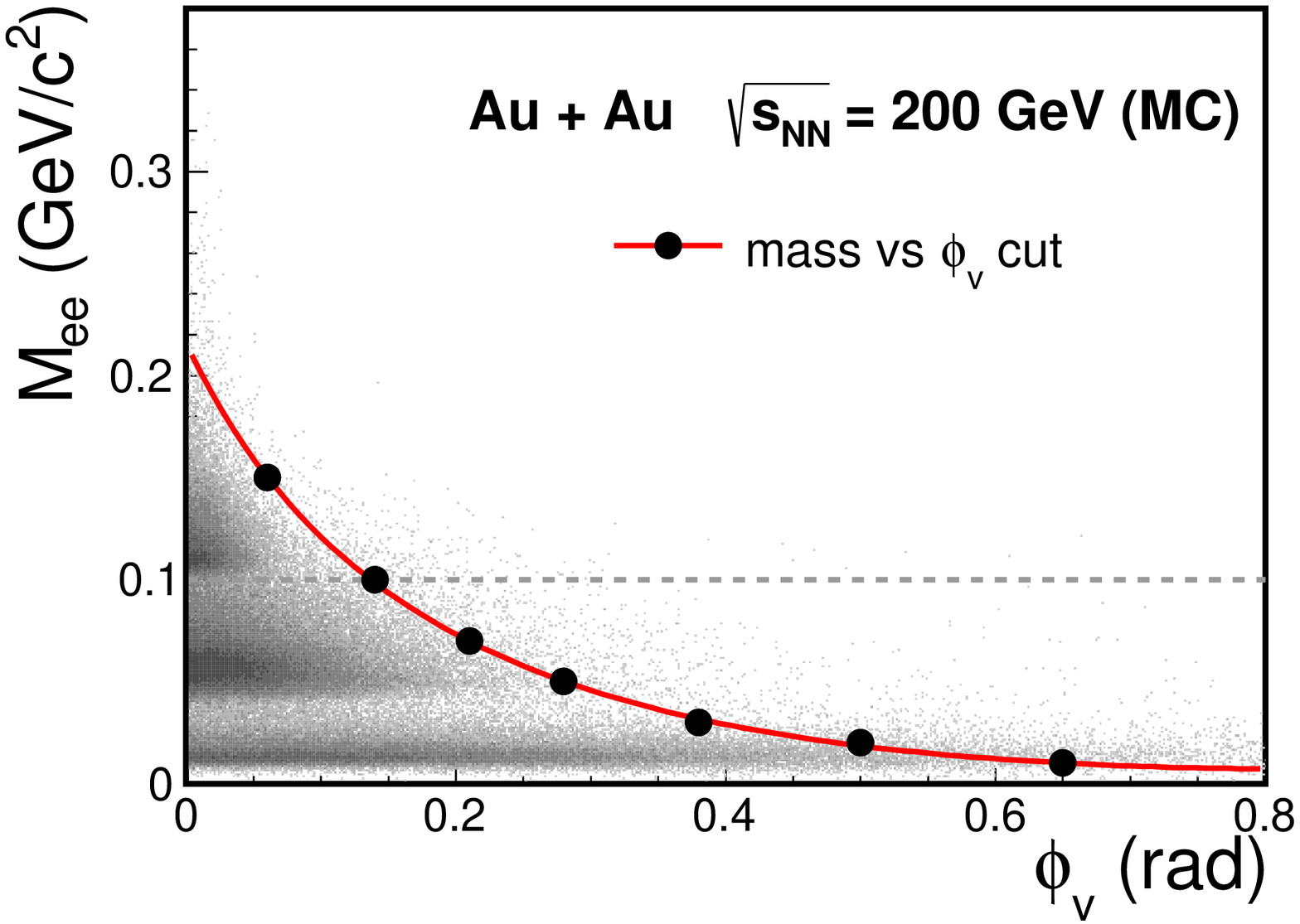}}
\centering{
\includegraphics[width=0.5\textwidth]{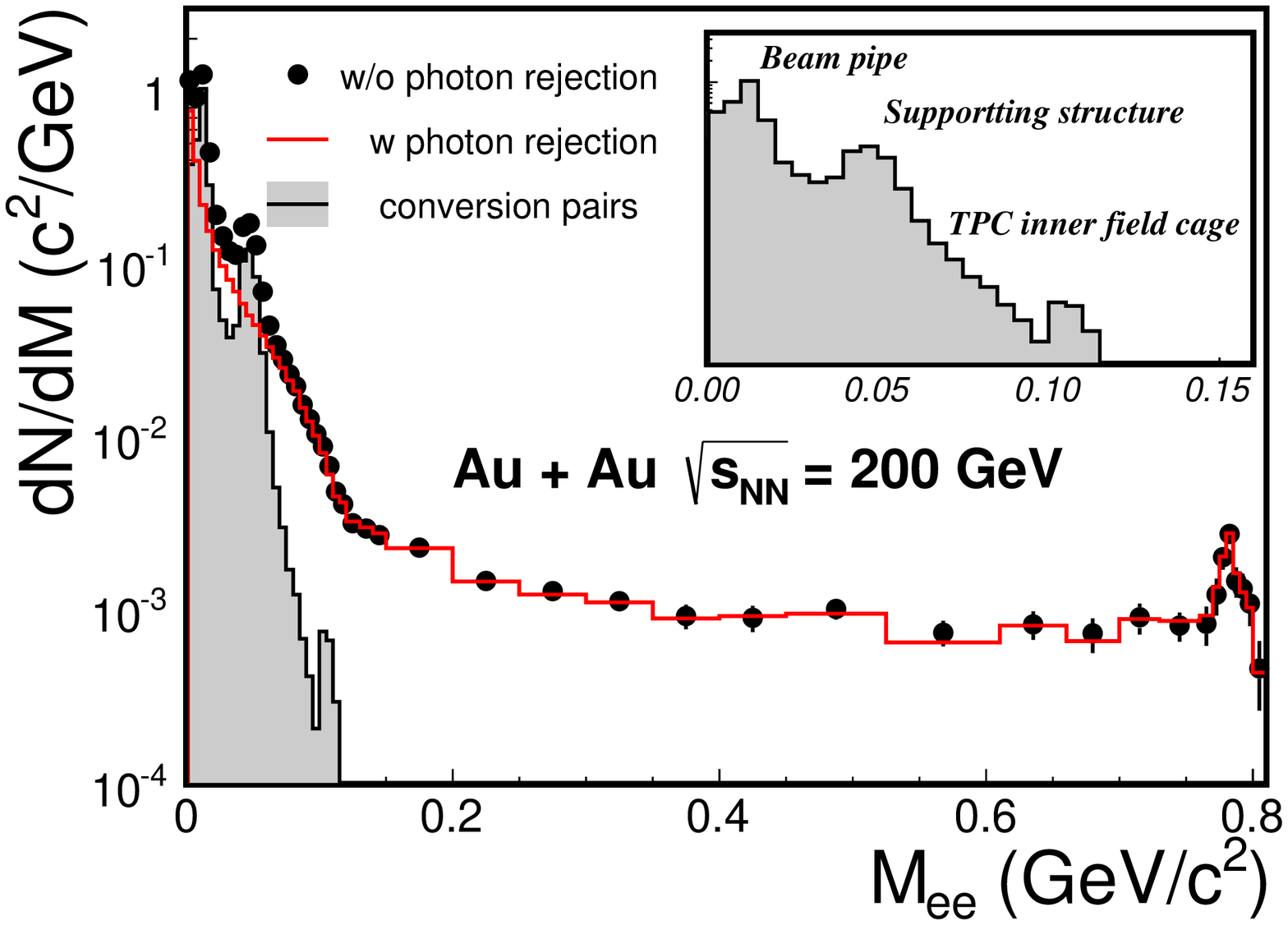}}
\caption[]{Upper panel: \phiV\ {\it vs.}\ mass distributions for photon
  conversion electron pairs form the full {\sc Geant} MC simulation. The solid red
  line depicts the mass dependent \phiV\ cut that was used to remove these
  conversion pairs. Bottom panel: photon conversion contribution in 200~GeV Au$+$Au minimum bias
  collisions. The insert plot shows the structures from the beam pipe, the supporting
  bars of the inner cone, and the TPC inner field cage.}
\label{phiv}
\end{figure}

For pairs that originate from photon conversions \phiV\ should be zero. It has no
preferred orientation for combinatorial pairs, and only very weak dependence for
$e^{+} e^{-}$ pairs from hadron decays. 
The electron pair mass versus \phiV\ for conversion electron pairs from the full {\sc Geant}
simulation of the STAR detector~\cite{geant:321} is shown in the upper panel of Fig.~\ref{phiv}.
The populated bands at different mass positions depict the conversion electron
pairs from different detector materials. 
The reconstructed masses are shifted from zero as the electrons are assumed to originate 
from the primary vertex during the final track reconstruction.
As a result, the three main bands from low to high masses correspond to the
conversions from the beam pipe (at a radius $r\sim$ 4~cm), inner cone
support structure ($r\sim$ 20~cm), and TPC inner field cage (IFC) ($r\sim$ 46~cm).
In order to remove these conversion pairs, we define a mass-dependent \phiV\ selection which is shown as the red line 
in the upper panel of Fig.~\ref{phiv}.
We estimated that more than 95$\%$ conversion pairs are removed by this selection criterion. 

The signal pair invariant mass spectra before and after this photon conversion cut 
are shown in the bottom panel of Fig.~\ref{phiv}; their difference is shown as the filled histogram. 
Like-sign background subtraction was used to obtain these distributions. 
Almost all conversions appear in the mass region below 0.1\,GeV/$c^2$.


The cut removing the photon conversion pairs was applied only in
the very low mass region ($M_{ee}$$<$ 0.2~GeV/$c^2$).
The effect of the cut on the mixed-event distribution normalization is negligible
as that determination is done at a much higher mass region.


\subsubsection{Event mixing}

The event mixing technique was used to reproduce the combinatorial background with improved statistical precision.
In order to make the mixed-event distributions
close to that from real events, we have only selected events with similar 
properties for the mixed-event calculation. The full sample is divided into 
different pools according to the following event level properties:
multiplicity, vertex position, event plane angle, and magnetic field direction. 
The sorting by event multiplicity and vertex position 
ensures electrons are mixed between events with similar detector acceptance and
efficiency. This technique has been widely used in many other STAR analyses for
the reconstruction of the combinatorial backgrounds~\cite{STARKstarPhiD}. The small
signal-to-background ratio requires a very good understanding of the mixed-event distribution in the dielectron analysis. 
Its dependence on the event pool division for event-plane angle and magnetic
field direction were studied in detail and are presented here.

\begin{figure}
\centering{
\includegraphics[width=0.5\textwidth]{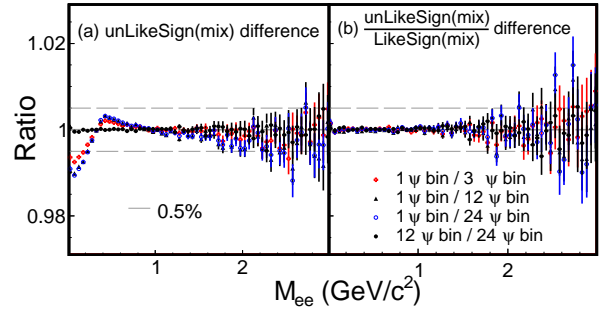}}
\caption[]{(Color online) Difference in mixed-event distributions when using
  different number of event pools in the reconstructed event-plane angle. The left
  plot shows differences of unlike-sign distributions with different divisions
  and the right plot shows differences of ratios of unlike-sign to like-sign
  distributions.}
\label{poolmb}
\end{figure}

Elliptic flow measurements~\cite{FlowMethod} in 200~GeV Au$+$Au collisions have shown that 
the momentum phase space distribution of particles produced in the event is approximately elliptical. 
Therefore, we only mix events with similar event-plane direction to ensure the events have similar 
momentum phase space alignment,
and further guaranteed by the multiplicity assortment to ensure the events have similar momentum phase space distributions. 
The event plane was reconstructed with a conventional method using 
tracks in the TPC ($0.1<p_T<2$\,GeV/$c$ and $|\eta|<1$) in order to obtain the
second-order event-plane angle $\Psi$~\cite{FlowMethod}. 
In Fig.~\ref{poolmb}, results of a study using minimum bias Au$+$Au
collisions in which mixed-event unlike-sign and like-sign distributions are compared
using different numbers of event pools in event-plane angle are shown.
The figure illustrates several scenarios from 1 up to 24 event pools. 
The dashed lines at $\pm$0.5\% corresponds to a 100\% change in the yield where the signal-to-background ratio is 1/200.
This study shows the importance of doing the division in event-plane angle 
in order to avoid distortion of the mixed-event distributions.
The distortion is quite clear in 
the low mass region ($<$ 1~GeV/$c^2$) and not negligible in the intermediate
mass region (1$-$3~GeV/$c^2$). 
The differences become negligible when the number of event pools is 12 in 200~GeV Au$+$Au minimum-bias collisions, comparable to the TPC $2^\mathrm{nd}$ order event plane resolution. 

A similar study of the centrality dependence for background distributions was carried out.
As a result, to ensure the minimal difference in all centrality bins studied, we
choose 24 event pools in the event-plane angle in our analysis.

The data samples used in this analysis were taken under two different magnetic
field configurations of similar magnitude but opposite direction. 
The acceptance for oppositely charged tracks in the two magnetic field configurations 
is not exactly the same due to a slight offset of the beam
line with respect to the center of STAR detector system. 
Only electrons from events with the same magnetic field configuration were mixed when 
constructing total mixed-event distributions.

The final number of event pools used in track multiplicity, vertex position, event
plane angle, and magnetic field configuration is
$16\times10\times24\times2$ for this analysis of the 200~GeV minimum-bias Au$+$Au data.

\begin{figure}
\centering{
\includegraphics[width=0.5\textwidth]{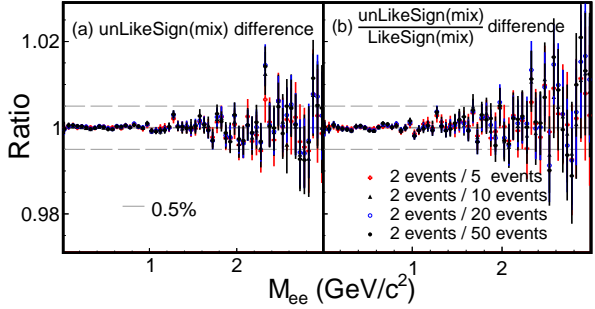}}
\caption[]{(Color online) 
Ratios between mixed-event distributions using different number of events to be mixed with in event buffers.
}
\label{mixbuffer}
\end{figure}

The statistics in the mixed-event distributions depend on the number of events
chosen for the calculation.
In practice, however, the calculation can be done to sufficient precision for every event pool 
with a sizable number of event pools under the limitations of the number of events during the calculation.
The differences between mixed-event distributions with different number of events 
in the buffers are shown in Fig.~\ref{mixbuffer}.
We observe no distortions beyond statistics in our calculation using a buffer of 50 events per event pool. 
With this choice, the statistical uncertainties in the mixed-event background are negligible compared to 
the same event distributions.

\subsubsection{Mixed-event normalization}

The unlike-sign and like-sign pair distributions in the same event ($N_{+-}$, $N_{++/--}$), and in the mixed-event ($B_{+-}$, $B_{++/--}$) were constructed in two dimensions ($M_{ee}$, $p_T$).   
The mixed-event unlike-sign distribution ($B_{+-}$) 
provides the shape of the uncorrelated combinatorial background, with an overall normalization factor determined separately.
The normalization factor was evaluated from the like-sign pair distribution using the technique described in Ref.~\cite{PHENIX}.
This technique is susceptible to a systematic bias if correlated pairs exist in the the like-sign sample. Therefore, the kinematic region used to evaluate the normalization is carefully selected where such correlated backgrounds are negligible.

The procedure to obtain the normalized combinatorial background $B_{+-}^{\rm comb}$
is described in Ref.~\cite{PHENIX} and also shown in the following
Eq.~\ref{EQmix}:

\begin{equation}
\begin{split}
&A_{+}=\frac{\int_{\rm N.R.} N_{++}(M, p_{\rm T}) dMdp_{\rm T}}{\int_{\rm N.R.} B_{++}(M, p_{\rm T}) dMdp_{\rm T}}\\ 
&A_{-}=\frac{\int_{\rm N.R.} N_{--}(M, p_{\rm T}) dMdp_{\rm T}}{\int_{\rm N.R.} B_{--}(M, p_{\rm T}) dMdp_{\rm T}}\\ 
&B_{++}^{\rm norm}=\int_{0}^{\infty} A_{+}{B_{++}(M, p_{\rm T})} dMdp_{T}\\ 
&B_{--}^{\rm norm}=\int_{0}^{\infty} A_{-}{B_{--}(M, p_{\rm T})} dMdp_{\rm T}\\
&B_{+-}^{\rm comb}(M,p_{\rm T})=\frac{2\sqrt{B_{++}^{\rm norm}\cdot B_{--}^{\rm norm}}}{\int_{0}^{\infty} B_{+-} dMdp_{T}}B_{+-}(M,p_{\rm T})\\
\end{split} 
\label{EQmix}
\end{equation}
N.R. denotes the integral calculated in a certain kinematic region, {\it i.e.} the
normalization region. Table~\ref{aNorTb} lists the total like-sign pairs in the
normalization region for each centrality class and the corresponding
statistical uncertainties of the normalization factors.

\bt
\caption{Total Like-sign pairs in the normalization region (N.R.) in each
  centrality class and the corresponding statistical uncertainties of the normalization factors.} 
\centering
\begin{tabular}{c|c|c}
\hline
	\hline
	Centrality  & Like-sign pairs in N.R.  & Statistical un.   \\ 
	\hline
	0-80\%           & $4.2\times10^6$              & $4.8\times10^{-4}$ \\ 
	\hline
	0-10\%           & $8.9\times10^6$              & $3.3\times10^{-4}$ \\ 
	\hline                                           
	10-40\%          & $2.3\times10^6$              & $6.5\times10^{-4}$ \\ 
	\hline                                           
	40-80\%          & $8.0\times10^5$              & $1.1\times10^{-3}$ \\ 
	\hline                               
	\hline
\end{tabular}
\label{aNorTb}
\et

\begin{figure}
\centering
\bmn[b]{0.5\textwidth}
\includegraphics[width=1.0\textwidth]{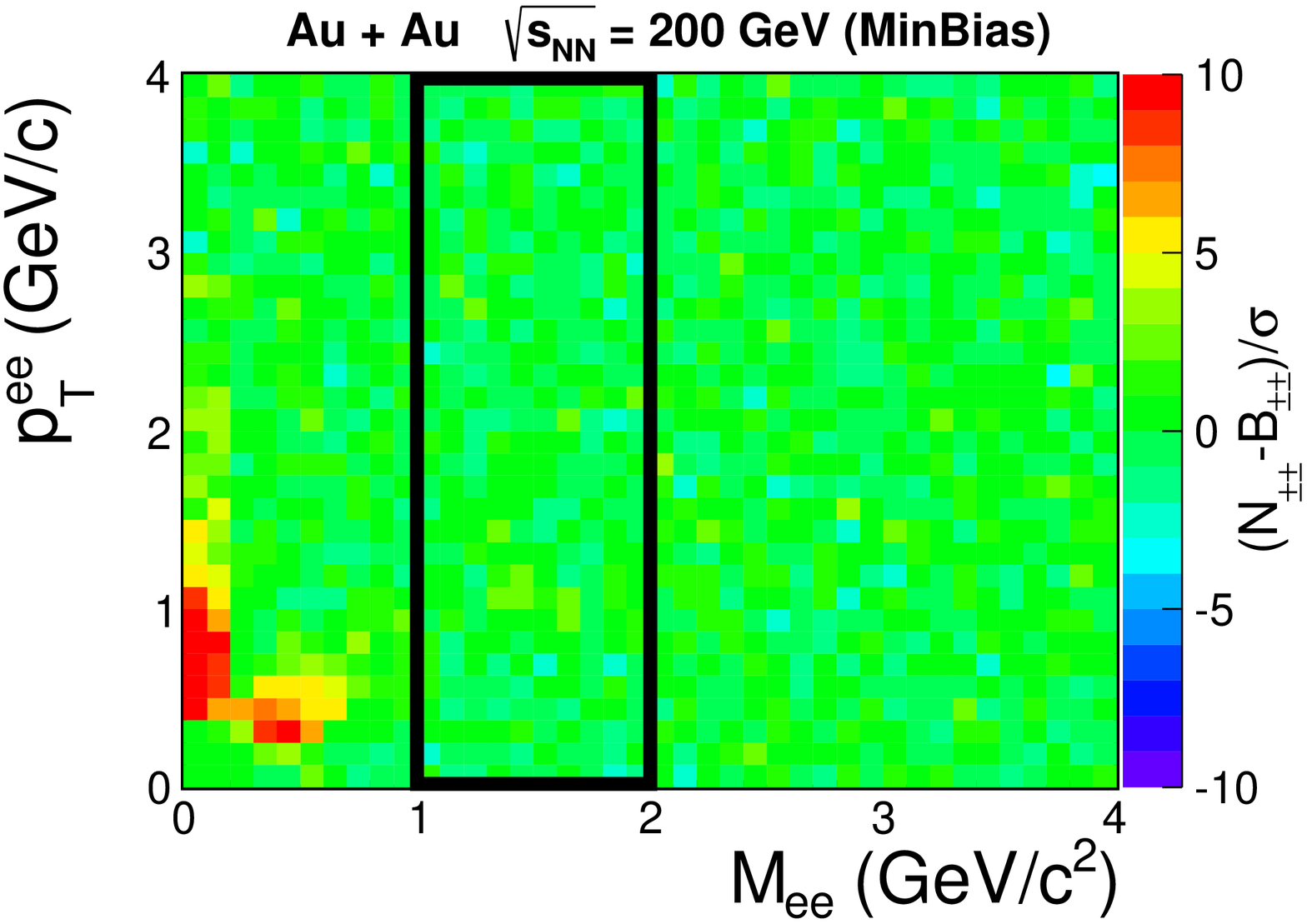}
\emn \\ 
\centering
\bmn[b]{0.5\textwidth}
\includegraphics[width=1.0\textwidth]{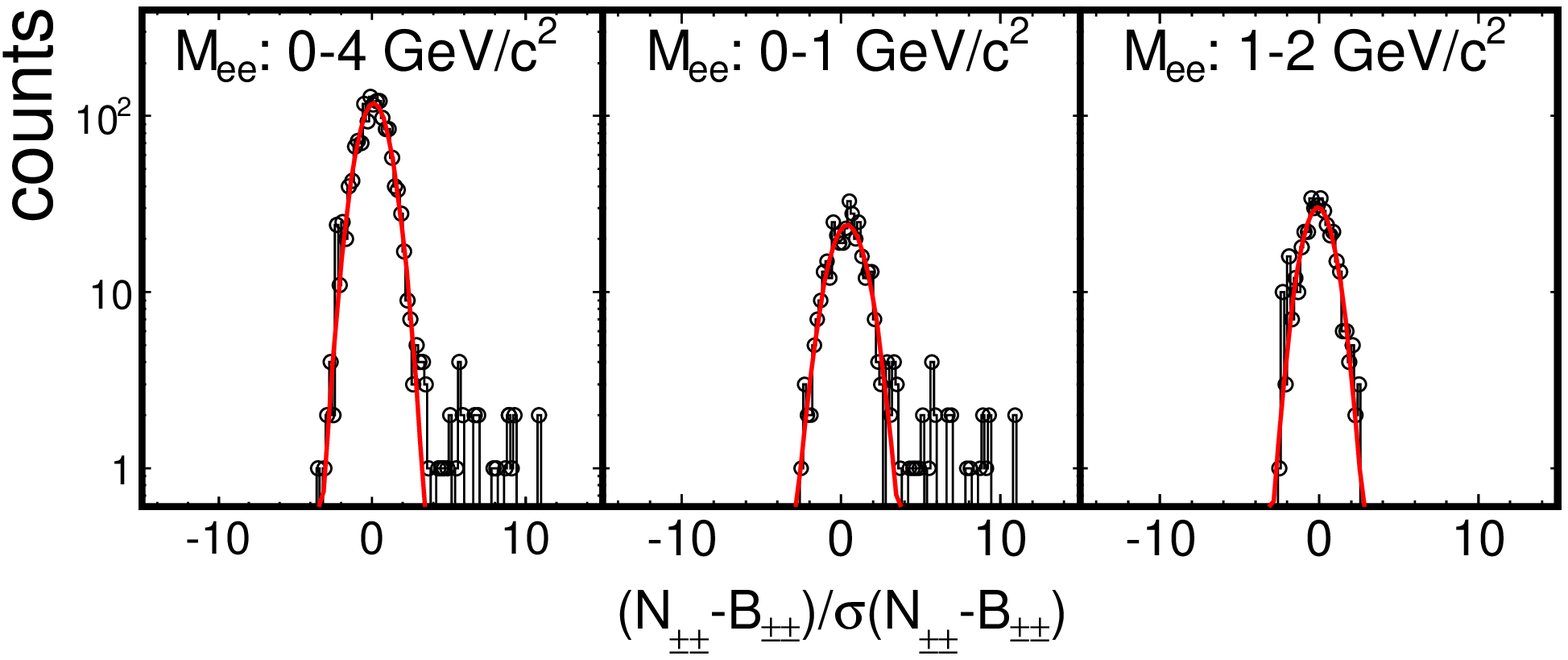}
\emn \\
\caption{(Color online) Upper panel: Residual differences between same-event and
  mixed-events like-sign distributions divided by its standard deviation. The
  black box indicates the default normalization region. Bottom panel: Residual
  difference distributions for all the entries in different mass regions.} 
\label{bgnor}
\end{figure}

The residual difference between same-event like-sign $N_{++,--}$ and the normalized mixed-event $B_{++,--}^{\rm norm}$
as a function of $M_{ee}$ and $p_{\rm T}$ is shown in the upper panel of Fig.~\ref{bgnor}.
The difference is normalized by the expected statistical error in each kinematic bin.  
The residual difference distributions for all entries in different mass regions are shown in the bottom panel of Fig.~\ref{bgnor}.
In the black box  in the upper panel of Fig.~\ref{bgnor}, the normalized residuals follow the
statistical fluctuation.  
We then chose this area $1<M_{ee}<2$\,\GeVcsq as the
normalization region in our analysis. 
The systematic uncertainty introduced by the selected normalization region was studied by varying the selection as will be discussed in more detail in Section III-H.

\begin{figure}
\centering{
\includegraphics[width=0.5\textwidth]{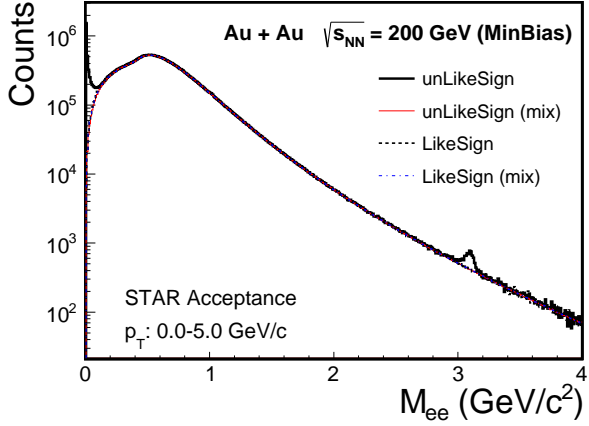}}
\caption[]{(Color online) Raw pair mass distributions for 200\,GeV Au + Au
  collisions. The mixed-event unlike-sign and like-sign distributions are
  normalized in the mass region from 1.0 to 2.0\,GeV/$c^2$.} 
\label{QA1}
\end{figure}

\begin{figure*}
\centering{
\includegraphics[width=0.5\textwidth]{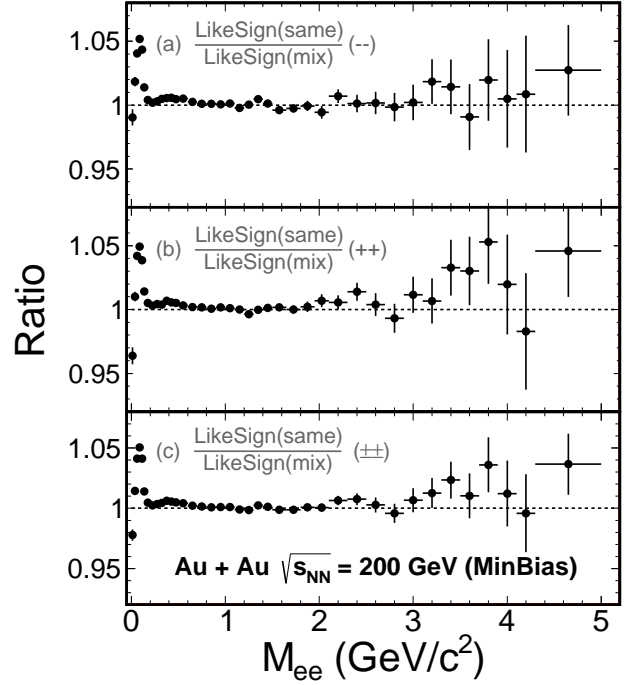}}
\caption[]{Panel (a), (b), (c): Ratios between same event and mixed-event like-sign
  distributions. 
}
\label{QA2}
\end{figure*}

\begin{figure*}
\centering
\bmn[b]{0.5\textwidth}
\includegraphics[width=0.95\textwidth]{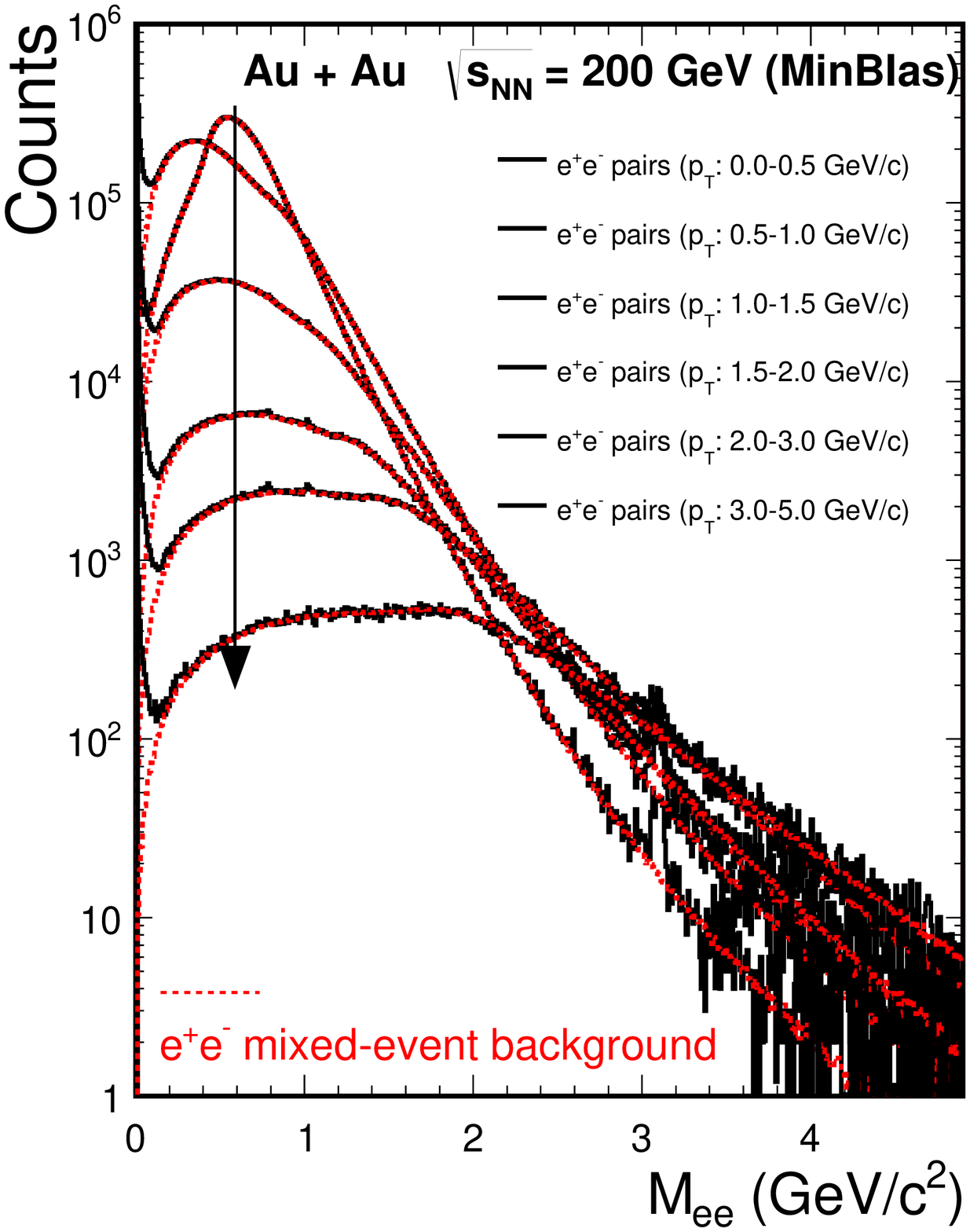}
\emn 
\centering
\bmn[b]{0.5\textwidth}
\includegraphics[width=0.95\textwidth]{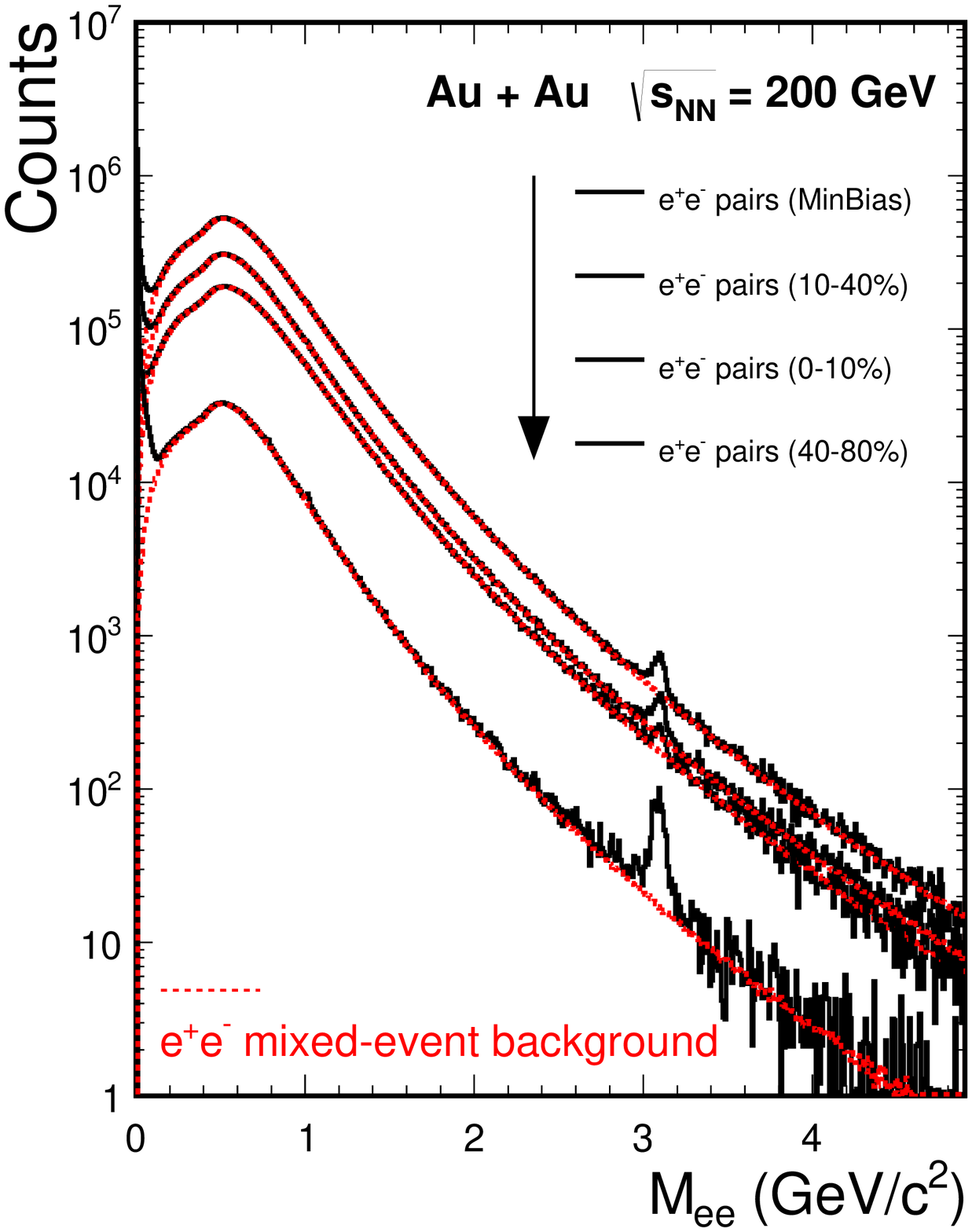}
\emn \\
\caption{(Color online) $p_{T}$ (left panel) and centrality (right panel) dependence of the same-event inclusive unlike-sign distributions (histograms) and the normalized unlike-sign mixed-event background distributions (red lines).}
\label{bg_pt_cent}
\end{figure*}

In Fig.~\ref{QA1}, the raw mass distributions of mixed-event like-sign and unlike-sign pairs in
the full \pT\ region are plotted together with the same event
distributions. To further investigate any residual differences between these
distributions, the ratios between them are plotted in Fig.~\ref{QA2}. Panels (a-c)
show that in the normalization region the residuals are negligible.
The  slight increasing trend in the higher mass region can be attributed to the possible jet-related
correlated background~\cite{PHENIX}. This will be discussed further in
Section III-E.5.
The \pT\ and centrality dependence of the inclusive unlike-sign and the normalized mixed-event mass distributions are shown in Fig.~\ref{bg_pt_cent}.

\renewcommand{\floatpagefraction}{0.75}

\subsubsection{Like-sign and unlike-sign acceptance difference correction}

\begin{figure}
\centering{
\includegraphics[width=0.5\textwidth]{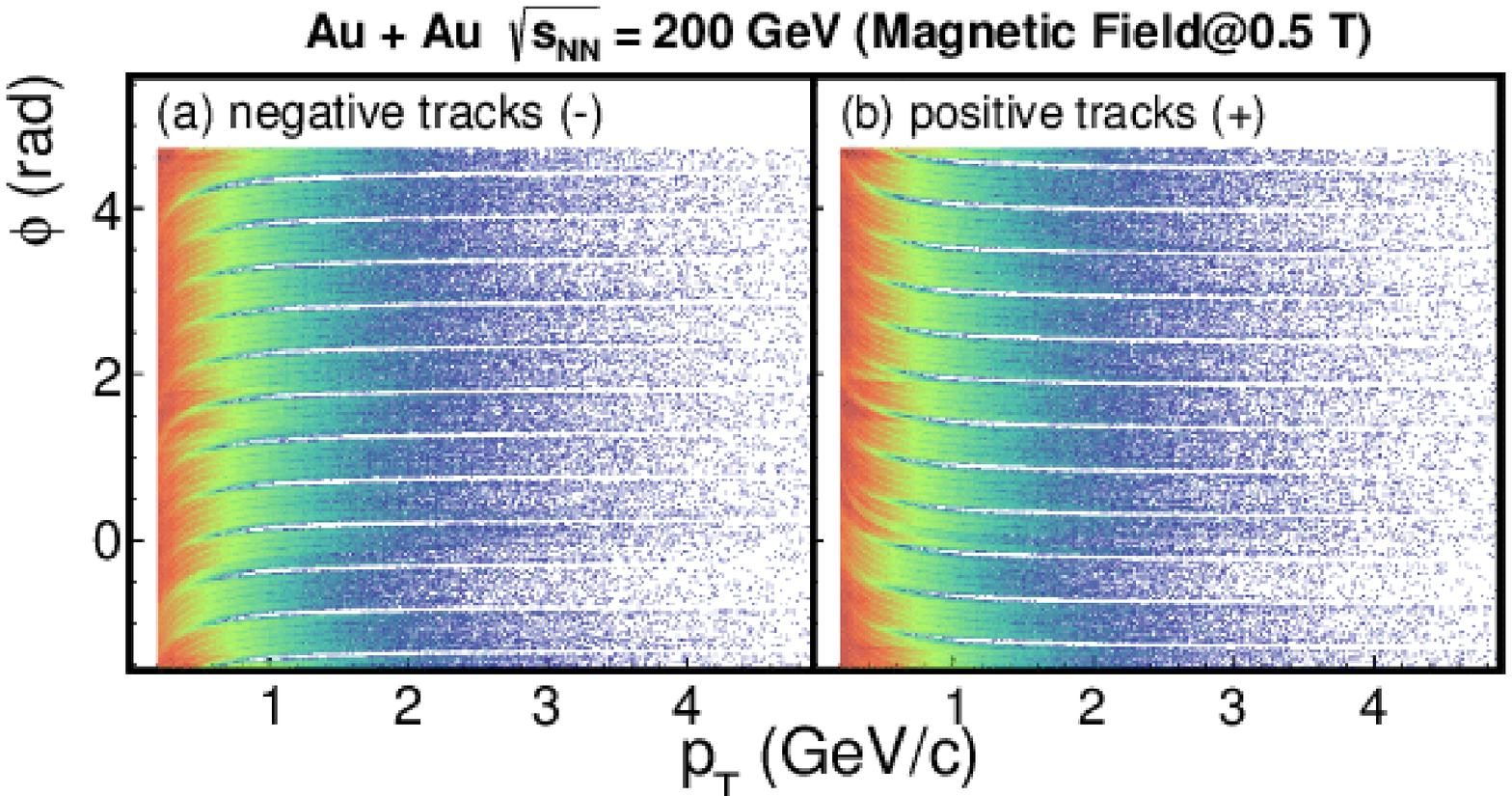}}
\caption[]{(Color online) $\phi$ {\it vs.} \pT\ for all negative (left panel) and positive
  (right panel) tracks from a single magnetic field configuration. The blank areas are due to the TPC sector
  boundaries, which shows the different acceptances between positive and negative tracks particularly in the low \pT\ due to the magnetic field.} 
\label{Acc}
\end{figure}

The like-sign distribution is taken as the best estimate for the background
in the inclusive unlike-sign distribution. 
However, the acceptances for like-sign and unlike-sign pairs differ in the STAR detector due to
the magnetic field. The observed candidate $e^+$ and $e^-$ tracks $\phi$ versus \pT\ are shown in Fig.~\ref{Acc}.
The empty strips along the $\phi$ direction are due to the TPC read-out sector boundaries. These acceptance boundaries and 
local inefficiencies or acceptance holes in the active detecting area will 
results in different acceptances for like-sign and unlike-sign pairs. We used the
mixed-event technique to calculate these acceptance differences. 

The correction factor for the acceptance difference between like-sign and
unlike-sign pairs is obtained as a ratio of the like-sign and unlike-sign distribution from mixed-event.
The ratio was calculated in each ($M_{ee}$, \pT)
bin, and the corresponding correction applied in this 2D plane. The geometric mean from the two like-sign charge
combinations $++$, $--$ describes the background in the unlike-sign $+-$ combinations in total pairs in spite of any detecting efficiency~\cite{PHENIX}.
When calculating the combined like-sign
pair in each kinematic bin, we use both the geometric mean and the direct sum of
$++$ and $--$ pairs in the calculation to estimate the impact of potentially different
detecting efficiencies for positive and negative tracks, shown in Eqs.~\ref{EQlike} and ~\ref{EQlike1}.

\begin{widetext}
\begin{equation}
\begin{split}
N_{\pm\pm}^{\rm corr}(M, p_{\rm T}) = 2\sqrt{N_{++}(M,p_{\rm T}) \cdot
  N_{--}(M,p_{\rm T})}\;\cdot \boxed{\frac{B_{+-}(M,p_{\rm
      T})}{2\cdot\sqrt{B_{++}(M,p_{\rm T}) \cdot B_{--}(M,p_{\rm T})}}} \\
\end{split} 
\label{EQlike}
\end{equation}

and

\begin{equation}
\begin{split}
& N_{\pm\pm}^{\rm corr}(M,p_{\rm T}) = a[N_{++}(M,p_{\rm T})+N_{--}(M,p_{\rm T})]\cdot \boxed{\frac{B_{+-}(M,p_{\rm T})}{b\cdot [B_{++}(M,p_{\rm T})+B_{--}(M,p_{\rm T})]}} \\  
& a = \frac{\int_{0}^{\infty} 2\cdot\sqrt{N_{++}(M,p_{\rm T})\cdot N_{--}(M,p_{\rm T})} dMdp_{\rm T}}{ \int_{0}^{\infty} [N_{++}(M,p_{\rm T})+N_{--}(M,p_{\rm T})] dMdp_{\rm T} }\\ 
& b = \frac{\int_{0}^{\infty} 2\cdot\sqrt{B_{++}(M,p_{\rm T})\cdot B_{--}(M,p_{\rm T})} dMdp_{\rm T}}{ \int_{0}^{\infty} [B_{++}(M,p_{\rm T})+B_{--}(M,p_{\rm T})] dMdp_{\rm T} }
\end{split} 
\label{EQlike1}
\end{equation}
\end{widetext}
where $N_{++}$, $N_{--}$, $B_{++}$, and $B_{--}$ denote the distributions of like-sign $(++)$ and $(--)$ from the same event and mixed-event calculation, respectively. $B_{+-}$ denotes the unlike-sign distribution from mixed-event calculations.
$N_{\pm\pm}^{\rm corr}$ denotes the acceptance-corrected like-sign background distribution.

\begin{figure}
\centering{
\includegraphics[width=0.5\textwidth]{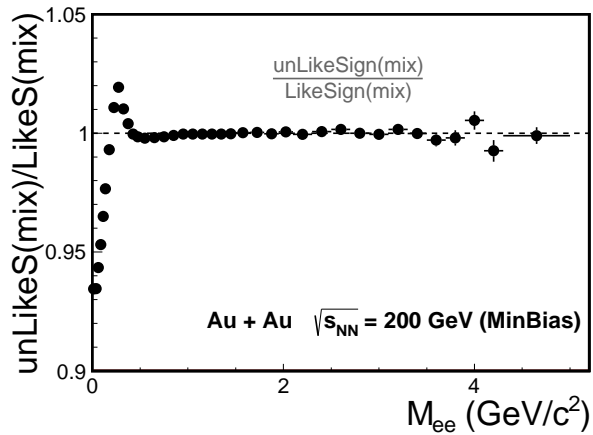}}
\caption[]{Acceptance correction factor for unlike-sign and like-sign pair
  difference from 200\,GeV Au$+$Au minimum-bias collisions.}
\label{Accratio}
\end{figure}

In Fig.~\ref{Accratio}, the ratio of mixed-event unlike-sign and like-sign
distributions is shown as a function of the pair mass integrated over \pT. 
The structures observed in the ratio at low mass are caused by local inefficiencies and acceptance holes.
This ratio has a dependence on the pair
\pT\ and a correction is applied to the like-sign distributions in the 2D ($M_{ee}$, $p_{\rm T}$) plane.

There are additional inefficiencies from merging effects that are different 
for like-sign and unlike-sign pairs in a magnetic field.
These inefficiencies can originate from TPC-track merging or TOF-hit merging.
We use two-particle correlations to study this acceptance loss due to
the TPC-track merging. We calculate the $\Delta\eta$ and $\Delta\phi$
correlations of like-sign and unlike-sign pairs in both same and mixed events. As
a conservative estimation, we artificially remove a significant amount of
the detection area near ($\Delta\eta$,$\Delta\phi$)= $(0,0)$, and correct the 
background-subtracted spectra with the cut efficiency which was estimated by the mixed events. 
The difference in the final mass spectrum was $<$1\%. 
The actual TPC hit resolution is around 1~mm, for which the expected acceptance hole 
due to the merging is significantly smaller than the estimate that is used. 
As a result, we conclude that effects due to track merging in the TPC are negligible.

Signal loss can also occur when two TPC tracks point to the same TOF read-out 
cell (size 6$\times$3~cm$^2$ at a typical radius of about 215~cm). 
The TOF matching algorithm removes any TPC-TOF association in this situation since 
it cannot resolve the timing of two close hits. 
To evaluate such losses, pairs are artificially removed for which the TPC tracks 
pointed to neighboring TOF cells, 
thereby increasing the acceptance hole by a factor of about 9. 
The impact on the final acceptance correction factor is $\sim$0.05\% and limited to two particular mass 
regions ($\sim$0.35~GeV/$c^2$ due to unlike-sign pairs, $\sim$0.1~GeV/$c^2$ due to like-sign pairs).


\subsubsection{Correlated background}

In this analysis, the like-sign distribution is used as 
the best estimate of the background in the inclusive unlike-sign distribution. 
The properly normalized mixed-event unlike-sign distributions were taken as the combinatorial background contribution.
The difference between the like-sign and the mixed-event unlike-sign was 
used to understand the correlated background contributions.

\begin{figure}
\centering{
\includegraphics[width=0.5\textwidth]{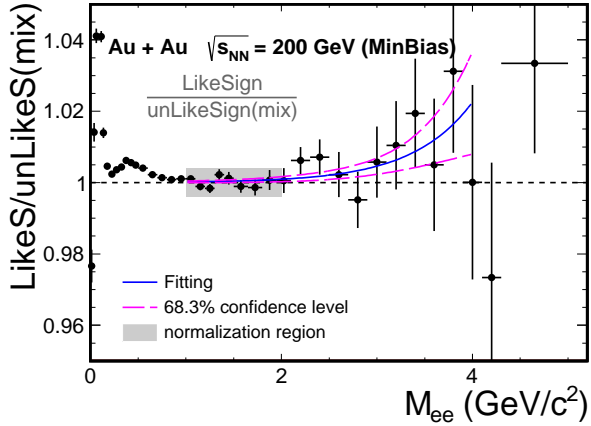}}
\caption[]{Ratio of the same-event like-sign to the mixed-event unlike-sign distributions. The gray
  area indicates the normalization region. The solid and dashed lines depict an empirical fit to the distribution in the mass region of 1$-$4\,\GeVcsq and the fit uncertainties, respectively.
  } 
\label{QA3}
\end{figure}

The ratio of the acceptance corrected like-sign to the
mixed-event unlike-sign distributions is shown in Fig.~\ref{QA3}.
In the low-mass region ($<$1~GeV/$c^2$), 
the difference is due to the cross-pair contributions such as $\pi^{0} \rightarrow e^{+} e^{-}\gamma$, 
followed by $\gamma Z \rightarrow e^{+}e^{-}Z^{*}$. 
In the intermediate and high-mass regions, the like-sign and
mixed-event distributions generally agree within our current precision, 
but also show a trend of an increasing excess with increasing mass. 
This trend is expected to be mostly due to back-to-back jet correlations.

We use a data-driven method to estimate the correlated background
contribution. We fit the ratio in Fig.~\ref{QA3} in the mass region above
1\,\GeVcsq with two different empirical functions: a second order polynomial and an
exponential function.
The small difference from unity in these fits is assigned as residual correlated background.
We use the 68.3\% confidence limits from the
fit~Eq.~\ref{FitR} (indicated by the dashed lines in the figure) as the systematic
uncertainty on the correlated background. The lower limit of this uncertainty
is consistent with unity, indicating that the like-sign background is consistent with the
mixed-event unlike-sign background.
\begin{eqnarray}
R(M) = 1 + e^{(M-a)/b}
\label{FitR}
\end{eqnarray}

\renewcommand{\floatpagefraction}{0.75}
\begin{figure*}
\centering
\bmn[b]{0.5\textwidth}
\includegraphics[width=1.0\textwidth]{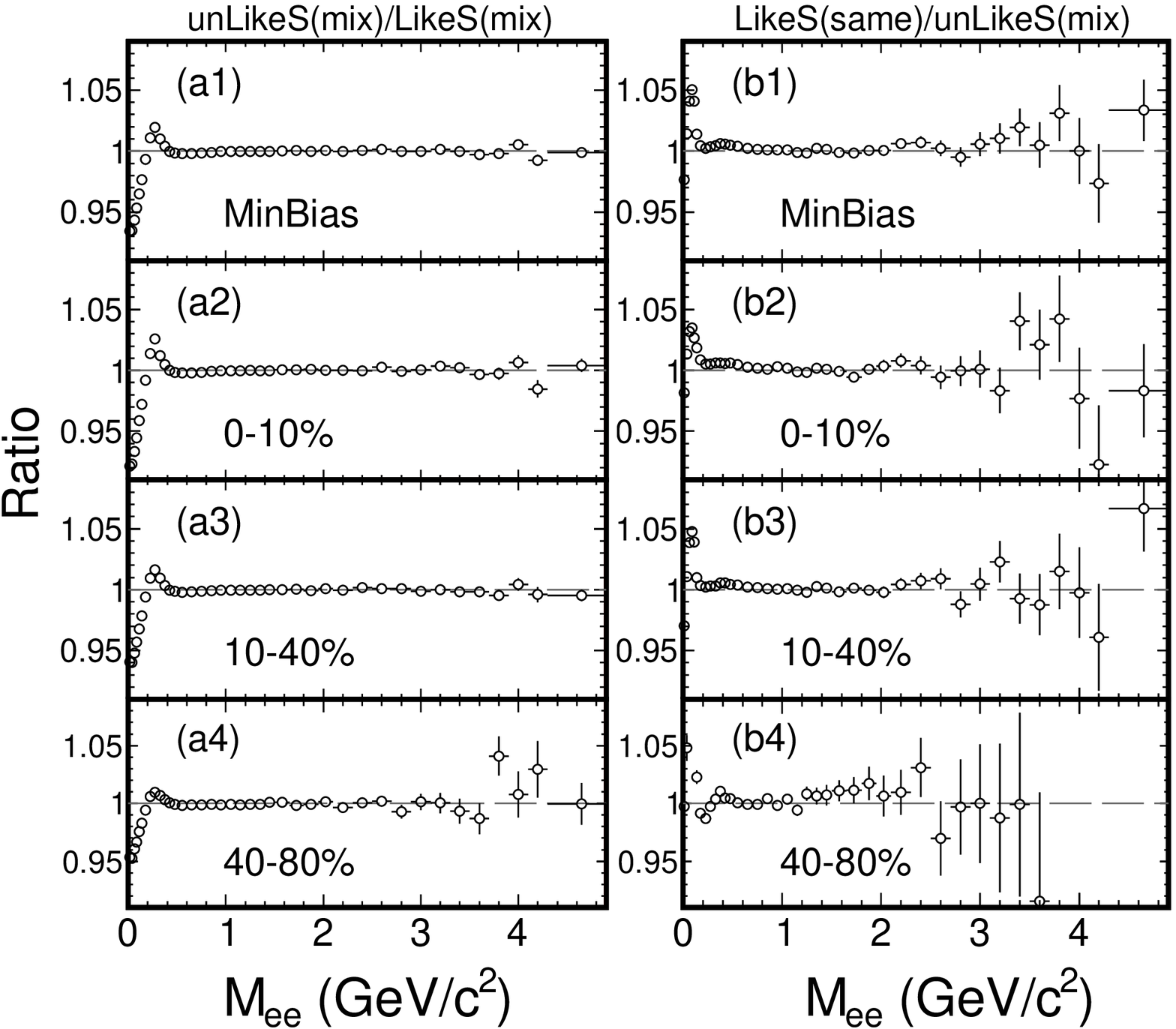}
\emn 
\centering
\bmn[b]{0.5\textwidth}
\includegraphics[width=1.0\textwidth]{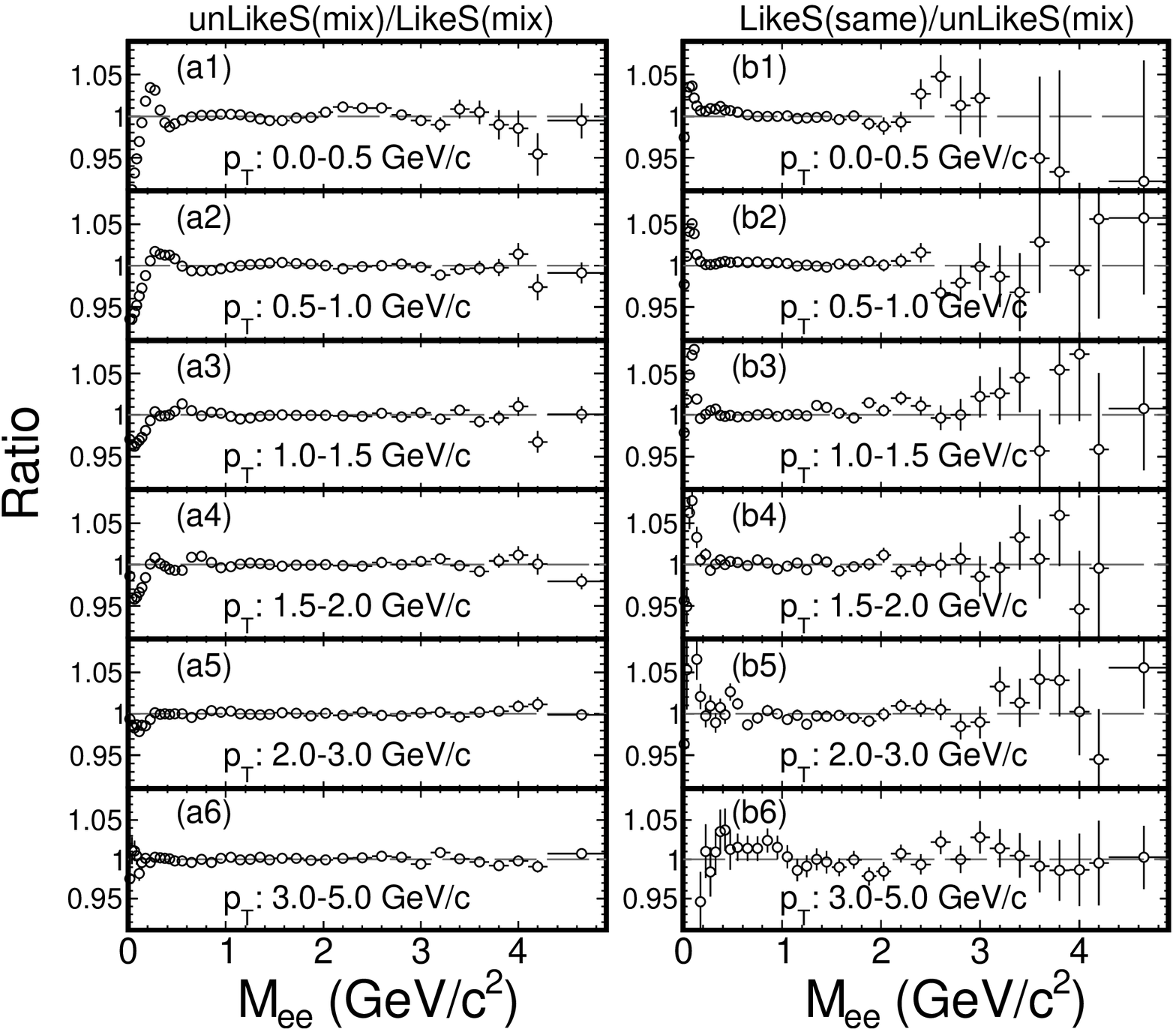}
\emn \\
\caption{Ratios of unlike-sign to like-sign mixed-event distributions (denoted as the acceptance-difference correction factor) and
  acceptance corrected like-sign background to mixed-event unlike-sign
  background distributions for different centralities and \pT\ regions.}
\label{Acc_pt_cent}
\end{figure*}

This residual background has been studied in different \pT\ and different centrality bins.  
The acceptance correction factors, which are estimated via the ratio between unlike-sign and like-sign mixed-event
distributions are shown in Fig.~\ref{Acc_pt_cent}. 
In Fig.~\ref{Acc_pt_cent}, the ratios of acceptance corrected like-sign backgrounds to
mixed-event unlike-sign distributions are also shown for various \pT\ and centrality selections.
The acceptance correction factor shows a slight centrality dependence as the
number of electron candidates is different in each centrality. 
On the other hand, it shows a strong \pT\ dependence due to the varying 
track curvatures in the magnetic field for tracks as a function of \pT.
At sufficiently high \pT, tracks are nearly straight, 
and the acceptance of like-sign and unlike-sign pairs is expected to be similar. 
A data-driven procedure was used to estimate the correlated background in each \pT\ and centrality bin.

\vspace{0.2 in}

\subsubsection{Signal extraction}
In this analysis, the dielectron signal for invariant masses of $M_{ee}$$<$1.0~GeV/$c^2$ 
is obtained by subtracting the same-event like-sign background from the inclusive unlike-sign distribution.
In the higher mass region, we first subtract the combinatorial background using
the mixed-event unlike-sign pairs for better statistical precision. The residual correlated background 
is evaluated by the data-driven method described in the previous
subsection and subtracted together with the combinatorial background.
The signal extraction evaluated over the entire invariant mass region reported here is described as follows:

\begin{widetext}
\begin{equation}
\begin{split} 
S_{+-} (M,p_{\rm T}) = \begin{cases} N_{+-}(M,p_{\rm T}) - N_{\pm\pm}^{\rm corr}(M,p_{\rm T}) & \mbox{for } M < M_{\rm th} \\
                                                                 N_{+-}(M,p_{\rm T}) - B_{+-}^{\rm comb}(M,p_{\rm T})\times[1+r(M,p_{\rm T})] & \mbox{for } M \ge M_{\rm th}
                               \end{cases}
\label{signal}
\end{split} 
\end{equation}
\end{widetext}
where $r(M,p_{\rm T})$ is the correlated background contribution normalized to
the mixed-event combinatorial background and $M_{\rm th}$ is 1.0~GeV/$c^2$ in our
default calculations. We vary this transition mass point between
1.0$-$2.0\,\GeVcsq and find the difference in the final mass spectrum to be 
negligible ($<$0.05\%).

\begin{figure}
\centering{
\includegraphics[width=0.5\textwidth]{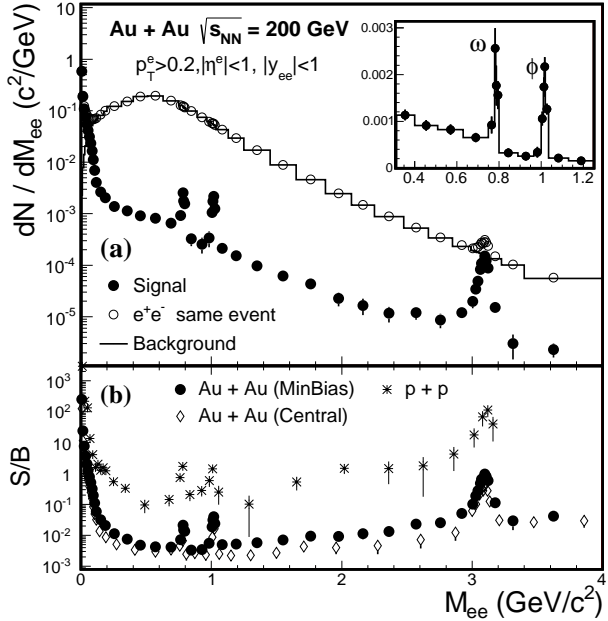}}
\caption[]{Panel (a): $e^+e^-$ invariant mass pair distributions of
  signal pairs compared to the inclusive unlike-sign (open symbols) and reconstructed background
  pairs (black line) in 200~GeV Au$+$Au minimum-bias collisions. The insert shows the signals
  of the $\phi$ and $\omega$ vector mesons. Panel (b): Signal-to-background ratios in $p$$+$$p$ and Au$+$Au collisions.}
\label{bg1}
\end{figure}

The raw signal invariant mass spectrum, $S_{+-}(M, p_{\rm T})$, for 200~GeV Au$+$Au minimum-bias collisions obtained by applying Eq.~\ref{signal} is shown in the top panel of Fig.~\ref{bg1} along with the inclusive unlike-sign and background distributions.
The bottom panel shows the signal-to-background ratio ($S/B$)
in $p$$+$$p$~\cite{STARpp} and Au$+$Au collisions. For the latter, the $S/B$
at $M_{ee}$ = 0.5~GeV/$c^2$ is about 1/200 in minimum-bias and 1/250 in 0-10\%
central collisions.

\subsection{Efficiency and Acceptance Correction}
The raw dielectron signal yields must be corrected for the detector
efficiency and acceptance loss. In this section, we discuss separately the single-electron efficiencies and electron pair efficiencies.

\subsubsection{Single-electron efficiency}

The single-electron efficiency is determined by the product of the TPC tracking efficiency $\varepsilon_{\rm TPC}$, the TOF matching efficiency $\varepsilon_{\rm TOF}$, and the electron identification efficiency $\varepsilon_{\rm eID}$:

\begin{equation}
\varepsilon_e = \varepsilon_{\rm TPC}\times\varepsilon_{\rm TOF}\times\varepsilon_{\rm eID}
\end{equation}

The TPC tracking efficiency, $\varepsilon_{\rm TPC}$, was evaluated via the
standard STAR embedding technique. In the embedding process, simulated electron tracks
with a certain phase space definition were generated and then passed through the
STAR detector geometry for the 2010 (2011) configuration using the {\sc Geant}
model. Next, the simulated detector signals were mixed with real data to have a realistic detector occupancy environment.
The mixed signals were processed with the same offline reconstruction software
that was used for the real data production. The tracking efficiency was studied by
comparing the reconstructed tracks with the simulated input tracks. The input number
of simulated tracks (5\% of total event multiplicity) were constrained to prevent a sizable impact on the final
single-track efficiency.

The electron track TOF-match efficiency, $\varepsilon_{\rm TOF}$, was
obtained from real data samples. Due to the limited pure electron statistics,
we first used a pure pion sample in order to deduce the TOF-match efficiency.
Pure pion samples were selected based on a TPC $dE/dx$ cut. We assume the
TOF-match efficiencies for different particle species are similar in the
\pT\ region where $dE/dx$ cannot distinguish different particle species. Pure
electron samples were selected to cross-check the efficiency scale differences between electrons 
and pions due to the decay loss of pions between the TPC and the TOF detectors as well as other effects. 
Electrons (or positrons) from photon conversion or $\pi^{0}$ Dalitz decays were identified by 
invariant mass and topological techniques and used as the high purity samples.

The TPC tracking and TOF matching efficiencies were calculated differentially
in three dimensions $(p_{\rm T}, \eta, \phi)$. The pion TOF matching
efficiency was also calculated in $(p_{\rm T}, \eta, \phi)$ 
while a same scaling factor, which accounts for the TOF matching efficiency difference between pions and electrons, was used for all $(\eta,\phi)$ bins due to limited statistics.
The choice of the binning in $(\eta,\phi)$ dimensions shows a
negligible effect in the $p_T^{ee}$-integrated final dielectron pair efficiency. 


\renewcommand{\floatpagefraction}{0.75}
\begin{figure*}
\centering
\includegraphics[width=0.7\textwidth]{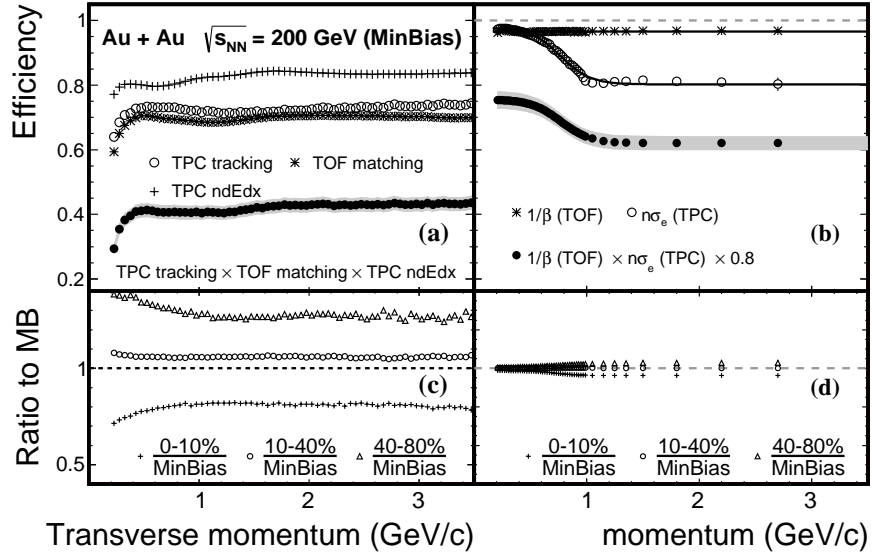}
\caption[]{The single-electron efficiency as a function of
  $p_{T}$ or $p$ in the pseudo-rapidity range of $|\eta|<1$ in Au$+$Au collisions at
  \sNN = 200~GeV. In the left panel solid points represent the combined TPC tracking, 
  TOF matching and TPC ndEdx efficiencies. In the right panel the solid points represent the
  combined TPC $n\sigma_e$ cut and TOF $1/\beta$ cut efficiencies. Gray bands indicate the systematic
  uncertainties. The bottom panels show these efficiencies in various
  centrality bins compared to the minimum-bias data.}
\label{singleEff}
\end{figure*}

The electron identification cut efficiency, $\varepsilon_{\rm eID}$, includes
two components: efficiency due to the TOF $1/\beta$ cut ($\varepsilon_{\rm \beta}$) and
efficiencies due to the $dE/dx$ PID selection criteria ($\varepsilon_{\rm dEdxPID}$).

\begin{equation}
\begin{split}
\varepsilon_{\rm eID}    = \varepsilon_{\rm \beta}\times\varepsilon_{\rm dEdxPID}  \\
\varepsilon_{\rm dEdxPID} = \varepsilon_{\rm ndEdx}\times\varepsilon_{\rm n\sigma_e}
\end{split}
\end{equation}

Pure electron samples were used to study the TOF $1/\beta$ distributions. In order to estimate the $1/\beta$ efficiency, 
$\varepsilon_{\rm \beta}$, we applied two methods to the $1/\beta$ distributions: a realistic function fit and direct counting. The difference in the results from the two methods was included in the systematic uncertainty.

The $dE/dx$ PID selection efficiency, $\varepsilon_{\rm dEdxPID}$, includes the efficiency due to the cut
on both the number of $dE/dx$ points and $n\sigma_e$ which is used to select the electron candidates. 
The cut efficiency on the number of $dE/dx$ points, $\varepsilon_{\rm ndEdx}$, 
was deduced using the pure pion samples in the real data. The results from the electron sample
were consistent with those from pions in the region allowed by the statistics of the samples used. 
Then the efficiency from the pion samples was used in the final efficiency calculation in three dimensions $(p_{\rm T}, \eta, \phi)$.
The $n\sigma_e$ cut efficiency, $\varepsilon_{\rm n\sigma_e}$, was deduced via the same steps as described in Section III-D for calculating the electron purity and hadron contamination. With the extracted $n\sigma_e$ gaussian mean position and width values, 
the PID cut efficiency was calculated under the selection criteria described in Section III-C.

In the upper left panel of Fig.~\ref{singleEff}, $\varepsilon_{\rm TPC}(p_T)$,
$\varepsilon_{\rm TOF}(p_T)$, $\varepsilon_{\rm ndEdx}(p_{T})$ and their product are shown 
for $e^{\pm}$ tracks in minimum-bias collisions.
These efficiencies are averaged over $|\eta|<1$ and $2\pi$ in azimuth. 
The ratios of  $\varepsilon_{\rm TPC}\times\varepsilon_{\rm TOF}\times\varepsilon_{\rm ndEdx}$ at different 
centralities are shown in the bottom left panel of Fig.~\ref{singleEff}. 
The $\varepsilon_{\rm \beta}$, $\varepsilon_{\rm n\sigma_e}$ and their product as a function of momentum are shown in the upper right panel of Fig.~\ref{singleEff}. The centrality dependence of $\varepsilon_{\rm \beta}\times\varepsilon_{\rm n\sigma_e}$ is shown in the bottom right panel.


\subsubsection{Electron pair efficiency}

The dielectron pair efficiency was evaluated from the single-electron efficiency in the following two ways:

\begin{itemize}
 \item Toy Monte Carlo simulation, which used the virtual photons as the input  and let them decay into dielectrons isotropically.
 \item Cocktail simulation, which used the hadronic cocktail (see Secttion G) as input including the correlated heavy-flavor decay electrons
   from {\sc Pythia} simulations~\cite{pythia}.
\end{itemize}

In the final dielectron spectra, we have experimental ambiguities in
separating heavy-flavor decayed dielectron yields 
from medium-produced dielectron yields (including contributions from both hadronic and partonic sources). 
Furthermore, the heavy-flavor decay dielectron production is not known in heavy-ion collisions due
to possible medium modifications of the heavy-flavor correlations when compared to those in $p$$+$$p$ collisions. 
We used these two methods to estimate our dielectron pair
efficiency. 
The single-electron efficiencies, described in the previous section, were folded in for each daughter track in a full 
three dimensional $(p_{\rm T}, \eta, \phi)$ momentum
space. The pair efficiency and acceptance was finally calculated in $(M_{ee}, p_{\rm T})$. 

\begin{figure}
\centering{
\includegraphics[width=0.5\textwidth] {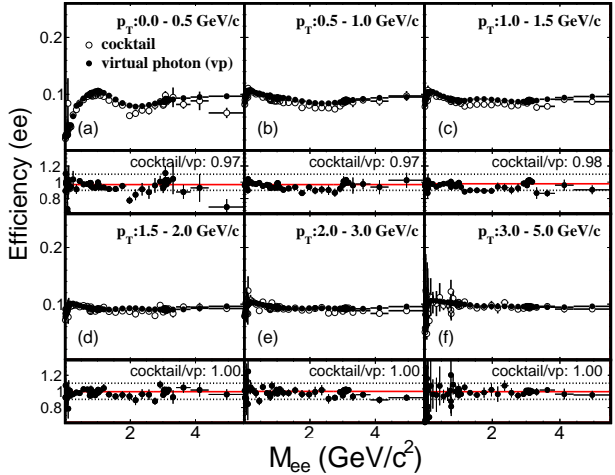}}
\caption[]{ $e^{+}e^{-}$ pair efficiency as a function of pair mass in different \pT\ regions calculated from two different methods.}
\label{Effpt_dependence}
\end{figure}

Shown in Fig.~\ref{Effpt_dependence} are the dielectron pair efficiencies in the STAR
acceptance ($p_{T}^{e}>0.2$~GeV/$c$, $|\eta^{e}|<1$) with $|y_{ee}|<1$.
The difference in pair efficiency in the STAR acceptance between these two methods is small, ranging from about 3\% at low \pT, down to about 1\% at high \pT. And due to statistical limits of the cocktail simulation for the dielectron from heavy-flavor decay, we use the pair efficiency calculated from the virtual photon decay in this analysis and include the difference between these two methods in the systematic uncertainty.

The $\phi_{\rm V}$ pair cut efficiency was evaluated using a $\pi^0$ embedding sample 
in which simulated $\pi^0$ particles with enriched Dalitz decays were embedded into the real data. 
The efficiency was calculated after re-weighting the input $\pi^0$ yield with a realistic \pT\ 
distribution (details in the next part). 
We also used a pure virtual photon decay convoluted with the detector resolution for this calculation. 
The difference was included as the systematic uncertainty of the $\phi_{\rm V}$ pair cut efficiency.

\begin{figure}
\centering{
\includegraphics[width=0.5\textwidth] {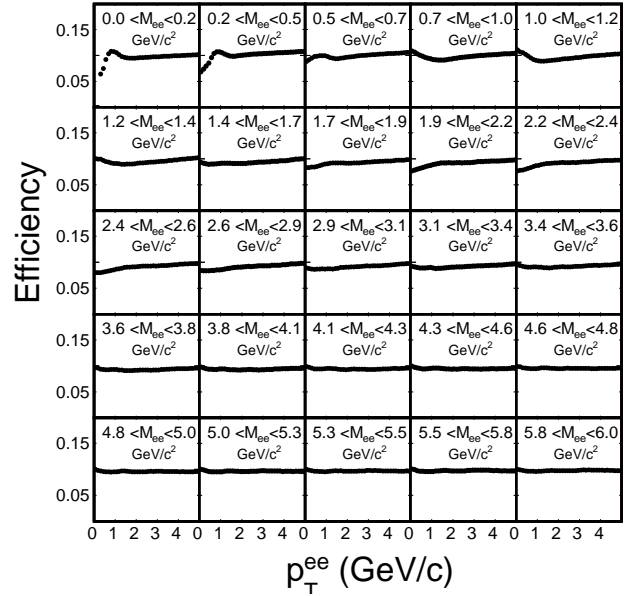}}
\caption[]{$e^{+}e^{-}$ pair efficiency as a function of pair $p_{T}$ for different mass regions. 
The dashed lines represent $\pm$10\% difference from the unit, the solid lines show a constant fit to the data. 
 } 
 \label{pairEffpt}
\end{figure}

\begin{figure}
\centering{
\includegraphics[width=0.5\textwidth] {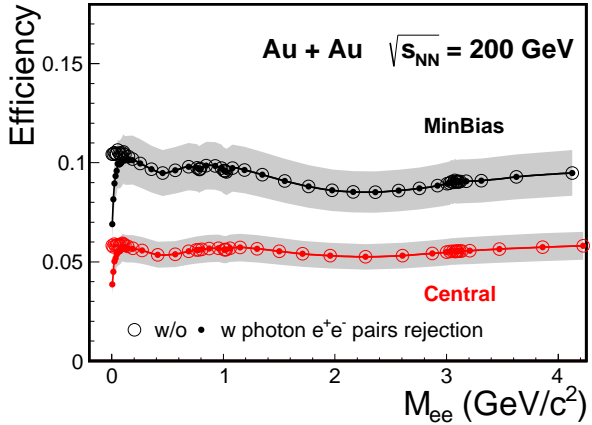}}
\caption[]{(Color online) The $p_{T}$-integrated $e^{+}e^{-}$ pair efficiency
  as a function of pair mass within the STAR acceptance in Au$+$Au minimum-bias
  (black) and central (red) collisions at \sNN = 200\,GeV. The open markers
  represent the corresponding efficiencies before the photon conversion
  rejection. The gray bands depict the systematic uncertainties.}
 \label{pairEffmass}
\end{figure}

In Fig.~\ref{pairEffpt}, the $e^{+}e^{-}$ pair efficiencies are shown as a function of
pair $p_{T}$ in different mass regions.
In the high $p_{T}$/mass region the efficiency is almost constant as the single track 
efficiency turns stable at high $p_{T}$ (see Fig.~\ref{singleEff}). 
The $p_{T}$-integrated $e^{+}e^{-}$ pair efficiencies as a function of pair mass within 
STAR acceptance in Au$+$Au collisions at \sNN = 200~GeV are shown in Fig.~\ref{pairEffmass}. 
The pair efficiency without the $\phi_V$ cut is also plotted, which contributes only
in the very low mass region.

\subsection{Hadronic cocktails}
\label{Sect:Cocktail}
Dielectrons as measured by the detector originate from all stages in the evolution of heavy-ion collisions. 
These pairs include the decay products of long-lived particles which typically decay after they have frozen out of the medium. 
The contributions in the final dielectron spectrum
can be evaluated as long as their yields at freeze-out are known. 

The simulation process for constructing the contributions from such decays in Au$+$Au collisions, often referred to as
the  hadronic cocktail, is similar to what has been done in $p$$+$$p$ collisions and reported in~\cite{STARpp}.
The cocktail simulations only contain the hadron form-factor decays in the vacuum at freeze-out.
Cocktails included in our calculation contain contributions from decays  
of $\pi^{0}$, $\eta$, $\eta^{\prime}$, $\omega$, $\phi$,
$J/\psi$, $\psi^{\prime}$, $c\bar{c}$, $b\bar{b}$ as well as from Drell-Yan (DY) production. 
A vacuum $\rho$ meson calculation is included separately when discussing the data
compared to cocktail with the vacuum $\rho$. For the hadron decay
calculations, the input rapidity distributions are assumed to be flat within
$|y|<1$. The input yields $dN/dy$ within this rapidity window as well as the
\pT\ distributions are discussed below. 

The charged pion yields at 200~GeV Au$+$Au collisions have been
accurately measured in the STAR acceptance~\cite{DataPion1,DataPion2}. 
The input $\pi^0$ spectrum is taken as the averaged yield between STAR's $\pi^+$ and $\pi^-$ measurements.
Other light hadron yields include the $\eta$ meson, measured by PHENIX for $p_{\rm T}>$
2\,GeV/$c$~\cite{DataEta}, and $\phi$ meson data from STAR~\cite{DataPhi}. These hadron spectra together
with hadron spectra ($K^{\pm}$, $K_S^0$, and $\Lambda$) measured by STAR and
PHENIX were simultaneously fit to a core-corona based Tsallis Blast-Wave (TBW)
model~\cite{TBW} where the core describes the Au$+$Au bulk production and the
corona describes the hard scattering contribution from $p$$+$$p$ like collisions.

\begin{figure}
\centering{
\includegraphics[width=0.5\textwidth] {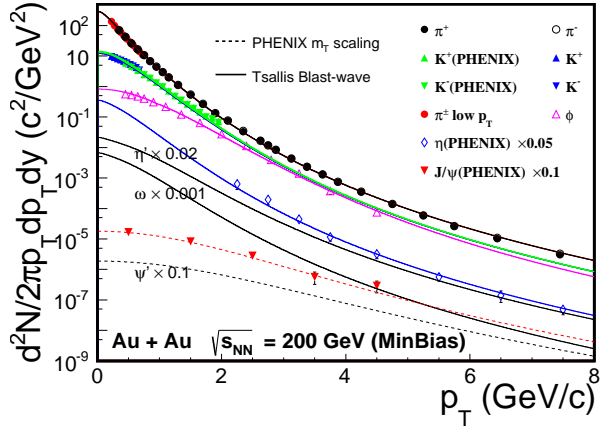}}
\caption[]{(Color online) Invariant yields of mesons in Au$+$Au collisions at \sNN =
  200\,GeV. The solid lines show the simultaneous TBW fit to the measured data points and the TBW predictions for $\eta$, $\eta^\prime$, and $\omega$  with the same set of fit parameters. The dashed lines show the same parametrization to the measured $J/\psi$ spectrum and the predicted $\psi^\prime$ spectrum as in~\cite{PHENIX}.}
 \label{hadronInput}
\end{figure}

In Fig.~\ref{hadronInput}, the simultaneous fit results for all input
hadron spectra are shown except for $J/\psi$. 
The $J/\psi$ contribution is not considered as a component of the bulk medium.
The cocktail input for $J/\psi$ was taken from the measurement by the PHENIX collaboration~\cite{DataJpsi}. For light hadrons, 
the TBW functions provide good parameterizations to these
measured spectra. For those hadron cocktail components without corresponding direct
measurements (e.g. low \pT\ $\eta$, $\eta^{\prime}$, $\omega$), we use the same
core TBW parameters obtained from the fit and predict the spectral shapes for
each of these unknown components, shown as solid curves in
Fig.~\ref{hadronInput}.  The low \pT\ $\eta$ spectrum
was fixed by requiring the match with the measured data points at \pT $>$
2\,GeV/$c$, while the $dN/dy$ of $\eta^{\prime}$ meson was taken
with the same values as used in the PHENIX publication~\cite{PHENIX}. 
The same set of TBW parameters from the simultaneous fit were used to generate the $\omega$ spectrum 
and the $dN/dy$ was tuned to match our dielectron yield in the $\omega$ peak region.



Additional corrections were applied to account for the differences
in centrality and rapidity windows between the input hadron spectra and our dielectron measurements.
The measured pion yields were calculated in the
rapidity window of $|y|<0.1$ in Ref.~\cite{DataPion1} and $|y|<0.5$ in
Ref.~\cite{DataPion2}. We used the pion rapidity distributions from the {\sc Hijing} calculations and scaled the measured pion yields down by 3\% to obtain
the \pT\ spectrum in the rapidity window of $|y|<1$. This correction factor was also included in  the uncertainty of the input $\pi^0$ $dN/dy$. 
The different centrality windows matter when taking the minimum-bias data from PHENIX measurements, done in 0-92\% centrality, and compare those to our results which are for 0-80\% centrality.
We corrected for this difference using the measured  $\pi^0$ $dN/dy$ values in 0-92\% and 0-80\% centralities by the PHENIX experiment~\cite{DataPion3}.

The correlated charm, bottom and Drell-Yan contributions were obtained from
{\sc Pythia} calculations~\cite{pythia} and scaled by the number of binary collisions
in Au$+$Au collisions for the default cocktail calculations. We used {\sc Pythia}
version 6.419 with parameter settings: MSEL=1, PARP(91) ($\langle
k_{\perp}\rangle$) = 1.0\,GeV/$c$ and PARP(67) (parton shower level) = 1.0.
This setting was tuned to match our measured charmed meson spectrum in $p$$+$$p$
collisions~\cite{DataCharm}. 
The input charm-pair production cross section per nucleon-nucleon collision 
was also taken from charm meson measurements~\cite{DataCharm,AuAuD0}.
We used the same {\sc PYTHIA} setting to calculate the dielectron yields from
correlated bottom decays and from the Drell-Yan production. The input bottom
and Drell-Yan production cross sections are: $\sigma^{\rm b\bar{b}}_{pp}= 3.7$~$\mu$b, $\sigma^{\rm DY}_{pp} =$ 42~nb.

The $\rho$ meson contribution is expected to be modified due to a strong coupling to the hot QCD medium created in heavy-ion collisions. 
Therefore, the $\rho$ meson was not included in our default cocktail calculations. In the
comparison between our measured dielectron spectra and the cocktail calculations
including the vacuum $\rho$, we used the $\rho$ meson measurements in peripheral
collisions by STAR~\cite{DataRho} and assumed a similar $\rho/\pi$ ratio in order to
extrapolate to other centrality selections. The mass spectrum of the vacuum
$\rho\rightarrow e^+e^-$ was taken the same line shape as reported in our
dielectron measurement in $p$$+$$p$ collisions~\cite{STARpp}.

Table~\ref{table1} summarizes all sources of the hadron cocktail
and their  decay branching ratios. The
TBW~\cite{TBW} parametrizations were used to describe the
input hadron \pT\ distributions, shown in Fig.~\ref{hadronInput}.
The resulting $e^+e^-$ pair mass distributions from the individual sources are 
normalized by the respective decay branching ratios and measured yields $dN/dy$.
Additional scaling parameters for various centrality bins are listed in Appendix \ref{Appendix:cocktail}.

\begin{table*}
\caption{Input yields of various cocktail components for 0-80\% minimum bias Au$+$Au collisions at \sNN = 200\,GeV.}
\begin{tabular}{c|c|c|c|c}
\hline \hline
source & B.R. & $dN/dy$ or $\sigma$ & Uncertainty & Reference \\
\hline
$\pi^{0} \rightarrow \gamma ee$ & 1.174 $\times 10^{-2}$   & 98.5      & 8\%   & STAR \cite{DataPion1,DataPion2}  \\
$\eta \rightarrow \gamma ee$ & 7 $\times 10^{-3}$          & 7.86      & 30\%  & PHENIX \cite{DataEta,PHENIX}\\
$\eta^{\prime} \rightarrow \gamma ee$ & 4.7 $\times 10^{-4}$ & 2.31                  & 100\% & PHENIX \cite{PHENIX}, STAR \cite{STARpp}\\
$\rho \rightarrow ee$ & 4.72 $\times 10^{-5}$              & 16.7                  & 42\%  & STAR \cite{DataRho}\\
$\omega \rightarrow ee$ & 7.28 $\times 10^{-5}$            &                                   &       &              \\
$\omega \rightarrow \pi^{0} ee$ & 7.7 $\times 10^{-4}$     & 9.87                  & 33\%  & STAR \cite{Dataomega}\\
$\phi \rightarrow ee$ & 2.95 $\times 10^{-4}$ &                                                &       &              \\
$\phi \rightarrow \eta ee$ & 1.15 $\times 10^{-4}$         & 2.43                  & 10\%  & STAR   \cite{DataPhi}\\
$J/\psi \rightarrow ee$ & 5.94 $\times 10^{-2}$            & 2.33 $\times 10^{-3}$ & 15\%  & PHENIX \cite{DataJpsi}\\
$\psi\prime \rightarrow ee$ &  7.72 $\times 10^{-3}$       & 3.38 $\times 10^{-4}$ & 27\%  & PHENIX \cite{Datapsi1,Datapsi2}\\ 
\hline
$c\bar{c} \rightarrow ee$ & 1.03 $\times 10^{-1}$ &  $d\sigma^{\rm c\bar{c}}/dy = 171 \mu{\rm b}$  & 15\%  & STAR \cite{AuAuD0}\\
$b\bar{b} \rightarrow ee$ & 1.08 $\times 10^{-1}$ & $\sigma^{\rm b\bar{b}}_{pp}$ = 3.7 $\mu$b    & 30\%  & {\sc Pythia}\cite{pythia}\\
$DY \rightarrow ee$ & 3.36 $\times 10^{-2}$     & $\sigma^{\rm DY}_{pp}$ = 42 nb    & 30\%  & {\sc Pythia}\cite{pythia}\\
\hline \hline
\end{tabular}
\label{table1}
\end{table*}

The mass spectra reported in this paper are not corrected for the STAR detector
resolution. It is very challenging to precisely reproduce the momentum
resolution in the STAR TPC simulation package in the high luminosity RHIC
environment due to various distortion effects in the TPC detector. Instead a
data-driven method was used to obtain the dielectron mass line shape in the cocktail simulation. 

Based on the full detector simulation, the reconstructed electron $p_{\rm T}^{\rm
  rec}$ probability distribution at a given input $p_{\rm T}^{\rm MC}$ was
parametrized with a double Crystal Ball function~\cite{dCB}, defined as:

\begin{eqnarray}
 P(p_{\rm T}^{\rm rec}, p_{\rm T}^{\rm MC}) \propto \left\{
\begin{array}{lll}
A\times (B-R)^{-n},  & R<-\alpha \\
e^{\frac{-R^{2}}{2}},  & -\alpha < R < \beta \\
C\times (D+R)^{-m},  & R>\beta 
\end{array}
\right.
\label{crystalball}
\end{eqnarray}

with 

\begin{equation}
\begin{split}
A & = \left(\frac{n}{|\alpha |}\right)^{n} \times e^{\frac{-\alpha ^{2}}{2}} \\
B & = \frac{n}{|\alpha |} - |\alpha | \\
C & = \left(\frac{m}{|\beta |}\right)^{m} \times e^{\frac{-\beta ^{2}}{2}} \\
D & = \frac{m}{|\beta |} - |\beta | \\
R & = \left(\frac{p_{\rm T}^{\rm rec}-p_{\rm T}^{\rm MC}}{p_{\rm T}^{\rm MC}} - \mu\right) / \frac{\sigma_{p_{\rm T}}}{p_{\rm T}}
\end{split}
\end{equation}
where $n = 1.29$, $\alpha = 1.75$, $m = 2.92$, and $\beta = 1.84$.
The value of $\mu = -0.001$ is slightly shifted from 0 due to the electron energy loss in the
detector material as STAR tracking accounts for the energy loss
assuming all tracks are pions.

The \pT\ resolution is taken as $\sigma_{p_{\rm T}}/p_{\rm T}$ and assumed to follow the form:

\begin{equation}
  \left(\frac{\sigma_{p_{T}}}{p_{T}}\right)^{2} = (a \times p_{T})^{2} + \left(\frac{b}{\beta}\right)^{2};  ~~ \beta = \frac{p}{E}\sim \frac{p_{T}}{\sqrt{p_{T}^{2} + m^{2}}}.
  \label{momRes}
\end{equation}
For electrons $\beta\sim1$.

We used the $J/\psi$ signal which has the most statistics and tuned the
 parameters $a$ and $b$ in the Eq.~\ref{momRes} to get the best match to
the $J/\psi$ signal distribution. The two parameters were found to be $a=
6.0\times 10^{-3}$~$c$/GeV and $b=8.3\times10^{-3}$. 


\subsection{Systematic uncertainties}
\label{Sect:SystematicUncertainties}
The major sources of systematic uncertainty that contribute to the final result
in this analysis include:

\begin{enumerate}
\item Normalization factor for mixed-event distributions
\item Residual correlated background
\item Like-sign/unlike-sign acceptance difference correction
\item Hadron contamination
\item Efficiency and acceptance corrections
\end{enumerate}


The systematic uncertainty of the background of dielectron pairs was further 
separated in two mass regions, where we chose different background subtraction
methods (see Eq. ~\ref{signal}).

In the mass region of $M\ge$ 1.0\,GeV/$c^2$, we obtained the signal by
subtracting the mixed-event unlike-sign background plus the residual correlated
background. The normalization of the combinatorial background, applied to the mixed-event unlike-sign distribution, is determined by comparing the like-sign same-event and
mixed-event distributions. The statistics of the total like-sign pair in the
normalization region is the dominant systematic uncertainty. We
also chose different normalization ranges varying between the mass range of 1.2$-$
2.0\,GeV/$c^2$. Other sources that we considered include the normalization method 
and the slight asymmetry between the total number of mixed-event unlike-sign and like-sign pairs.
For the normalization method, we chose a different method compared to what was described in Section III-E, 
in this way we normalize the mixed-event unlike-sign distribution to the acceptance-corrected same-event like-sign distribution. 
Table~\ref{sysNorTb} summarizes the contributions for each of the individual components 
to the systematic uncertainty of the normalization factors 
in minimum bias as well as for various other centralities from 200~GeV Au$+$Au collisions.


The uncertainty in the residual correlated background was already mentioned in
Section III-E.5. In the data-driven approach, the statistical
uncertainty in determining the ratio of like-sign and mixed-event unlike-sign
$r(M_{ee},p_{\rm T})$ was used as the systematic uncertainty. The contribution to the final
dielectron mass spectrum in minimum-bias collisions is about 10\% from
1\,GeV/$c^2$ to 3\,GeV/$c^2$.

In the low mass region, $M_{ee}<$ 1.0\,GeV/$c^2$, we obtained the signal by
subtracting the acceptance corrected like-sign background, 
in which the acceptance difference correction between like-sign and unlike-sign pairs was 
calculated using mixed-event distributions. 
Different event mixing methods by varying the different event categories and event
pool sizes were chosen, and the largest deviations between these methods are used in the uncertainty calculation.  
The acceptance correction done in the 2D ($M_{ee}$, $p_{\rm T}$) plane may suffer from limited statistics. The difference between the results calculated using the 2D acceptance correction and using the 1D ($M_{ee}$ only) acceptance correction
was included in the systematic uncertainty as well.


\begin{table*}
\caption{Systematic uncertainties on normalization factors of mixed-event
  distributions for minimum-bias collisions and various centralities. The total
  number of $e^{+}e^{-}$ pairs in minimum-bias collisions is $\sim3.7\times10^{7}$,
  and for central collisions is $\sim7.0\times10^{7}$ (2010 data).}
\centering
\begin{tabular}{l|c|c|c|c|c}
\hline
\hline
 &  Like-sign pairs & Choice of N.R. & Norm.\ method & LS/US pair difference & Total \\ \hline
 MinBias &  $4.9\times10^{-4}$  & $2.1\times10^{-4}$  & $1.0\times10^{-4}$  &  $2.4\times10^{-5}$  & 0.05\%  \\
 0-10 \% &  $3.4\times10^{-4}$  & $1.4\times10^{-4}$  & $5.6\times10^{-5}$  &  $1.7\times10^{-5}$  & 0.04\% \\
 10-40\% &  $6.6\times10^{-4}$  & $3.2\times10^{-4}$  & $1.2\times10^{-4}$  &  $3.1\times10^{-5}$  & 0.07\% \\
 40-80\% &  $2.2\times10^{-3}$  & $5.2\times10^{-3}$  & $5.2\times10^{-4}$  &  $9.8\times10^{-5}$  & 0.56\% \\

\hline							     
\hline							     
\end{tabular}
\label{sysNorTb}
\end{table*}

The electron candidates contain a small amount of hadron contamination, which may be correlated with other particles ({\it e.g.} from resonance decays) and thus contribute to the final signal spectrum.
To estimate this contribution, we first selected pure pion, kaon and proton samples with stringent TOF
$m^2$ limits. We randomly picked hadrons from these pure samples according to
the estimated hadron contamination levels in both the total amount and the \pT\
differential yields, creating a hadron contamination candidate pool. 
The analysis procedure used in the dielectron analysis was applied to that pool to estimated the $e-h$ and $h-h$ correlated contributions. 

\begin{figure}
  \centering{
    \includegraphics[width=0.5\textwidth] {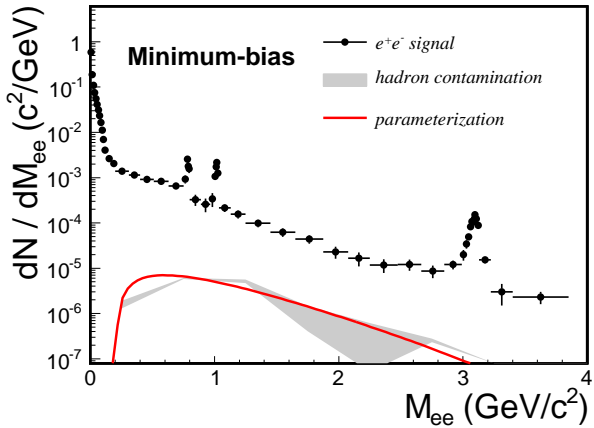}}
  \caption[]{Estimated hadron contamination from $e-h$ and $h-h$ contributions due to the finite contaminated hadrons 
	  in the elecron sample compare to the $e^+e^-$ pair signal in 200\,GeV Au$+$Au collisions.
	} 
  \label{hadroncontamination}
\end{figure}

The estimated hadron comtamination evaluated from $e-h$ and $h-h$ correlated contributions 
compared to the dielectron signal is shown in Fig.~\ref{hadroncontamination}.  
Overall, the relative contribution to the final
spectrum is $<$5\% between 1\,GeV/$c^2$ and 3\,GeV/$c^2$.

\begin{figure}
\centering{
\includegraphics[width=0.5\textwidth]{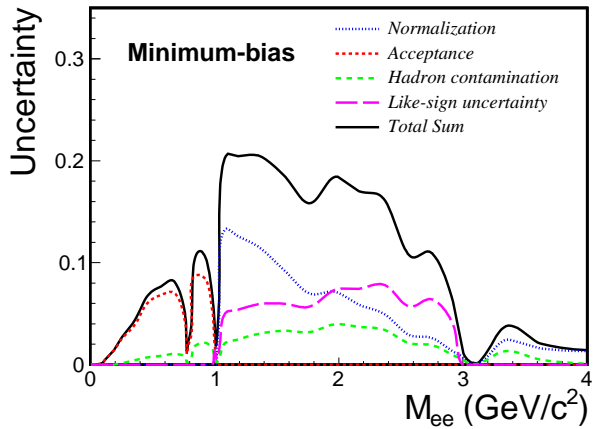}}
\caption[]{(Color online) Systematic uncertainties of raw dielectron invariant-mass spectrum in
	Au$+$Au minimum-bias collisions from various contributing sources.
	The direct sum of each individual component was taken as the total systematic uncertainty, shown as the solid curve.
}
\label{sysFinal}
\end{figure}

The systematic uncertainties on the raw dielectron invariant-mass spectra for minimum-bias collisions are summarized in Fig.~\ref{sysFinal}.
As a conservative estimation, we took the sum of each individual component as the
total systematic uncertainty.

For the reported dielectron yields in the STAR acceptance, the systematic
uncertainty due to the efficiency correction includes
uncertainties on the single-track efficiency, the pair efficiency evaluation
method, and the pair cut ($\phi_V$) efficiency. Table~\ref{effsys} summarizes
each individual component and their contributions to the total uncertainty of
the single-track efficiency. 
The individual component contributions were determined
by varying track selection cuts and comparing distributions between
data and MC for the uncertainty on the TPC tracking efficiency (nHitsFits,
DCA, etc.). The uncertainties on the TOF matching efficiency, the TOF PID cut
efficiency, and the ndEdxFits cut efficiency were evaluated by comparing the
results obtained from the pure electron samples from photon conversion and $\pi^{0}$ Dalitz decay.
The difference between a realistic function fit and direct counting methods of the TOF $1/\beta$ distribution was 
also included in the uncertainty of the TOF PID cut efficiency. 
The electron pair efficiency evaluated from single tracks was described in Section III-F.2.
Due to the unknown relative contributions between the correlated
charm decays and the medium contribution, 
two extreme calculations were used as conservative estimates for the systematic uncertainty. 
This uncertainty is mostly constrained to the intermediate and high mass regions 
of the mass spectrum and ranges from about 3\% at low \pT\ down to 1\% {\color{blue}at} high \pT.
The systematic uncertainty of the $\phi_V$ pair cut efficiency was evaluated by 
taking the difference between the calculations from the $\pi^0$ embedding sample 
and the virtual photon decay sample, which is about  3\% 
at $M_{ee}<0.05$\,GeV/$c^{2}$. 
The systematic uncertainty of the pair efficiency due to different methods is 5\%.
Finally the total systematic uncertainty of the electron pair efficiency is $\sim 13\%$.

\bt
\caption{Systematic uncertainties on single-track efficiency.}
\centering
\begin{tabular}{llr}
\hline
\hline
& component   \;\;\;            & uncertainty     \\                                
\hline
\multirow{4}{*}{TPC \;\;} & nHitsFits  & $4.0\%$         \\                                  
                          & DCA        & $2.5\%$         \\  
                          & ndEdxFits              & $  1.0\%$         \\ 
                          & n$\sigma_{e}$           & $  2.0\%$         \\  \hline
\multirow{2}{*}{TOF \;\;} & matching               & $1.0\%$         \\                                  
                          & $1/\beta$              & $3.0\%$         \\                                  
\hline
Total                     &                        & $6.1\%$         \\ 
\hline                               
\hline                               
\end{tabular}
\label{effsys}
\et

\section{Results and Discussion}

\subsection{Dielectron mass spectrum in minimum-bias collisions}

\begin{figure*}
\centering
\bmn[b]{0.33\textwidth}
\includegraphics[width=1.0\textwidth]{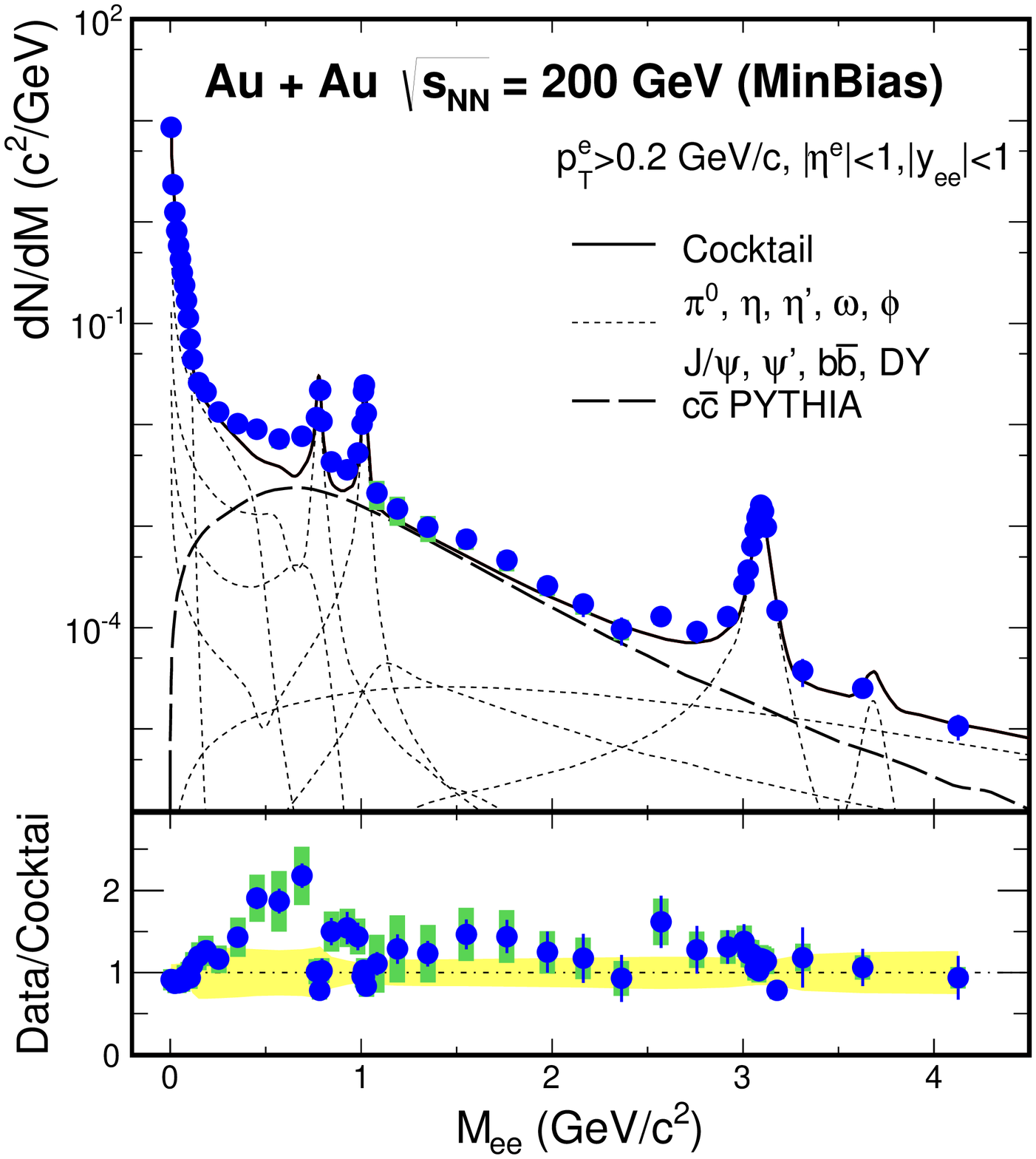}
\emn 
\centering
\bmn[b]{0.33\textwidth}
\includegraphics[width=1.0\textwidth]{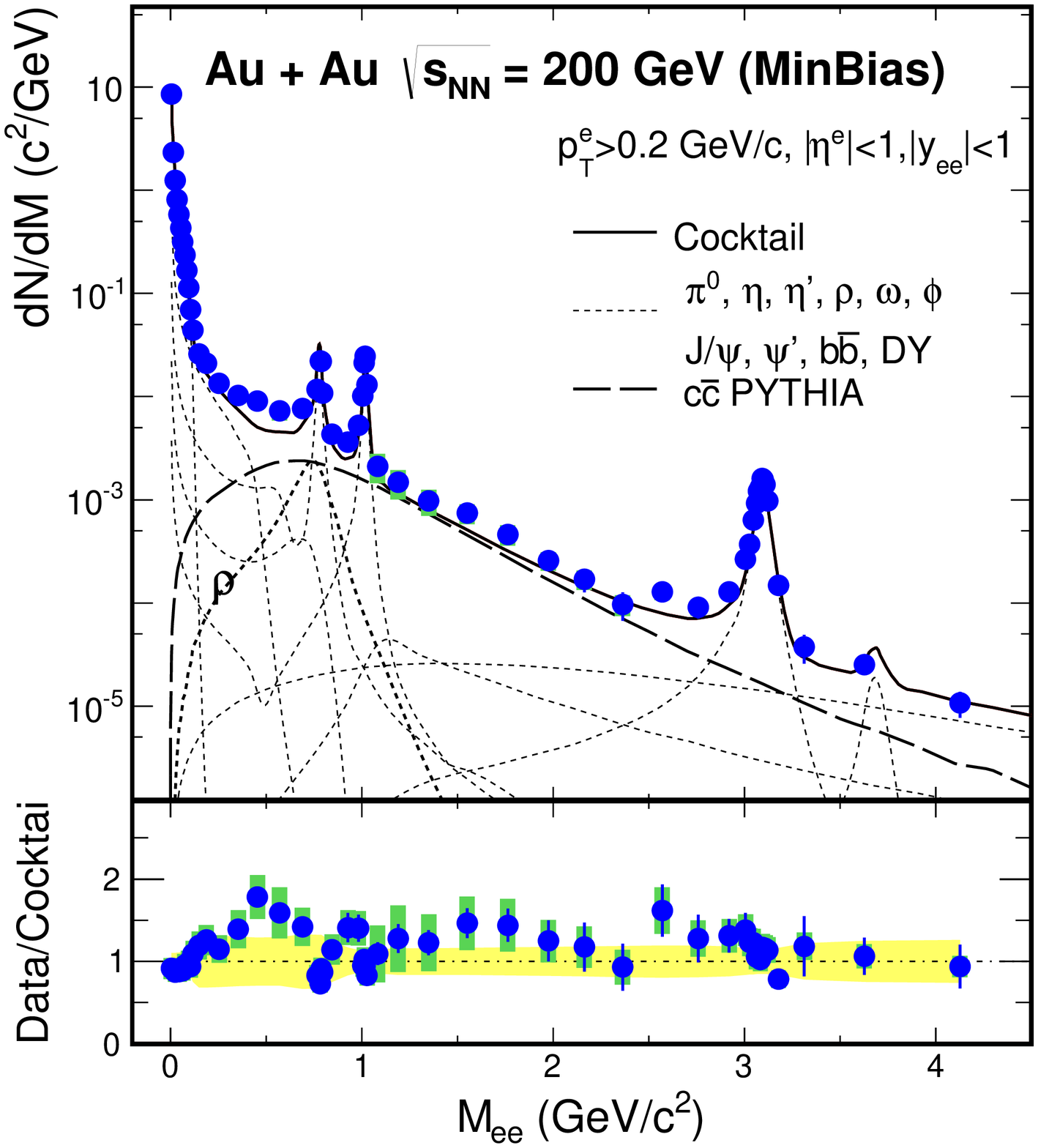}
\emn 
\centering
\bmn[b]{0.33\textwidth}
\includegraphics[width=1.0\textwidth]{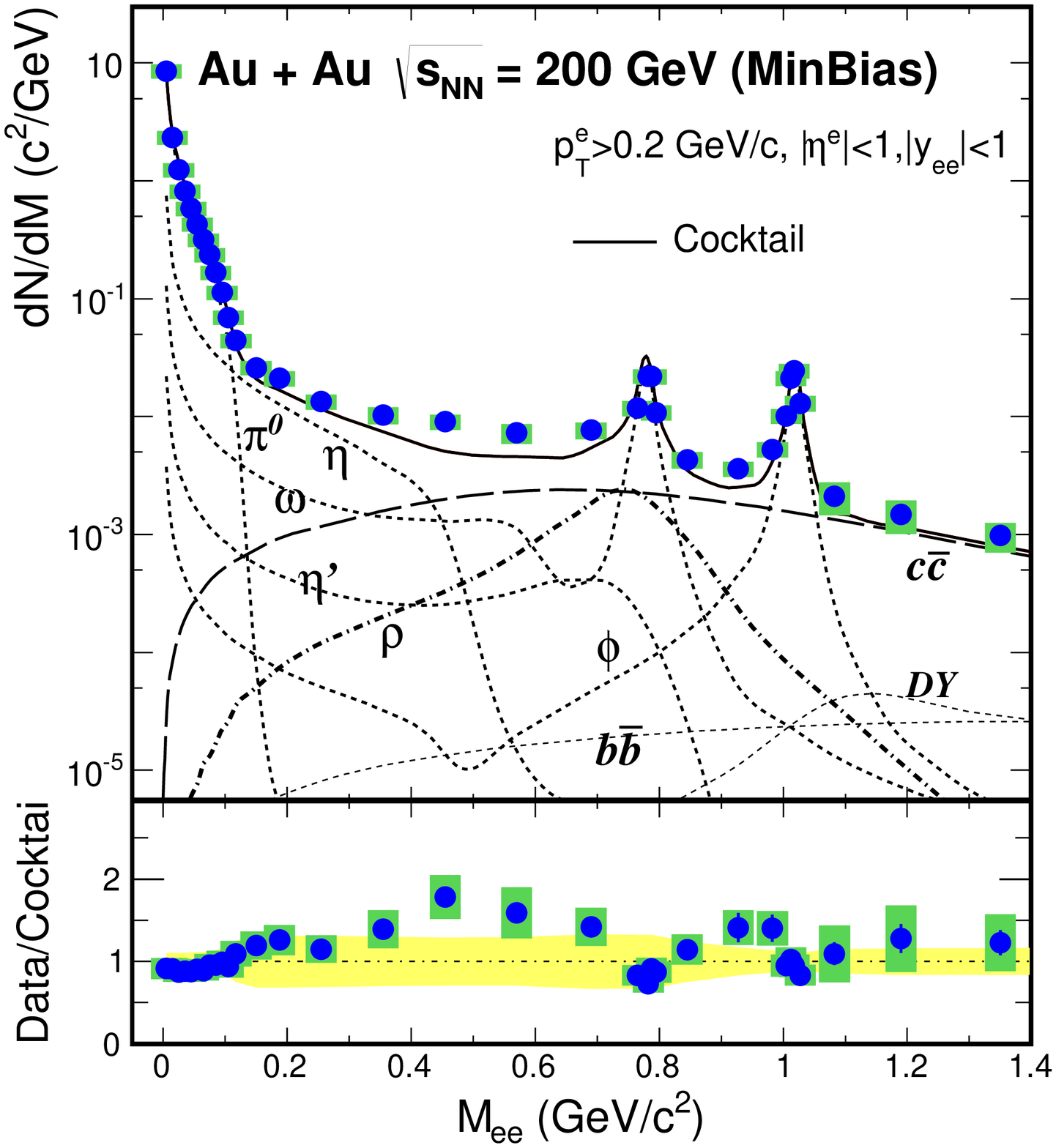}
\emn 
\caption{(Color online) Invariant mass spectrum in the STAR acceptance ($p_T^e>0.2$\,GeV/$c$,
$|\eta^e|<1$, and $|y_{ee}|<1$) from \sNN = 200\,GeV Au$+$Au
  minimum-bias collisions. The mass spectrum is compared to the hadronic
  cocktail simulations without (upper left panel) and 
  with (upper middle) the vacuum $\rho$ contribution 
  (upper right is an expanded version of upper middle below $M_{ee}$ of 1.4 $GeV/c^2$).
  The vertical bars on data points depict the statistical
  uncertainties, while the systematic uncertainties are shown as green boxes.
  Yellow bands in the bottom panels depict the systematic uncertainties on the
  cocktail. The dashed line indicates the charm decay dielectron contribution
  from {\sc Pythia}~\cite{pythia} calculations and scaled with $N_{\rm bin}$.}
\label{invMass}
\end{figure*}

The dielectron yields measured in the STAR acceptance ($p_T^e>0.2$\,GeV/$c$,
$|\eta^e|<1$, and $|y_{ee}|<1$) have been corrected for the dielectron
reconstruction efficiencies. The efficiency correction was done in $p_{\rm
  T}^{ee}$ and $M_{ee}$. The \pT-integrated efficiency-corrected dielectron
mass spectrum $dN/dM$ at midrapidity $|y_{ee}|<1$ in the STAR acceptance from
0-80\% Au$+$Au minimum collisions at \sNN = 200\,GeV is shown in
Fig.~\ref{invMass}. The data are compared to the hadronic cocktail simulations
without (upper left panel) and with (upper middle and upper right panels) the vacuum $\rho$
contribution. The vertical bars on the data points depict the statistical
uncertainty, while the green boxes represent the systematic uncertainty. The
ratios of the data over the cocktail simulations are shown in each of the bottom panels. The
yellow band around unity indicates the uncertainties on the cocktail
calculations. Those are mainly determined by the uncertainties on the $dN/dy$
and the decay branching ratios for each of the individual sources.

\renewcommand{\floatpagefraction}{0.75}

A few more remarks about cocktail calculations are in order:

\begin{itemize}
\item Since the $\rho$ mesons are strongly coupled to the medium in Au$+$Au collisions, 
      their contribution is considered part of the medium dilepton emission and depends on the 
      properties of the medium. We only included the vacuum $\rho$ contribution as a reference here.
      In the default hadronic cocktail calculations, the $\rho$ contribution is omitted 
      in order to allow for possible in-medium $\rho$ contributions depicted 
      by model calculations.

\item Correlated charm contributions included in the cocktail are the number-of-binary collisions ($N_{\rm bin}$) 
      scaled $p$$+$$p$ results calculated from {\sc Pythia}. 

\item Other hadron contributions are described in Section III-G.
\end{itemize}

Comparing the measured data points to the hadronic
cocktail calculations in the LMR,  an enhancement can be observed in the mass region between
0.30 and 0.76\,GeV/$c^2$. This enhancement cannot be fully explained by the
expected vacuum $\rho$ meson contribution as shown in the right plot of Fig.~\ref{invMass}. 
The data, integrated in the mass region of 0.30$-$0.76\,GeV/$c^2$, is a factor of 
1.76\,$\pm$\,0.06\,(stat.)\,$\pm$\,0.26\,(sys.)\,$\pm$\,0.29\,(cocktail) larger than the model cocktail 
without the vacuum $\rho$ contribution. This enhancement factor is significantly lower than what 
has been reported from the dielectron measurement in the PHENIX detector acceptance~\cite{PHENIX}. 

Detailed comparisons of the differences between the STAR and PHENIX experimental acceptances and cocktail simulations are unable to account for the measured enhancement difference. These details are described in Appendices~\ref{Appendix:acceptance} and \ref{Appendix:PHENIXvSTARcocktails}.

In the intermediate mass region (IMR), the cocktail simulations are dominated by correlated charm 
pair decays which are calculated from {\sc Pythia} simulations. 
The simulations generally describe the data but run slightly below the data points, allowing for additional source contributions.
The uncertainty on the charm production cross section $d\sigma^{c\bar{c}}/dy$ at mid-rapidity, 
which is used for the normalization of this contribution, is around 15\%. More precise
measurements in this mass region of both the total charm cross-section as well as the correlation
in Au$+$Au collisions are needed to either verify or rule out significant contributions from other sources, 
such as QGP thermal radiation.



%

\subsection{Comparison to models}

One major motivation for measuring dileptons is the study of
chiral symmetry properties of the QCD medium created in the heavy-ion
collisions. Restoration of the spontaneously broken chiral symmetry will lead
to modification of the vector meson (short-lived $\rho$ meson in particular) spectral
functions, which are accessible via dilepton measurements. 
There are two chiral symmetry restoration scenarios commonly used in calculations:
(a)the drop of the pole mass or degeneracy of vector and axial-vector mesons due to
the reduced $\langle q\bar{q} \rangle$ condensate~\cite{BrownRho}; and (b)
broadening of the spectral function due to many body collisions in the
vector-meson dominance~\cite{RappWambach,RappWambach2,Eletsky}. Both scenarios will
introduce an enhancement in the mass region below the $\rho$ mass compared to
the spectral function in vacuum. Precision measurement from the NA60 experiment
demonstrated that the broadened $\rho$ scenario can reproduce the low mass
dilepton enhancement data at SPS energy~\cite{NA60}, while the dropping mass scenario failed 
to describe the data. It is anticipated that the hadronic medium at top RHIC
energy is similar to that created at SPS energy, 
thus the dilepton production in the LMR region are comparable between SPS and RHIC. 

The QGP contribution to the dilepton spectra has often been calculated perturbatively via 
Born $q+\bar{q}$ annihilation at leading order. Various approaches have
been studied to take into account high-order contributions at finite
$T-\mu_B$~\cite{Braaten}. The QGP contribution is expected to
become sizable for $M>1.5$\,\GeVcsq at  top RHIC energies due to a well
established partonic phase. 

There have been many model calculations for dielectron production at RHIC,
with particular focus on the low mass region. 
We group these models into two categories and describe their features and predictions separately below.


\paragraph*{Category I: Macroscopic effective many-body theory models.}

In these models, the dilepton production in the hadronic medium is calculated via
electromagnetic correlators based on the Vector-Meson Dominance Model (VDM)
approach. Assuming a thermal equilibrated hadronic medium, dilepton rates are
determined by the $\rho$ meson propagator in the medium, which depends on the
interactions of the $\rho$ with mesons and baryons in this medium at finite $T$
and $\mu_B$. It has been shown that the resulting broadened $\rho$ spectral
function is mostly due to the interactions with the baryons rather than
the mesons~\cite{RappWambachvanHees,HJXu,Vujanovic}. Thus, the medium total-baryon
density, and not the net-baryon density or $\mu_B$, is the  critical factor in
determining the dielectron yield in the heavy-ion collisions at these energies.

Dilepton production in the partonic phase is mostly calculated via perturbative
$q\bar{q}$ annihilation with some improved corrections. It has been
demonstrated in these calculations that the dilepton rates from the hadronic
medium, extrapolated bottom-up to $T_{c}$, should be equivalent to the rates from the
partonic medium, extrapolated top-down to $T_{c}$. This is referred to as the 
``parton-hadron" duality~\cite{RappWambachvanHees}. The final resulting
dielectron yields for observation are calculated via the integral over the
full space-time evolution for this medium.
We have chosen one model calculation from Rapp~\cite{RappPriv} from this category in
the following comparisons to our data. Some of the key ingredients in this model
calculation are listed below:

\begin{itemize}
\item Initial spectral functions were fixed using the measurements from
  $e^+e^-$ collision data~\cite{RappWambach2}.
\item Space evolution was chosen to be a cylindrical expanding
  fireball~\cite{Rapp4SPS}
\item The latest Lattice QCD Equation-of-State (EOS) was used, in particular lower
  $T_{\rm c}$ (170 MeV) and $T_{\rm ch}$ (160 MeV) values were chosen in the
  calculations shown here, which are slightly different compared to previous
  calculations from this same model.
\item QGP radiation from the partonic phase was updated as well, and 
  using the choice of the latest Lattice QCD EOS.
\end{itemize}

There are several other model calculations available in this category: some models 
chose different initial spectral functions~\cite{Vujanovic}, and several of
them used the space-time evolution obtained from either ideal or viscous
hydrodynamic model calculations~\cite{HJXu,Vujanovic}. Calculations from
these models show similar results compared to Rapp's model and provide reasonable descriptions
of the low-mass excess observed in our dielectron data in 200~GeV Au$+$Au minimum-bias collisions.



\paragraph*{Category II: Microscopic transport dynamic models.}

We chose the Parton-Hadron String Dynamic (PHSD) transport model from this
category when comparing to our data in the following sections. The PHSD
transport approach incorporates the relevant off-mass-shell dynamics of the vector
mesons and an explicit partonic phase in the early hot and dense reaction
region as well as the dynamics of hadronization~\cite{PHSD}. It allows for a
microscopic study of the various dilepton production channels in non-equilibrium
matter. In the hadronic sector, PHSD is equivalent to the HSD transport
approach that has been used for the description of $p$$+$A and A$+$A collisions
from SIS to RHIC energies. It reproduces fairly well the measured hadron yields,
rapidity distributions, and transverse momentum spectra~\cite{HSD}. The dilepton
radiation by the constituents of the strongly interacting QGP is produced via:
(i) basic Born $q+\bar{q}$ annihilation, (ii) gluon Compton
scattering ($q/\bar{q} + g \rightarrow \gamma^{*} + q/\bar{q}$), and
(iii) quark/anti-quark annihilation with the gluon bremsstrahlung in the final state
($q + \bar{q} \rightarrow g + \gamma^{*}$). Dilepton production in these
partonic channels is calculated with off-mass-shell partons using a phenomenological
parametrization for the quark and gluon propagators in the QGP.

The PHSD model has been used to calculate the dielectron yields in the STAR
acceptance and it shows a fair agreement with our preliminary
data~\cite{PHSDEE}. 




\renewcommand{\floatpagefraction}{0.75}
\begin{figure*}
\centering
\bmn[b]{0.5\textwidth}
\includegraphics[width=1.0\textwidth]{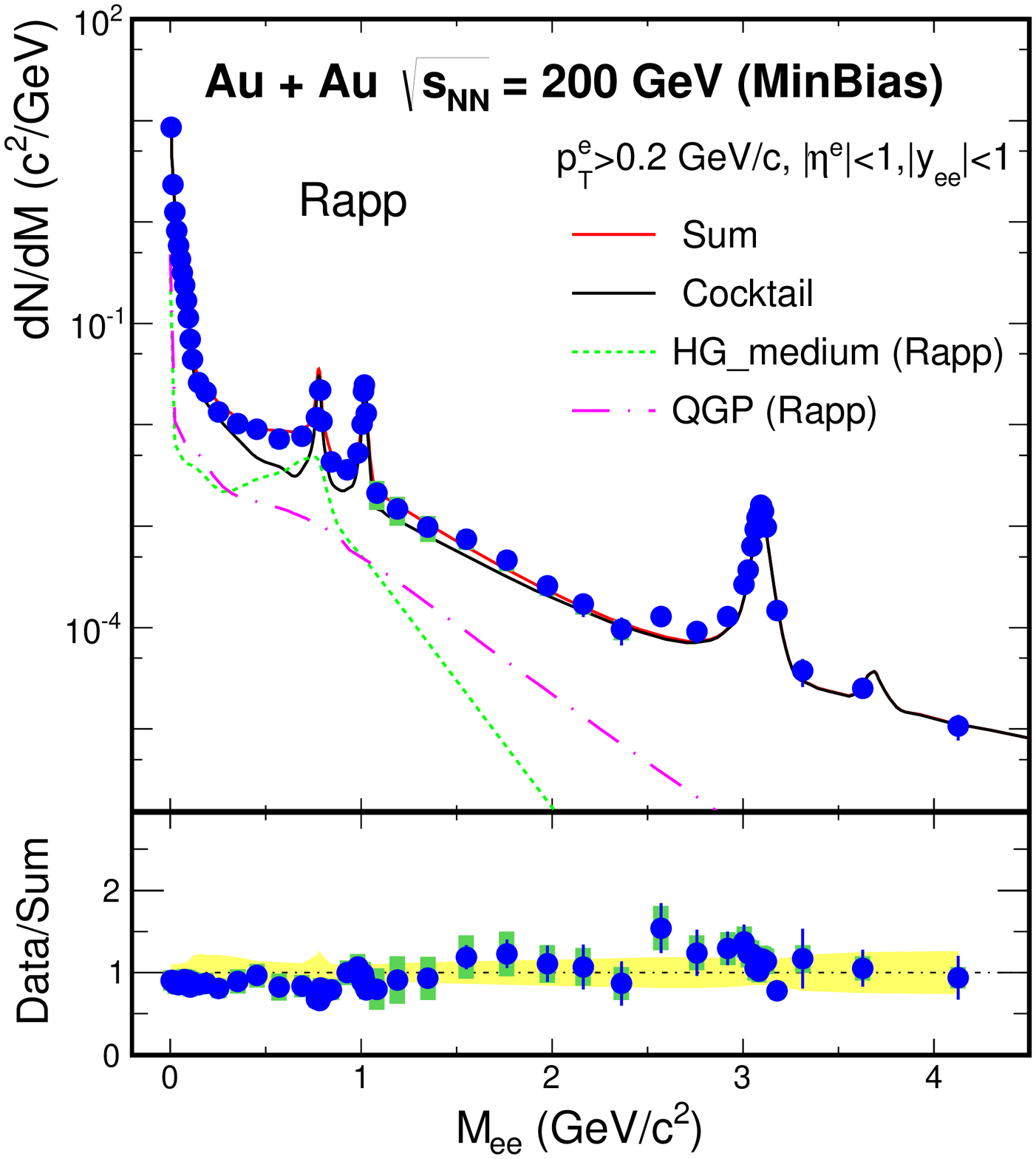}
\emn 
\centering
\bmn[b]{0.5\textwidth}
\includegraphics[width=1.0\textwidth]{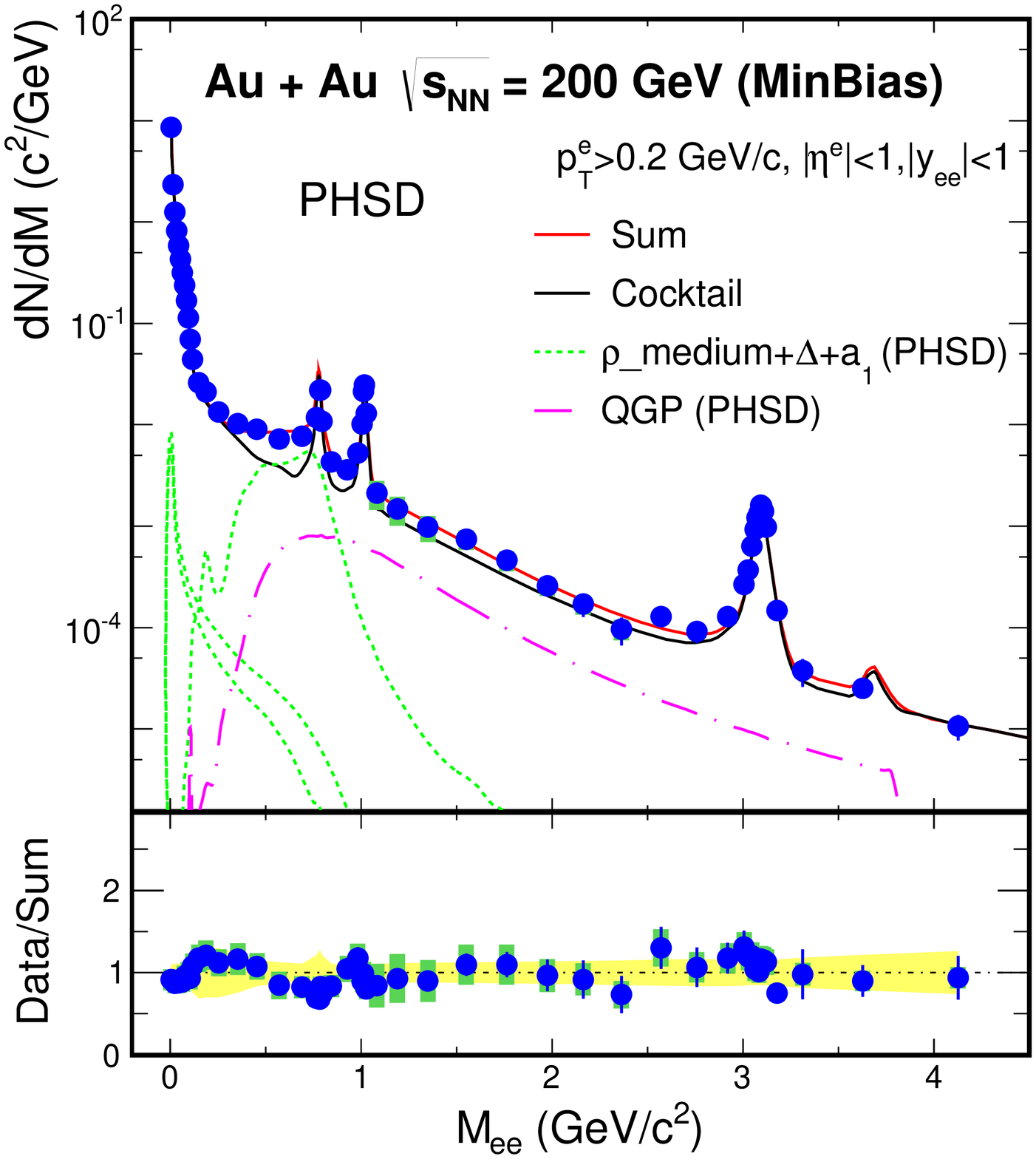}
\emn \\
\caption{(Color online) Dielectron mass spectrum in 200\,GeV minimum-bias Au$+$Au
  collisions compared to the hadron cocktail plus the hadronic medium and
  partonic QGP contributions calculated from Rapp (upper left panel) and PHSD (upper right panel)
  models. Yellow bands in the bottom panels depict systematic uncertainties on the cocktail.} 
\label{FullMasswModels}
\end{figure*}

Detailed comparisons of the model calculations with the data are shown in Fig.~\ref{FullMasswModels}. In the LMR, the data and model calculations are in a fairly good agreement. In the IMR, the charm contribution is the most
important component of $e^{+}e^{-}$ spectrum. We discuss  the
effect of possible modification of this component in Section IV-F.

\subsection{\pT\ dependence}

To gain more insight into dielectron production, we studied the \pT\ dependence of
the dielectron yields in comparison to the hadron cocktail and
model calculations. 
In different \pT\ regions, comparisons to
hadron cocktails require precise knowledge of the light hadron production in a
wide \pT\ region. Details of the cocktail calculations on the \pT\ shape of input
particle are described in Section III-G.

\begin{figure*}
  \centering{
    \includegraphics[width=0.7\textwidth] {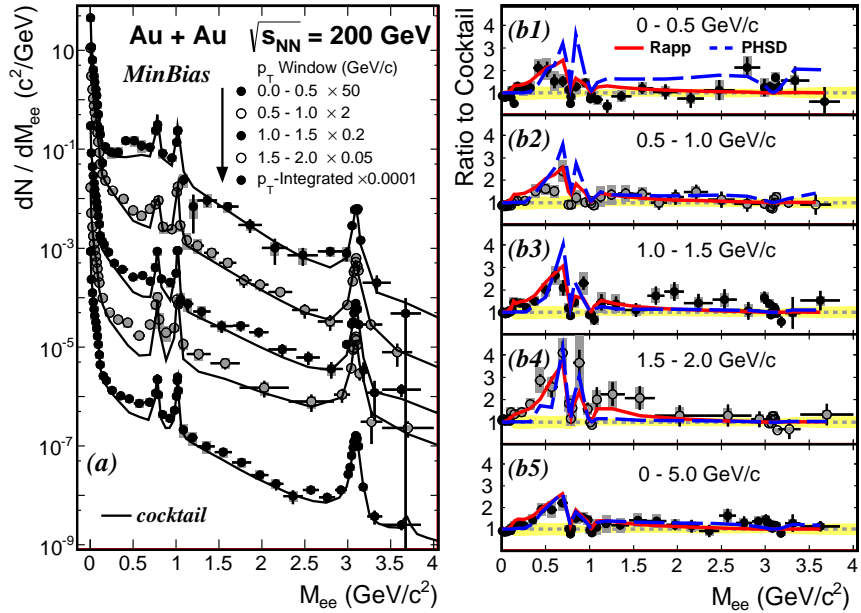}}
  \caption[]{(Color online) Left panel: Invariant mass spectra from \sNN
    = 200\,GeV Au$+$Au minimum-bias collisions in different $p_{T}$
    ranges. The solid curves represent the cocktail of hadronic sources
    and include the charm-decayed dielectron contribution calculated by
    {\sc Pythia} scaled by $N_{\rm bin}$. Right panels: The ratio of dielectron
    yield over cocktail for different $p_{T}$ bins, and the comparison with model calculations. The gray boxes show
    the systematic uncertainties of the data.  Yellow bands depict systematic uncertainties on the cocktail.
  }
  \label{invMassPt}
\end{figure*}

The measured dielectron yields within STAR acceptance
in each individual \pT\ region as well as the total expected hadron cocktail
contributions are shown in the left panel of Fig.~\ref{invMassPt}.
Note that the correlated charm contributions, which become very important in the mass region from 0.5$-$3.0\,GeV/$c^2$, 
were all taken from the $N_\mathrm{bin}$ scaled {\sc Pythia} calculations.
The ratios of data over cocktail calculations as a function of $M_{ee}$ for several transverse momentum 
ranges are shown in the right panels of Fig.~\ref{invMassPt}. For comparison, the theoretical
model calculations in each \pT\ window are included as well. The enhancement factor with respect to the hadronic cocktail
does not change significantly in these  \pT\ bins. Both
theoretical models are able to reasonably describe the LMR excess in all \pT\
bins.

\begin{figure*}
  \centering{
    \includegraphics[width=0.7\textwidth] {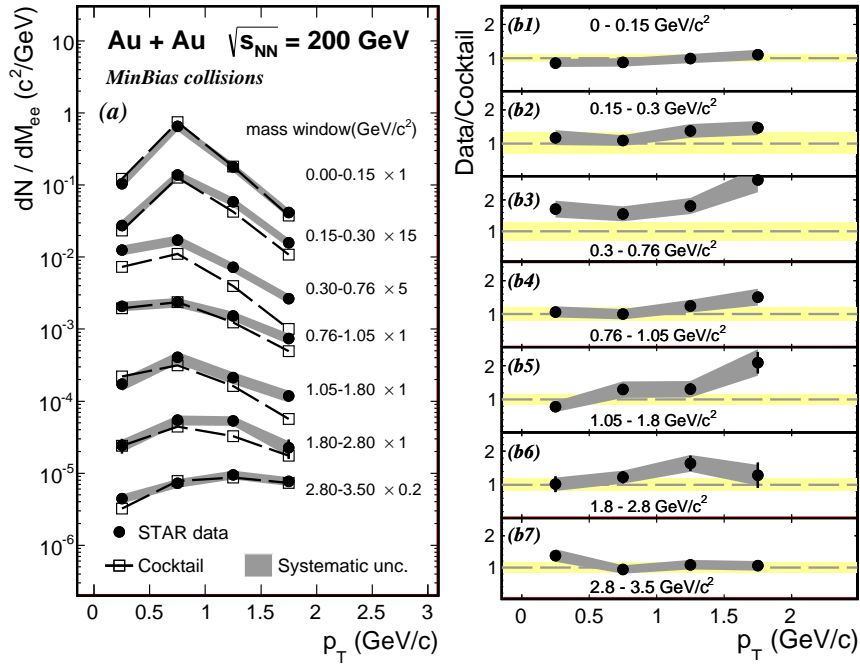}}
        \caption[]{(Color online) Left panel: The integrated dielectron yield
          as a function of pair $p_{T}$ in different invariant mass ranges compared with
          cocktail. The solid lines represent the yield of the cocktail in
          different mass ranges , while the gray bands show the systematic
          uncertainties of the data. Right panels: The ratio of dielectron yield
          over cocktail for different mass ranges as a function of pair
          $p_{T}$. The yellow bands show the systematic uncertainties of the
          cocktail. The gray bands show the systematic uncertainties of the
          data.} 
  \label{enhancePt}
\end{figure*}

We quantify the \pT\ dependence by comparing the measured dielectron yields with the cocktail in
each mass window within the STAR acceptance, the results from Au+Au 0-80\%
minimum-bias collisions at 200\,GeV are shown in Fig.~\ref{enhancePt}. The left panel
shows the measured data points (markers) together with cocktail
calculations (dashed lines). The ratios of the data over the cocktail are shown in
the right panels for different mass windows. The data points and the cocktail
calculations are in good agreement throughout the measured \pT\ range up to
2\,GeV/$c$ in the mass regions of the $\pi^0$ (up to 0.15\,GeV/$c^2$), the $\omega$/$\phi$
(0.76$-$1.05\,GeV/$c^2$), and the $J/\psi$ mesons (2.8$-$3.5\,GeV/$c^2$).
In the LMR region, particularly in the mass region of 0.30$-$0.76\,GeV/$c^2$, we
see that the relative enhancement in the data compared to the cocktail has no
significant \pT\ dependence. 
Table~\ref{tableLMRpT} summarizes the enhancement factors for each \pT\ bin. 
In the IMR region, cocktail calculation can describe the data reasonably well. 
Due to the large uncertainty on the correlated charm contribution there is little constraint 
on other possible dilepton contributions, {\it e.g.}, QGP thermal radiation.

\renewcommand{\floatpagefraction}{0.75}

\renewcommand{\floatpagefraction}{0.75}

\bt
\caption{The \pT\ dependence of dielectron yields, measured in the STAR acceptance, and the
  enhancement factor with respect to the hadronic cocktail in the invariant mass region
  of 0.30$-$0.76\,\GeVcsq.}
\centering
\begin{tabular}{c|c|c}
\hline\hline
 \pT\ (GeV/$c$) & yield  ($\times 10^{-3}$) & yield/cocktail \\ \hline \hline

0 -0.5  & $115 \pm 0.09 \pm 0.20 $ & $ 1.71 \pm 0.12 \pm 0.29$ \\ \hline 
0.5-1.0 & $158 \pm 0.07 \pm 0.27 $ & $ 1.56 \pm 0.07 \pm 0.27$ \\ \hline 
1.0-1.5 & $066 \pm 0.03 \pm 0.11 $ & $ 1.81 \pm 0.09 \pm 0.29$ \\ \hline 
1.5-2.0 & $024 \pm 0.02 \pm 0.04 $ & $ 2.65 \pm 0.16 \pm 0.44$ \\ \hline

\hline
\end{tabular}
\label{tableLMRpT}
\et

\subsection{Centrality dependence}

\begin{figure*}
  \centering{
    \includegraphics[width=0.7\textwidth] {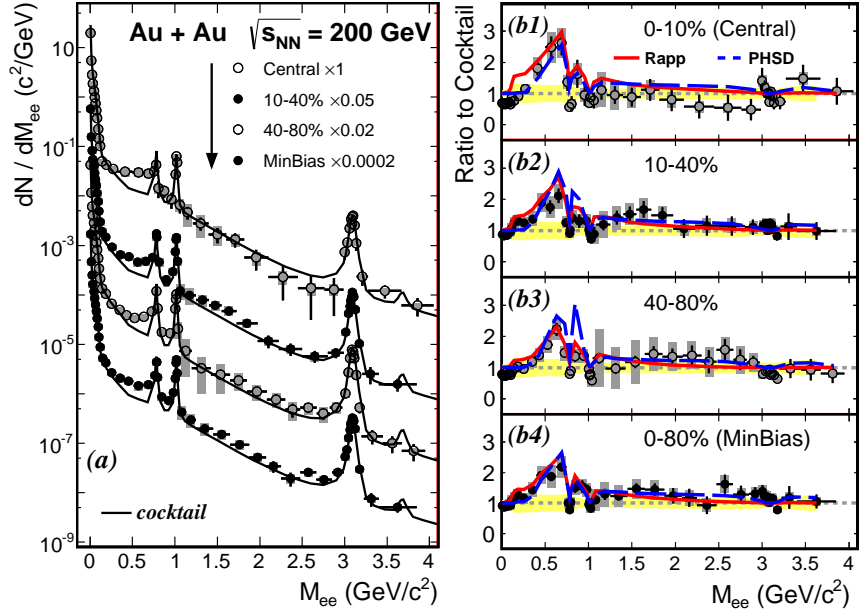}}
        \caption[]{(Color online) Left panel: Invariant mass spectra from \sNN
          = 200\,GeV Au$+$Au collisions in different centralities. The solid
          curves represent the cocktail of hadronic sources and include charm decay 
		  dielectron contribution from {\sc Pythia} scaled by
          $N_{\rm bin}$. Right panel: The ratio of dielectron yield over
          cocktail for different centrality. The gray boxes show the systematic
          uncertainties of the data. Yellow bands depict systematic uncertainties on the cocktail.}
  \label{invMassCent}
\end{figure*}

\begin{figure*}
  \centering{
    \includegraphics[width=0.7\textwidth] {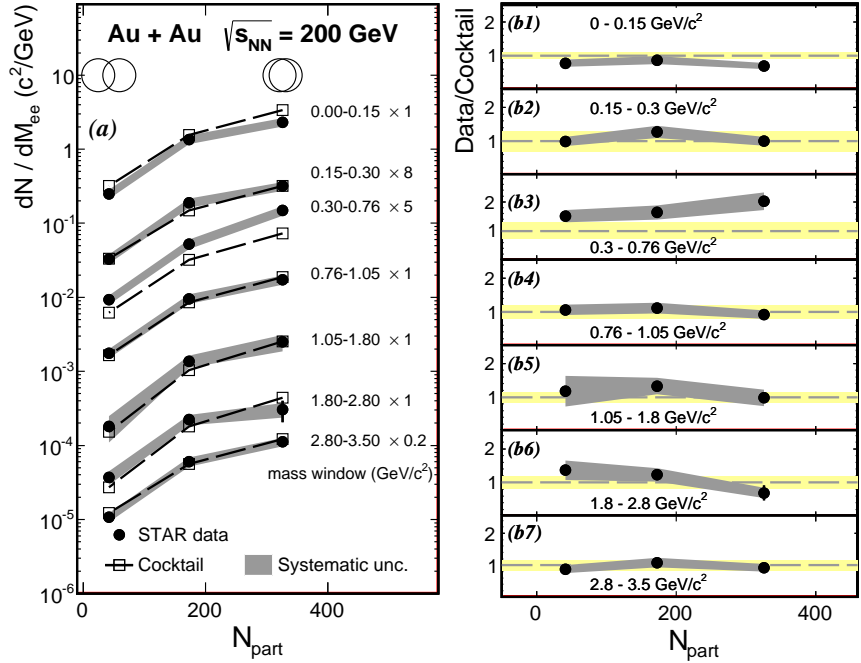}}
        \caption[]{(Color online) Left panel: The integrated dielectron yield
          in different centralities compared with
          the cocktail. The solid lines represent the yield of cocktail in
          different invariant mass ranges. The gray bands show the systematic
          uncertainties of the data. Right panel: The ratio of dielectron yield
          over cocktail for different centralities, and the comparison with model calculations.
          The yellow bands show the systematic uncertainties of the cocktail.
          The gray bands show the systematic uncertainties of the data.
	}
  \label{enhanceCent}
\end{figure*}

The dielectron spectra are studied in various centrality bins (0-10\%, 10-40\% and
40-80\% ). 
The left panel of Fig.~\ref{invMassCent} shows the dielectron spectra in
these centrality bins compared to cocktail calculations. 
The ratios of the data to the cocktail are presented in the right panels. Model calculations are also included in the right plots. 
In Fig.~\ref{enhanceCent}, we quantify the measured yields as a function of centrality by means of $N_\mathrm{part}$ for different mass windows.

In the LMR, particularly in the mass region 0.30$-$0.76\,GeV/$c^2$, the
observed enhancement factor of the dielectron yield with respect to the cocktail does
not show a significant centrality dependence 
within current uncertainty. 
Both theoretical models can reasonably reproduce the centrality
dependence of this observed enhancement in the LMR. 
Table~\ref{tableLMRcen} summarizes the enhancement factors for each centrality bin. 

\begin{figure}
  \centering{
    \includegraphics[width=0.5\textwidth] {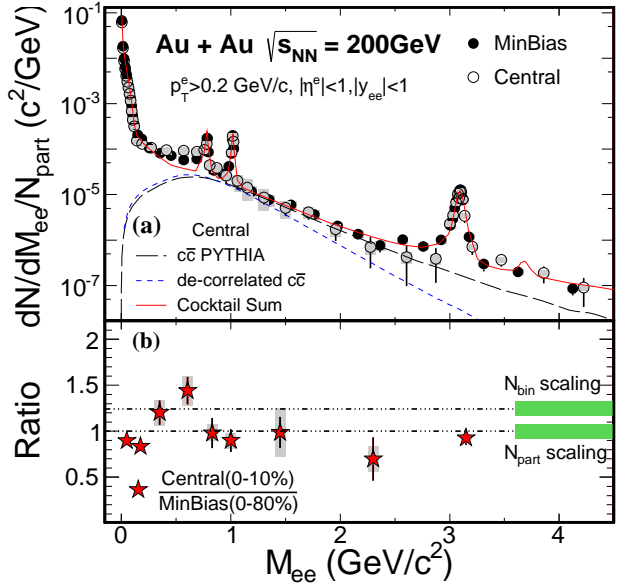}}
        \caption[]{(Color Online) Upper panel: Dielectron invariant mass
          spectra from minimum-bias and the most central (0-10\%) collisions that 
          we are able to achieve most statistics at present. For direct comparison, the
          spectra are scaled with the number of participant nucleons
          ($N_{\rm part}$). The solid line represents the cocktail for minimum-bias
          collisions. Bottom panel: The ratio of the $N_{\rm part}$-scaled dielectron
          yield between minimum-bias and the most central collisions. 
		  The gray boxes show the systematic uncertainties of the data.}
  \label{dielectronRcp}
\end{figure}

In Fig.~\ref{dielectronRcp}, we overlay the dielectron mass spectra from
minimum-bias and the most central (0-10\%) collisions for which we are able to achieve
sufficient statistics for direct comparisons. The $N_\mathrm{part}$-scaled spectra are plotted in the upper panel, 
and the ratio between them is plotted in the bottom panel. 
The measured ratio is consistent with unity in the $\pi^0$ invariant mass region, 
indicating that the production scales with $N_{\rm part}$. 
The ratio starts to increase in the mass around 0.5$-$1.0\,GeV/$c^2$. 
This observation is consistent with a picture in which the correlated charm
contribution starts to be a dominant source in this mass region
while charm quark production at RHIC is expected to rather scale with $N_{\rm bin}$. 
Additionally, in this invariant mass range the in-medium $\rho$ meson contribution
from the hadronic medium is expected to increase faster than \Npart when moving towards central
collisions based on model calculations~\cite{RappPriv}. In the IMR, the
data indicate there is potentially a systematic change in the mass spectra
when comparing the minimum-bias and central collisions. This is suggestive of a possible 
modification of charmed hadron production or other contributions such as thermal radiation. To quantify the difference, 
exponential fits were performed to the mass spectra in central and minimum-bias
collisions and the resulting exponential slopes differ by $\sim$1.5$\sigma$.

\renewcommand{\floatpagefraction}{0.75}

\renewcommand{\floatpagefraction}{0.75}

\bt
\caption{The centrality dependence of dielectron yields, measured in the STAR acceptance,
  and the enhancement factor with respect to the hadronic cocktail in the invariant mass
  region of 0.30$-$0.76\,\GeVcsq.}
\centering
\begin{tabular}{rcl|c|c}
\hline\hline
\multicolumn{3}{c|}{centrality} & yield ($\times 10^{-3}$) & yield/cocktail \\ \hline \hline
\hspace{0.05in}  

0 &-&10\% & $13.63 \pm 1.01 \pm 2.06 $ & $ 2.03 \pm 0.15 \pm 0.31$ \\ \hline 
10&-&40\% & $4.81  \pm 0.22 \pm 0.71 $ & $ 1.63 \pm 0.08 \pm 0.24$ \\ \hline 
40&-&80\% & $0.85  \pm 0.03 \pm 0.12 $ & $ 1.51 \pm 0.06 \pm 0.22$ \\ \hline 
0 &-&80\% & $3.87  \pm 0.13 \pm 0.57 $ & $ 1.76 \pm 0.06 \pm 0.26$ \\ \hline

\hline
\end{tabular}
\label{tableLMRcen}
\et

\subsection{Low mass excess yields}

\begin{figure}
  \centering{
    \includegraphics[width=0.45\textwidth] {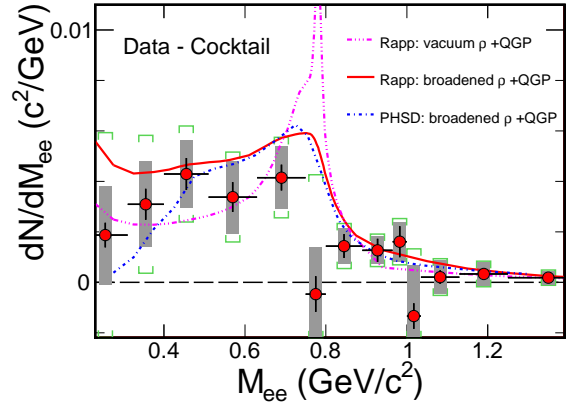}}
    \caption[]{(Color Online) Mass spectrum of the excess (data - cocktail) in the
	low-mass region in Au$+$Au minimum-bias collisions compared to model calculations. Green
	brackets depict the total systematic uncertainties including those from cocktails.
	}
  \label{excess_spectra}
\end{figure}

We subtracted the cocktail contribution from the measured dielecton mass spectrum to obtain the direct excess yields, shown in Fig.~\ref{excess_spectra} for the mass region of 0.3-1.4 GeV/$c^2$. The cocktail simulations used in the subtraction include the correlated charm contributions from {\sc Pythia} assuming the $N_{\rm bin}$ scaling. A possible charm de-correlation leads to a negligible modification of the cocktail spectra in the mass region around 0.5 GeV/$c^2$ as shown in Fig.~\ref{dielectronRcp}. The obtained excess spectra in Au$+$Au minimum-bias collisions are compared to model calculations in Fig. ~\ref{excess_spectra}.

The systematic uncertainty across all the data points are highly correlated. We utilized the modified $\chi^2$-test~\cite{newchi2} to quantify the comparison between the data and the model calculations; the results are summarized in Table~\ref{tableExcess}. 
The vacuum $\rho$ plus QGP scenario in Rapp's implementation cannot describe our data well.
The calculations, including the broadened $\rho$-meson scenario plus QGP contribution from both Rapp and PHSD, have reasonable agreements with our data. 

\begin{figure}
  \centering{
    \includegraphics[width=0.45\textwidth] {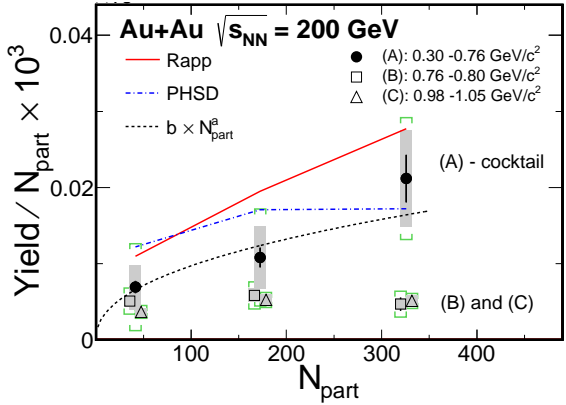}}
    \caption[]{(Color Online) The yields scaled by $N_\mathrm{part}$ for the
	$\rho$-like region with the cocktail subtracted, and the $\omega$-like 
	and the $\phi$-like regions without cocktail subtraction as a function of $N_\mathrm{part}$.
	Model calculations are included as red solid and dot-dashed curves, 
	while the dashed curve depicts a power-law 
    fit to the yield/$N_\mathrm{part}$ for the $\rho$-like region with the cocktail subtracted. 
	Systematic uncertainties from the data are shown as grey boxes, and the green brackets
	depict the total systematic uncertainties including those
	from cocktails. For clarity, the $\omega$-like and $\phi$-like data
	points are slightly horizontally displaced.}
  \label{excess_npart}
\end{figure}

Next, we studied the centrality dependence of the excess yields. In Fig.~\ref{excess_npart}, 
the integrated excess yields scaled by $N_{\rm part}$ as a function of centrality ($N_{\rm part}$) 
are shown in the $\rho$-like mass region (0.30-0.76 GeV/$c^2$). 
In the same figure, the $\omega$-like (0.76-0.80 GeV/$c^2$) and $\phi$-like (0.98-1.05 GeV/$c^2$) dielectron yields are plotted. 
For both sets, the yields were scaled by $N_{\rm part}$ and the cocktail subtraction was not applied in this range. 
The $\omega$-like and the $\phi$-like dielectron yields show an $N_{\rm part}$ scaling while the $\rho$-like 
dielectron excess yields increase faster than $N_{\rm part}$ as a function of centrality. 
The dashed curve depicts a power fit ($\propto N_{\rm part}^a$) to the $\rho$-like dielectron yields with the cocktail subtracted. 
The fit result shows $a=0.44\pm0.10$ (stat.+uncorrelated sys.), 
indicating the dielectron yields in the $\rho$-like region are sensitive to the QCD medium dynamics, 
as expected from $\rho$ medium modifications in theoretical calculations~\cite{RappPriv,RhoOmega}.

\bt
\caption{ Reduced $\chi^{2}$ for model calculations compared to the excess data in the invariant mass region of  0.3-1.0 \GeVcsq.}
\centering
\begin{tabular}{rl|c|c}
\hline\hline
 \multicolumn{2}{c|}{Model}      & $\chi^{2}/\mathrm{ndf}$ & $p$-value \\ \hline \hline
 Rapp : &vacuum $\rho$ + QGP     &  41.3/8  &  2.4$\times 10^{-7}$ \\ \hline 
 Rapp : &broadened $\rho$ + QGP  &  8.0/8   &  0.32                \\ \hline 
 PHSD : &broadened $\rho$ + QGP  &  16.5/8  &  0.040               \\ \hline 
\hline
\end{tabular}
\label{tableExcess}
\et

\subsection{Correlated charm contributions}

The correlated charm contributions start to play an important role in our
measured dielectron yields above 0.5\,\GeVcsq and dominate the cocktail in
the intermediate mass region. So far, no measurement of charm
correlation in the low transverse moment region at RHIC exists in either $p$$+$$p$ or A$+$A
collisions. Single-charm hadron spectra or their decay (``non-photonic") electron
spectra have been measured in $p$$+$$p$~\cite{DataCharm,DataNPESTAR} and Au$+$Au
collisions~\cite{DataNPEPHENIX, AuAuD0}. We relied on the {\sc Pythia} model to create the
correlated charm pairs and then calculate the decay-electron pair
distributions.

In $p$$+$$p$ collisions, with a tuned {\sc Pythia} setting: MSEL=1, PARP(91) ($\langle
k_{\perp}\rangle$) = 1.0\,GeV/$c$ and PARP(67) (parton shower level) = 1.0, we
have shown that this can reproduce the measured single $D$-meson \pT\ spectrum
from 0.6 $-$ 6\,GeV/$c$~\cite{DataCharm}. The dielectron mass spectrum
calculated with this {\sc Pythia} tune also showed a good agreement with our measurement
in the IMR in $p$$+$$p$ collisions at 200\,GeV. However, the limited statistics in
$p$$+$$p$ collisions do not allow us to determine whether {\sc Pythia} can produce the
correct $D-\bar{D}$ correlation.

In Au$+$Au collisions, we have observed that high-\pT\ electrons
are strongly suppressed compared to $p$$+$$p$ collisions. In the low \pT\
region, various model calculations indicate that the single-charm spectrum can
be modified due to interactions between charm quarks and the hot and dense
medium~\cite{HeGossiaux}. Consequently, the $D$$-$$\bar{D}$ correlation seen in
$p$$+$$p$ collisions will be modified, or even be completely washed
out~\cite{ZhuCharmCorr}. To study their impact on the dielectron spectrum, we
chose the following different configurations for the charm \pT\ spectra and
correlation functions to construct $D$$-$$\bar{D}$ pairs.

\begin{enumerate}[label=(\alph*)]
\item Keep the direct {\sc Pythia} calculation which was used in our default
  cocktail calculations.

\item Keep the momentum magnitude of charm decay electrons in {\sc Pythia}, but
  randomly select the azimuthal angle direction. In this case, the
  angular correlation between two electrons is completely washed out.


\item Randomly sample two electrons with the single electron $p_{T}$, $\eta$, $\phi$ distributions from {\sc Pythia} calculation. 
In this case, the correlation between the two electrons is completely washed out.

\item Based on (c), but sample the \pT\ of each electron track according to the modified \pT\
  distribution based on the non-photonic electron $R_{\rm AA}$ measurement in
  Au$+$Au collisions~\cite{DataNPEPHENIX}. The electron $R_{\rm AA}(p_{\rm T})$ was parametrized
  using the following function, with $p_{\rm T}$ in units of GeV/$c$.

\begin{equation}
R_{\rm AA} (p_{\rm T}) = \frac{4.70}{4.63 + e^{({\rm p_T}-0.62)/1.06}} - 0.22.
\label{NPERAA}
\end{equation}


\end{enumerate}

\begin{figure}
  \centering{
    \includegraphics[width=0.5\textwidth] {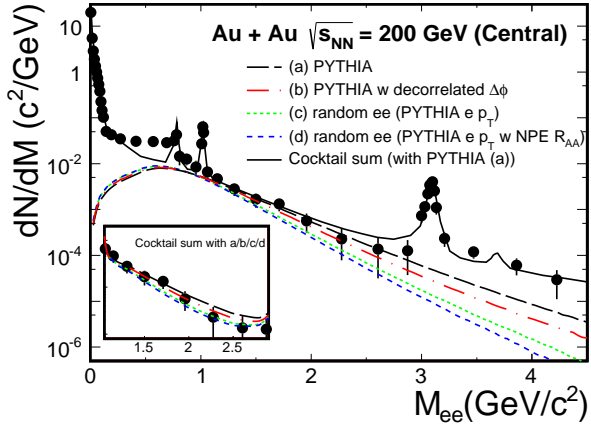}}
        \caption[]{(Color Online) Correlated charm contributions to the dielectron mass spectra for different
          assumptions of the correlation strength. The simulations are compared to the measured
          dielectron spectrum in the most central (0-10\%) Au$+$Au collisions. The total
          cocktail shown is calculated using the default {\sc Pythia}
          correlations. The insert plot shows the comparsion between the cocktail sums with above four different charm contribtuions and the measured spectrum in the IMR.}
  \label{yieldwCharm}
\end{figure}

\begin{figure}
  \centering{
    \includegraphics[width=0.5\textwidth] {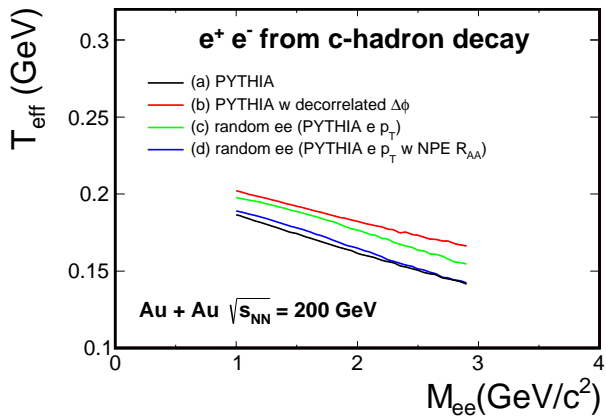}}
        \caption[]{(Color Online) Slope parameters $T_{\rm eff}$ versus invariant mass for
          dielectrons from charm hadron decays. Different lines show the
          outcome from {\sc Pythia} calculations assuming different levels of
          correlations between charm pairs.}
  \label{TeffwCharm}
\end{figure}

All these calculations were scaled with $N_{\rm bin}$ in each centrality bin to obtain the correlated charm mass spectra.
The correlated charm mass spectra for the above
four cases in the most central (0-10\%) Au$+$Au collisions, and a comparison with the
measured data, are shown in Fig.~\ref{yieldwCharm}. 
The total cocktail shown is still calculated based on the
default {\sc Pythia} correlations. The figure shows that both the modification in
electron momentum and the smearing in azimuthal angular correlation make the
dielectron mass distribution steeper. Calculations for case (d) seem to be
closer to the data points in the mass region of 1$-$3\,GeV/$c^2$, thus indicating
a possibly modification of charmed hadron production in central Au$+$Au collisions 
that is worthy of further experimental investigations. 
We also calculated the slope parameter in the transverse mass spectrum for
each of the aforementioned cases, as is shown in Fig.~\ref{TeffwCharm}.

\subsection{Low mass vector meson yields}

\begin{figure*}
  \centering{
    \includegraphics[width=0.45\textwidth] {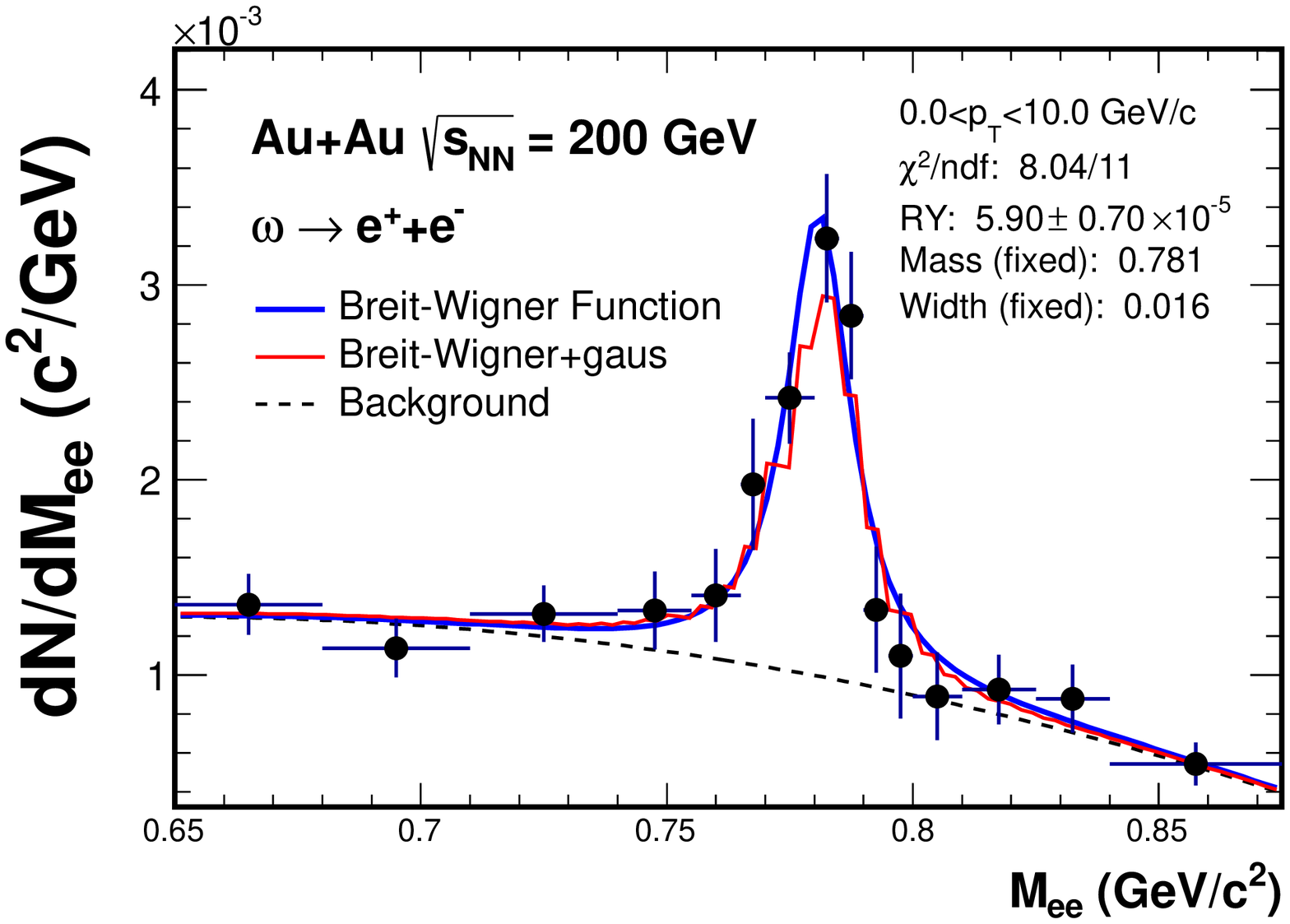} 
    \includegraphics[width=0.45\textwidth] {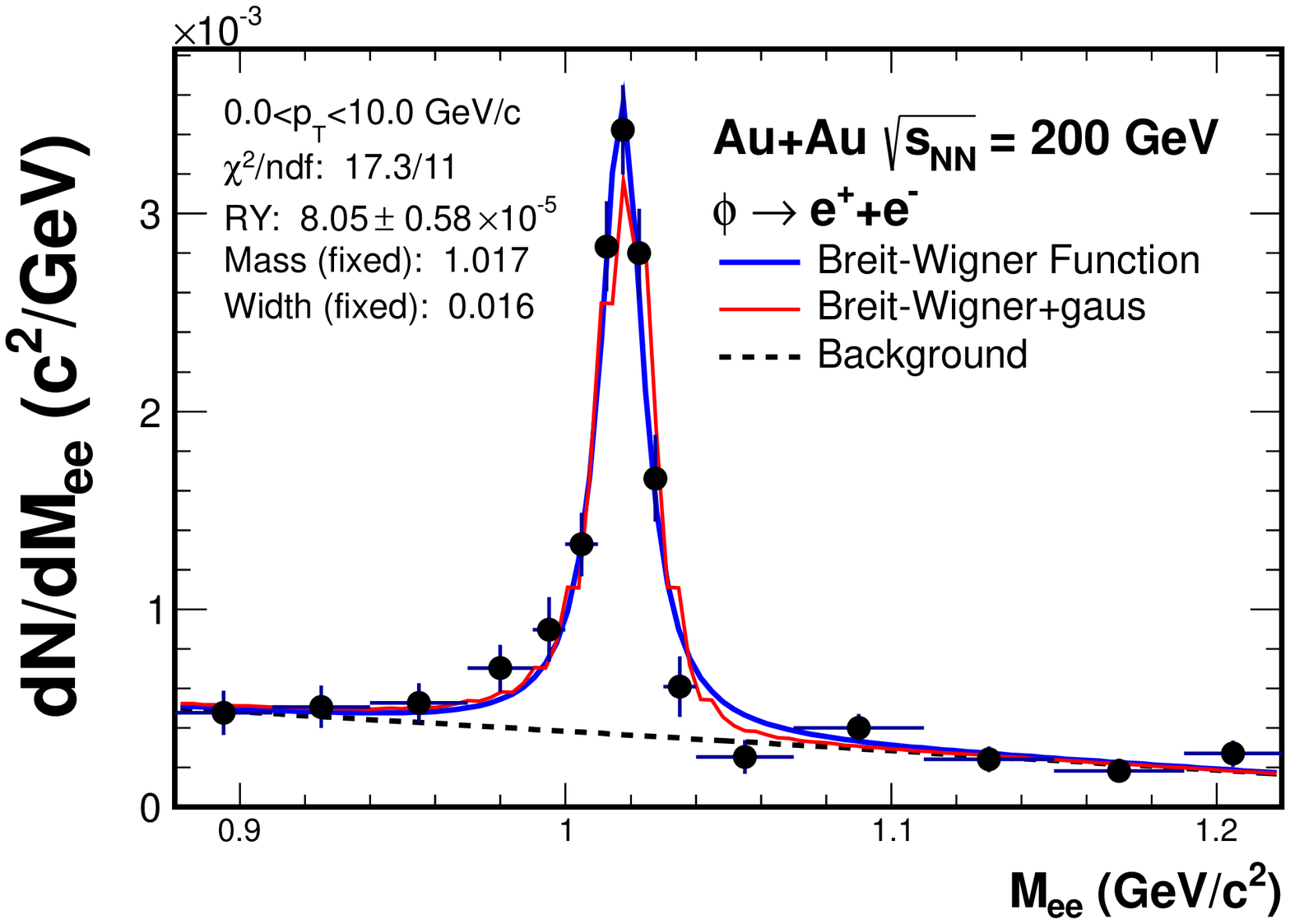}}
    \caption[]{(Color Online) $\omega$ and $\phi$ meson invariant mass distribution from \sNN = 200 GeV Au$+$Au minimum-bias collisions after subtraction of the combinatorial background  using the mixed-event method. The blue and red lines depict two functions used for the signals in the fit. A second-order polynomial function is used to describe the residual correlated background.}
  \label{vectormeson}
\end{figure*}

\begin{figure*}
  \centering{
    \includegraphics[width=1.0\textwidth] {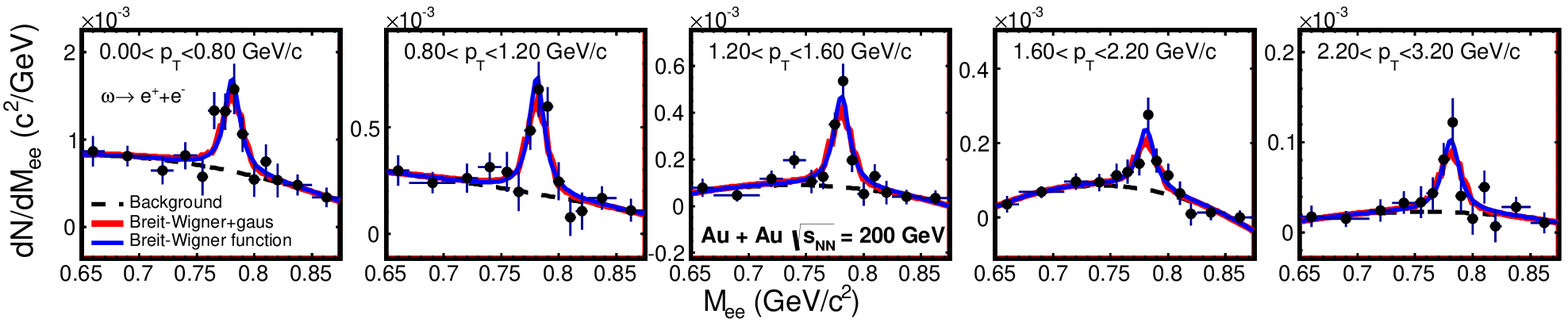} 
    \includegraphics[width=1.0\textwidth] {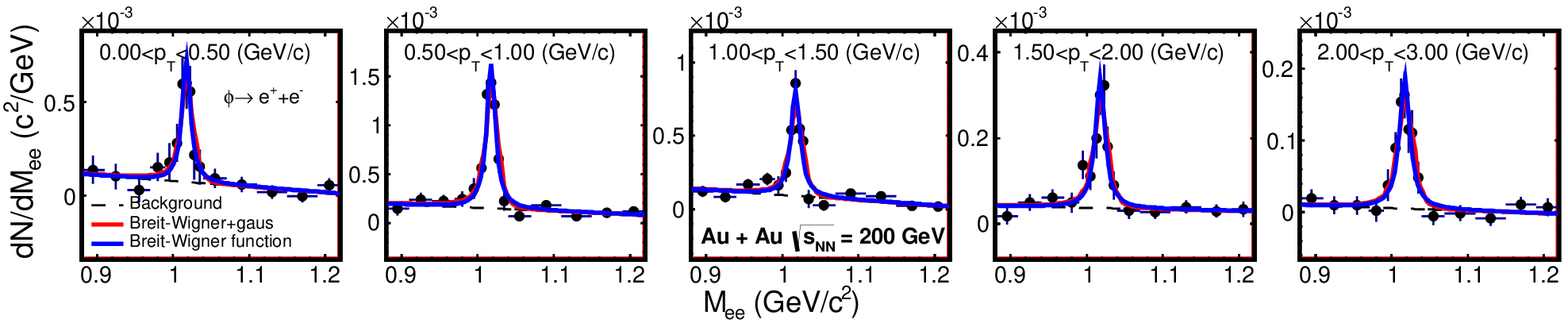}}
    \caption[]{(Color Online) $p_{T}$ dependence of the $\omega$ and $\phi$ meson invariant mass distributions from \sNN = 200 GeV Au$+$Au minimum-bias collisions.}
  \label{meson_pt}
\end{figure*}

The low mass vector meson ($\omega$ and $\phi$) yields have been extracted from the dielectron decay channel 
through this analysis. 
The results reported here are from combined data taken in RHIC year 2010 and 2011 runs. The measured $\phi$ yields are consistent with the results from a recent STAR publication~\cite{STARphi2ee}.
Figure~\ref{vectormeson} shows the invariant mass distributions of the vector 
mesons $\omega$ and $\phi$ from \sNN = 200~GeV Au$+$Au minimum-bias collisions. 
The signal spectra are reconstructed by subtracting the normalized mixed-event unlike-sign 
background (Section III-E3) from the inclusive same-event unlike-sign $e^{+}e^{-}$ distribution. 
A Breit-Wigner function plus a second order polynomial function are used to fit the invariant mass distributions. 
The second order polynomial function is used to describe the residual background. 
In addition, we use the vector meson $\omega$($\phi$) invariant mass distributions (line-shapes) directly 
from cocktail simulations (Breit-Wigner plus Gaussian functions) to fit the signal. 
As described in Section III-G, the detector momentum resolution in the cocktail simulation 
was estimated by tuning the simulation to match the $J/\psi$ signals in the data. 
The line shapes from this tuned simulation for the $\omega$ and $\phi$ mesons reproduce the signal well. 
The difference between these two methods is included in the systematic uncertainty of the raw yield. 
Figure~\ref{meson_pt} shows the $\omega$ and $\phi$ invariant mass distributions in different $p_{T}$ regions.  
    
\begin{figure}
  \centering{
    \includegraphics[width=0.45\textwidth] {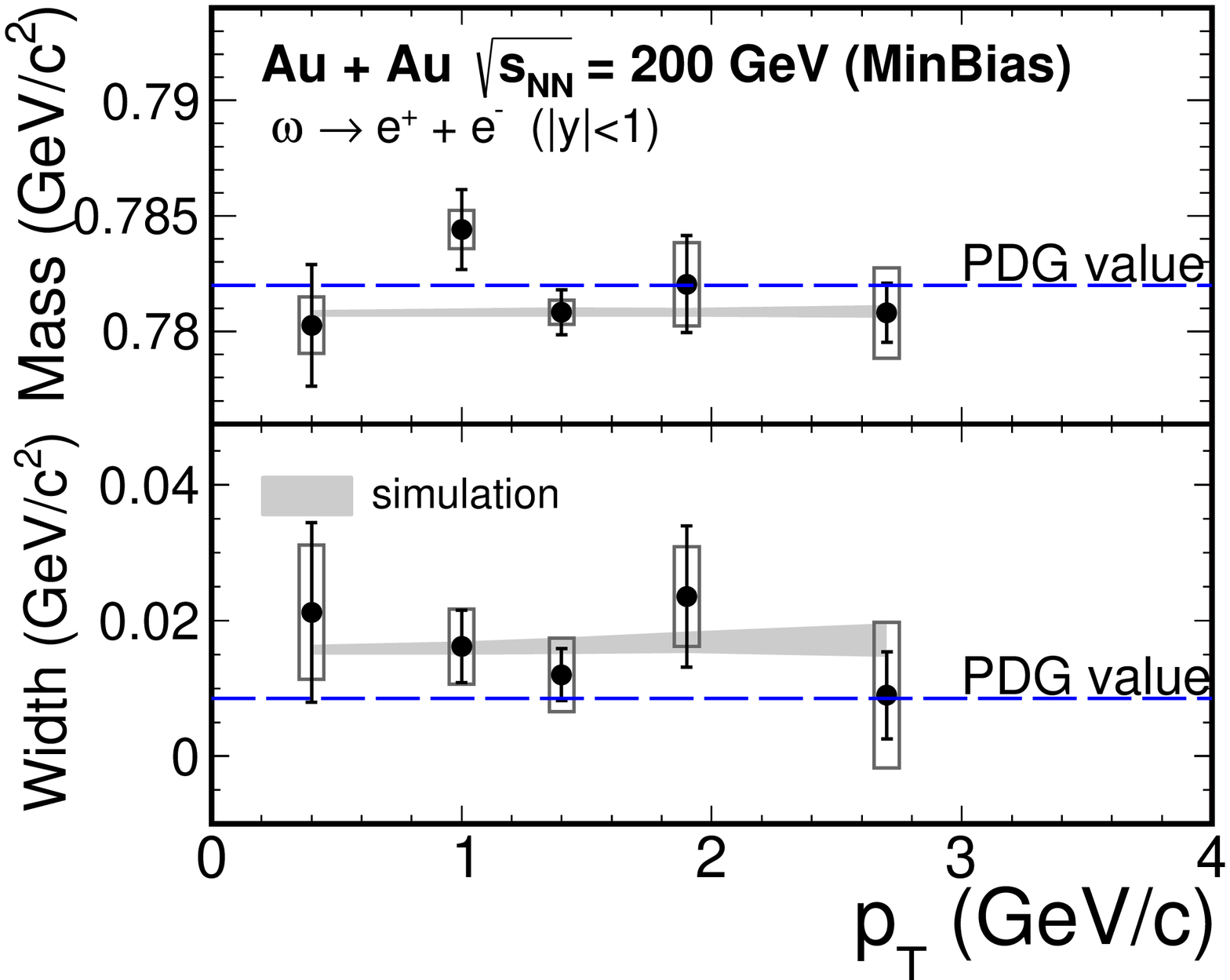} 
    \includegraphics[width=0.45\textwidth] {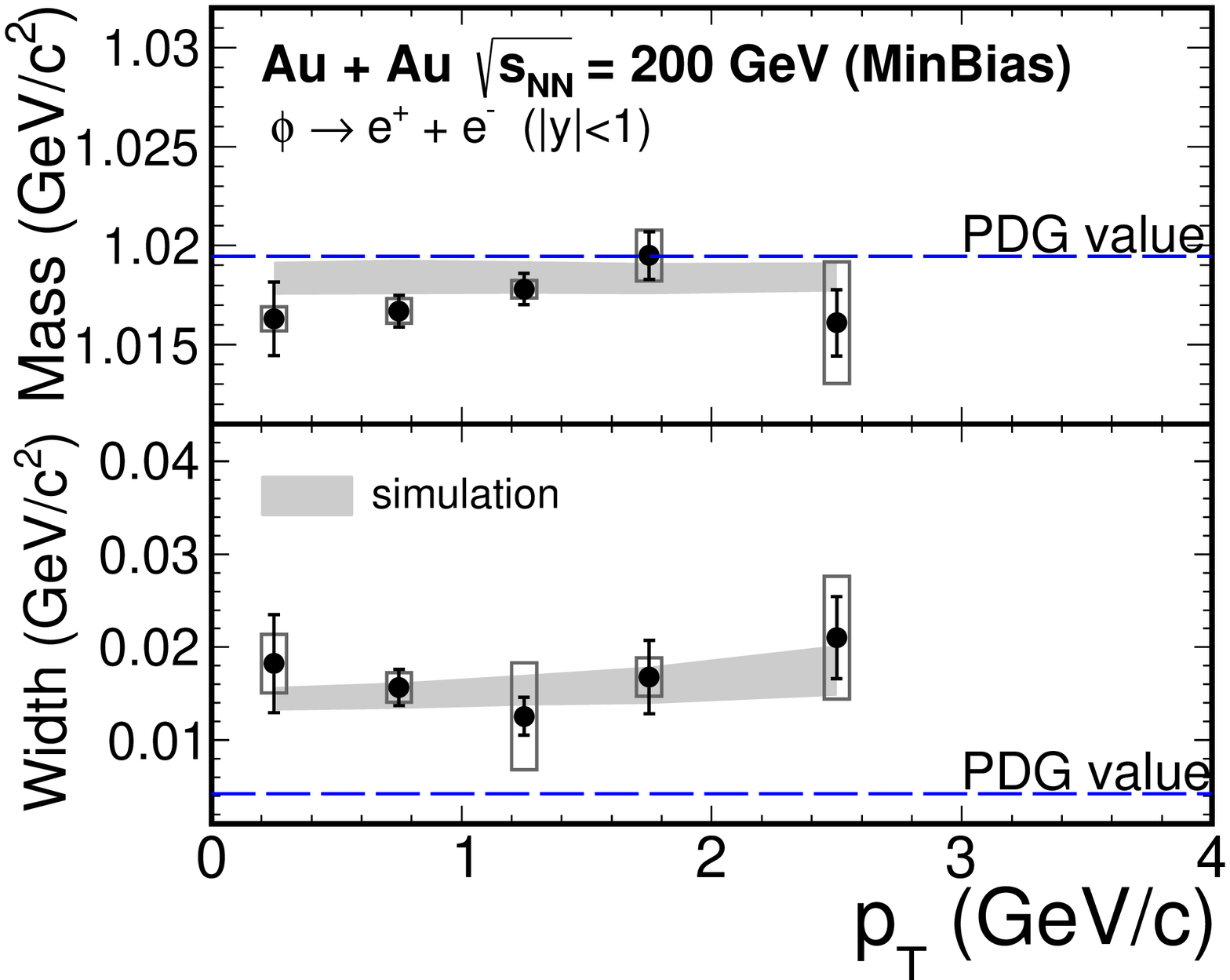}}
    \caption[]{(Color Online) The widths and mass positions of the $\omega$ and $\phi$ signals from data compared to the values from the PDG and the full {\sc Geant} simulation. Boxes on the data points depict the systematic uncertainties. Gray bands represent the uncertainty of the simulation.}
  \label{mass_width}
\end{figure}

Although the mass and width of vector mesons could be modified due to 
interactions with the hot and dense medium, 
the observed $\omega$ and $\phi$ spectra from the detector will have little sensitivity to such an effect. 
The lifetimes of $\omega$ and $\phi$ mesons are much longer than the typical lifetime of 
the medium created in high energy heavy-ion collisions. 
Therefore, the freeze-out $\omega$ and $\phi$ mesons will dominate the observed yields. 
We obtained the widths and mass positions of $\omega$ and $\phi$ signals from data and compared them to 
the values from the PDG as well as from our simulations, shown in Fig.~\ref{mass_width}.

The mass positions of the $\omega$ and $\phi$ mesons from our data generally agree with the PDG values, 
with a slight shift towards lower values. 
This is mainly because the STAR tracking algorithms account for the energy loss assuming pions. 
The observed mass shift (1-2 MeV/$c^2$) for $\omega$ and $\phi$ mesons are within the uncertainties 
of the particle energy loss correction in our GEANT simulations.
The widths of the mass distribution are larger than the PDG values as expected, due to detector resolution effects. 
A tuned simulation, using the $J/\psi$ mass distribution (described in Section III-G), 
can reproduce well the observed signal widths for $\omega$ and $\phi$ mesons in the full \pT\ region reported here. 
Because of uncertainty in the description of materials, 
including accessory components in the STAR detector system, 
we included the difference between the tuned simulation and the GEANT simulation 
in the width calculation as part of our systematic uncertainty.
Since the mass and width are well reproduced by the tuned simulation, 
we fixed the mass and width with the value from the tuned simulations when using 
the Breit-Wigner function to extract the $p_{T}$ differential yield.

\begin{figure}
  \centering{
    \includegraphics[width=0.45\textwidth] {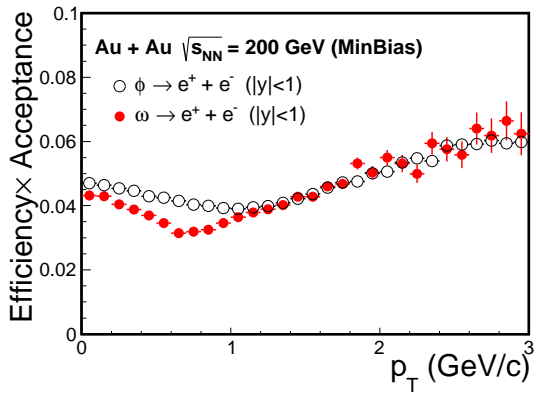}}
    \caption[]{(Color Online) The efficiency and acceptance correction factor as function of \pT\ for mid-rapidity $\omega$ and $\phi$ mesons.}
  \label{meson_eff}
\end{figure}

\begin{figure}
  \centering{
    \includegraphics[width=0.45\textwidth] {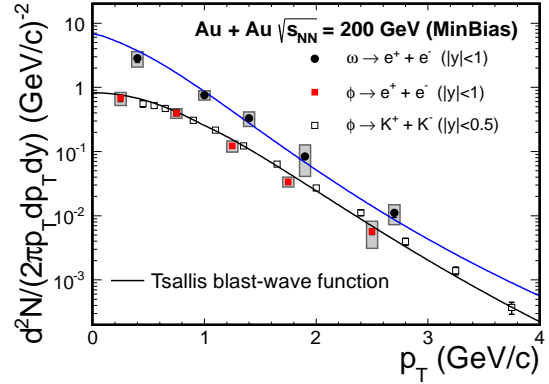}}
    \caption[]{(Color Online) The $p_{T}$ distributions of the $\omega$ and $\phi$ meson invariant yields 
	from \sNN = 200 GeV Au$+$Au minimum-bias collisions. The solid black and blue lines are 
	from the Tsallis Blast-Wave functions as described in Section \ref{Sect:Cocktail}.}
  \label{meson_spectra}
\end{figure}

In order to present the final $p_{T}$-differential invariant cross section, the raw vector meson yields are corrected for the detector's acceptance and efficiency. Figure~\ref{meson_eff} shows the total detector acceptance and efficiency as a function of $p_{T}$ for $\omega \rightarrow e^{+} + e^{-}$ and $\phi \rightarrow e^{+} + e^{-}$. In Fig.~\ref{meson_spectra}, the final $p_{T}$ differential invariant yields are shown for $\omega$ and $\phi$ from \sNN = 200\,GeV Au$+$Au collisions at mid-rapidity ($|y|<1$). The systematic uncertainties include the detector efficiency uncertainty and the raw signal extraction uncertainty. The latter is derived from changing the fit range, the function used for describing the background and the method used to extract the yields. The $\phi$ spectrum measured from $e^+e^-$ decays is consistent with the previous results measured from the hadronic decay channel ($\phi \rightarrow K^{+} +K^{-}$)~\cite{DataPhi}. Also included in the figure are the TBW-model~\cite{TBW} fit to the previous $\phi\rightarrow K^{+} +K^{-}$ data points as well as a prediction of the $\omega$ spectrum with the same set of parameters obtained from the simultaneous fit to all available light hadrons (see Section III-G). The TBW prediction describes the measured $\omega$ spectrum well. The measured $dN/dy$ for the $\omega$ meson is 8.46$\pm$0.67(stat.)$\pm$1.59(syst.), and for the $\phi$ meson is 2.20$\pm$0.10(stat.)$\pm$0.34(syst.).

\section{Summary}

We have reported STAR measurements of dielectron yields at mid-rapidity in Au$+$Au collisions at $\sqrt{s_{\rm NN}}$ = 200\,GeV. 
The measured dielectron yields within the STAR acceptance (defined by $p_T^e>0.2$\,GeV/$c$, $|\eta^e|<1$, 
and $|y_{ee}|<1$) show an enhancement when compared to hadronic cocktail calculations in the mass region below $M_{\phi}$. 
The enhancement factor, integrated over the mass region of 0.30 $-$ 0.76~GeV/$c^2$ and the full \pT\ acceptance, 
is 1.76\,$\pm$\,0.06\,(stat.)\,$\pm$\,0.26\,(sys.)\,$\pm$\,0.29\,(cocktail). 
Further systematic measurements show that this enhancement factor has a mild centrality and \pT\ dependence. 
A vacuum $\rho$ spectral function cannot fully describe the measured dielectron mass spectrum in this mass region. 
This enhancement factor is significantly lower than what has been reported by PHENIX.
We have compared the STAR and PHENIX cocktail simulations and applied PHENIX azimuthal acceptance.
We found that neither differences in the acceptance nor the cocktail simulations can explain
the difference in the enhancement factor measured by the two experiments.

We compared our results to model calculations including an effective many-body model (Rapp) and a microscopic transport model (PHSD). Both models invoked an in-medium modified $\rho$ spectral function through the interactions with mesons and baryons in the bulk medium. Both can reproduce the low-mass excess in our data reasonably well, including the \pT\ and centrality dependences.
A power-law fit to the excess yield in the $\rho$-like region as a function of $N_{\rm part}$ gives a power of 1.44$\pm$0.10.
We noted that the many-body model calculations have successfully explained the SPS low-mass dilepton data. 
These findings could indicate that the property of the hadronic medium that governs 
the dilepton production in the low-mass region is similar at top SPS and top RHIC energies despite the 
difference in center-of-mass energies of more than an order of magnitude. 
Dielectron measurements from the RHIC beam energy scan program will offer a unique opportunity 
to fill the energy gap between the SPS and RHIC and systematically evaluate the energy dependence of dielectron production. 

We also reported the measurement of $\omega$ meson production, and $\phi$ meson production through 
the dielectron decay channel in Au$+$Au collisions at $\sqrt{s_{\rm NN}}$ = 200\,GeV. 
The observed signal widths and mass positions are well reproduced in Monte Carlo simulations. 
The measured $\phi$ invariant yield spectrum through the $e^+e^-$ decay channel 
is consistent with the previously published STAR measurement based on the $K^+K^-$ decay channel.
The $\omega$ invariant yield spectrum can be well reproduced by Tsallis Blast-Wave model predictions 
which use the same set of parameters obtained from a simultaneous fit to all other available light hadrons. 
This indicates that the $\omega$ mesons behave much like the bulk medium, with similar radial flow velocity.

The understanding of the dielectron production in the mass region of 1$-$3\,GeV/$c^2$ is currently 
limited both statistically and systematically. 
We reported the inclusive dielectron yields which include the contribution from correlated charm decays. 
However, at this time we do not know the characteristics of the charm contribution in Au$+$Au collisions.
The dielectron data from 0-80\% minimum-bias collisions can be fairly well described by 
the number-of-binary-collisions scaled $p$$+$$p$ contribution based on {\sc Pythia} calculations. 
The ratio between the central and minimum-bias spectra in the mass region of 1-3\ GeV/$c^2$ 
shows a moderate deviation from the $N_{\rm bin}$ scaling (1.8$\sigma$ deviation for the data point at 1.8-2.8 GeV/$c^2$). 
This could be indicative of the modification of the correlated charm contribution or the existence of 
other contributing sources in Au$+$Au collisions.  The difference in the mass region 1-3 GeV/$c^2$, 
if confirmed in future measurements with a better precision, would constrain the magnitude of 
the de-correlating effect on charm pairs while traversing the QCD medium and/or other possible  
dielectron sources, {\it e.g.}, QGP thermal radiation, from central Au+Au collisions at RHIC.

\vspace{0.2 in}
We thank the RHIC Operations Group and RCF at BNL, the NERSC Center at LBNL, the KISTI Center in Korea, 
and the Open Science Grid consortium for providing resources and support. This work was supported in part 
by the Office of Nuclear Physics within the U.S. DOE Office of Science, the U.S. NSF, CNRS/IN2P3, FAPESP CNPq of 
Brazil,  the Ministry of Education and Science of the Russian Federation, 
the NNSFC, the MoST of China (973 Program No. 2014CB845400), CAS, the MoE of China, 
the Korean Research Foundation, GA and MSMT of the Czech Republic, FIAS of Germany, DAE, DST, and CSIR of India, 
the National Science Centre of Poland, National Research Foundation (NRF-2012004024), the Ministry of Science, 
Education and Sports of the Republic of Croatia, and RosAtom of Russia.

\newpage

\appendix{

\section{Centrality-dependent Cocktail Simulation Inputs}
\label{Appendix:cocktail}
When comparing to the measured spectra, the hadron cocktails were simulated for each of 
the corresponding centrality bins (0-10\%, 10-40\% and 40-80\%). 
The centrality dependence of the input hadron \pT\ distributions were obtained from the similar Tsallis
Blast-Wave function fit to the available data, including $\pi^{\pm}$, $\phi$, {\it etc}. 
For other hadrons with no available measurements, we use the Tsallis Blast-Wave predictions for the input \pT\ distributions. 
The input $dN/dy$ for all the components in each centrality bin were then scaled with the relative pion yields, $R_{\pi}$,
with respect to minimum-bias collisions. 
Correlated-charm contributions are scaled by the number of binary collisions for a given centrality.
All of these scale factors are summarized in Table~\ref{cocktailcen}.


\bt  
\caption{Scale factors for centrality dependent cocktails}
\centering
\begin{tabular}{c|c|c|c}
\hline
\hline
centrality & dN($\pi$)/dy  & $R_{\pi}$ & $\langle N_{\rm bin}\rangle$ \\ \hline
0-80\%  & 98.49 & 1     & $291.90\pm20.46$  \\ \hline
0-10\%  & 279.2 & 2.834 & $941.24\pm26.27$  \\ \hline
10-40\% & 131.1 & 1.331 & $391.36\pm30.21$  \\ \hline
40-80\% & 30.45 & 0.309 & $56.62\pm13.62$   \\
\hline
\hline
\end{tabular}
\label{cocktailcen}
\et

\section{Detector Acceptance Effect}
\label{Appendix:acceptance}

The STAR mid-rapidity detectors cover the full azimuth (0$<\phi<$ 2$\pi$) for
$|\eta|<$1 while the PHENIX central arms (used for the dielectron analysis)
cover about $2\times\pi/2$ at $|\eta|<$ 0.35. To investigate the impact of the
detector acceptance effect on the final dielectron mass spectrum, we tried to
narrow down the single-track acceptance cut to match the PHENIX acceptance as best
as possible. We acknowledge that fully reproducing  another
experiment's acceptance is virtually impossible due to subtle differences in detector structures and performances.
Instead, the STAR data is selected with the PHENIX azimuthal angle
acceptance cut. Due to the limited statistics, we cannot further reduce the
pseudorapidity window to match the respective PHENIX range. 
In addition, we also expect the physics is not 
significantly different between $|\eta|<0.35$ and $|\eta|<1$ rapidity ranges for
200\,GeV collisions. 

As a result of the magnetic field, the signal track $\phi$ acceptance varies
with $p_{\rm T}$. We used the kinematic acceptance cut presented in the PHENIX
publication~\cite{PHENIX}:


\begin{equation}
\begin{split}
	&\phi_{\rm min} \leq \phi + q\frac{k_{\rm DC\;}}{p_{T}} \;\; \leq \phi_{\rm max} \\ 
	&\phi_{\rm min} \leq \phi + q\frac{k_{\rm RICH}}{p_{T}} \leq \phi_{\rm max} \\ 
\end{split} 
\label{EQphenix}
\end{equation}
where $k_{\rm DC}$ = 0.206~rad$\cdot$\GeVc and $k_{\rm RICH}$ = 0.309~rad$\cdot$\GeVc represent the effective azimuthal bend to DC and
RICH, respectively. One arm
covers the region from $\phi_{\rm min} = \; $-$ \frac{3}{16} \pi$  to $\phi_{\rm max} =
\frac{5}{16} \pi$ , while the other arm covers from $\phi_{\rm min} = \frac{11}{16} \pi$ to
$\phi_{\rm max} = \frac{19}{16} \pi$.

The electron candidate occupancy distributions
for STAR data selected with the PHENIX $\phi$ acceptance cut are shown in Fig.~\ref{Acc_phenix_star}.
The upper panel shows the regular $\phi$ versus \pT\ for negative charged particles, while the
bottom panel shows 1/\pT\ versus $\phi$ for both charges. The plots show that while we can
capture the basic acceptance structure,  the inner fine structure within
this azimuthal angle acceptance may be slightly different due to different detector
structure for both experiments.


\begin{figure}
\centering{
\includegraphics[width=0.5\textwidth]{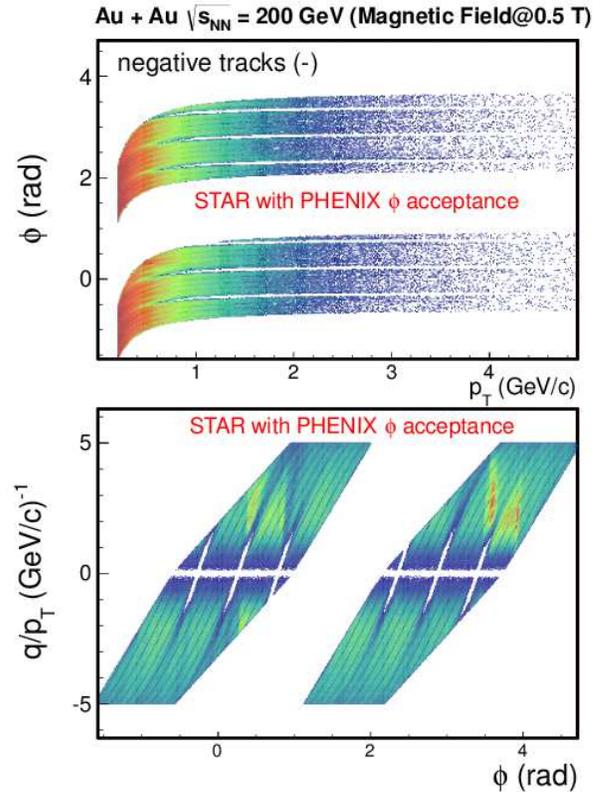}}
\caption[]{(Color online) Single electron/positron track density distributions using the
  STAR data selected within the PHENIX azimuthal angle acceptance.}
\label{Acc_phenix_star}
\end{figure}

With the electron candidates selected, we then carried out the same analysis
procedure as described in Section III. In Fig.~\ref{bg_phenix_star} panels (a), (b),
(c), the ratio is shown for like-sign distributions between same events and
mixed events from which we determine the normalization factor of the mixed-event
unlike-sign distribution for the combinatorial background. 

\begin{figure}
\centering{
\includegraphics[width=0.5\textwidth]{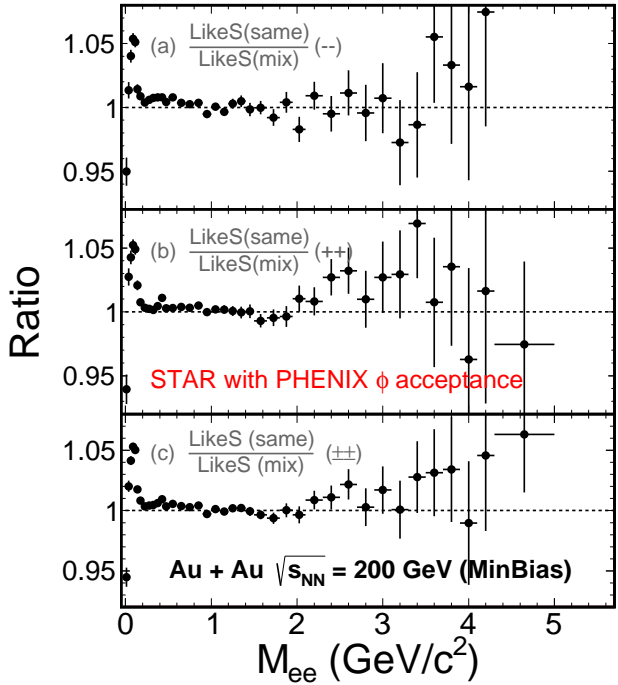}}
\caption[]{Ratios of pair distributions for electron candidates selected with
  the PHENIX $\phi$ acceptance. Panels (a), (b), (c): ratios of like-sign
  distributions between same event and mixed-event. 
}
\label{bg_phenix_star}
\end{figure}

We compared the acceptance difference correction factor between the results with
and without the PHENIX $\phi$ acceptance, as shown in
Fig.~\ref{acc_phenix_star}. One can clearly see that the $\phi$ acceptance
cut changes the pair acceptance between like-sign and unlike-sign pairs
significantly in the low-mass region, and the maximum of this ratio appears
around $M_{ee}\sim$ 0.5\,GeV/$c^2$. 

\renewcommand{\floatpagefraction}{0.75}
\begin{figure*}
\centering
\bmn[b]{0.5\textwidth}
\includegraphics[width=1.0\textwidth]{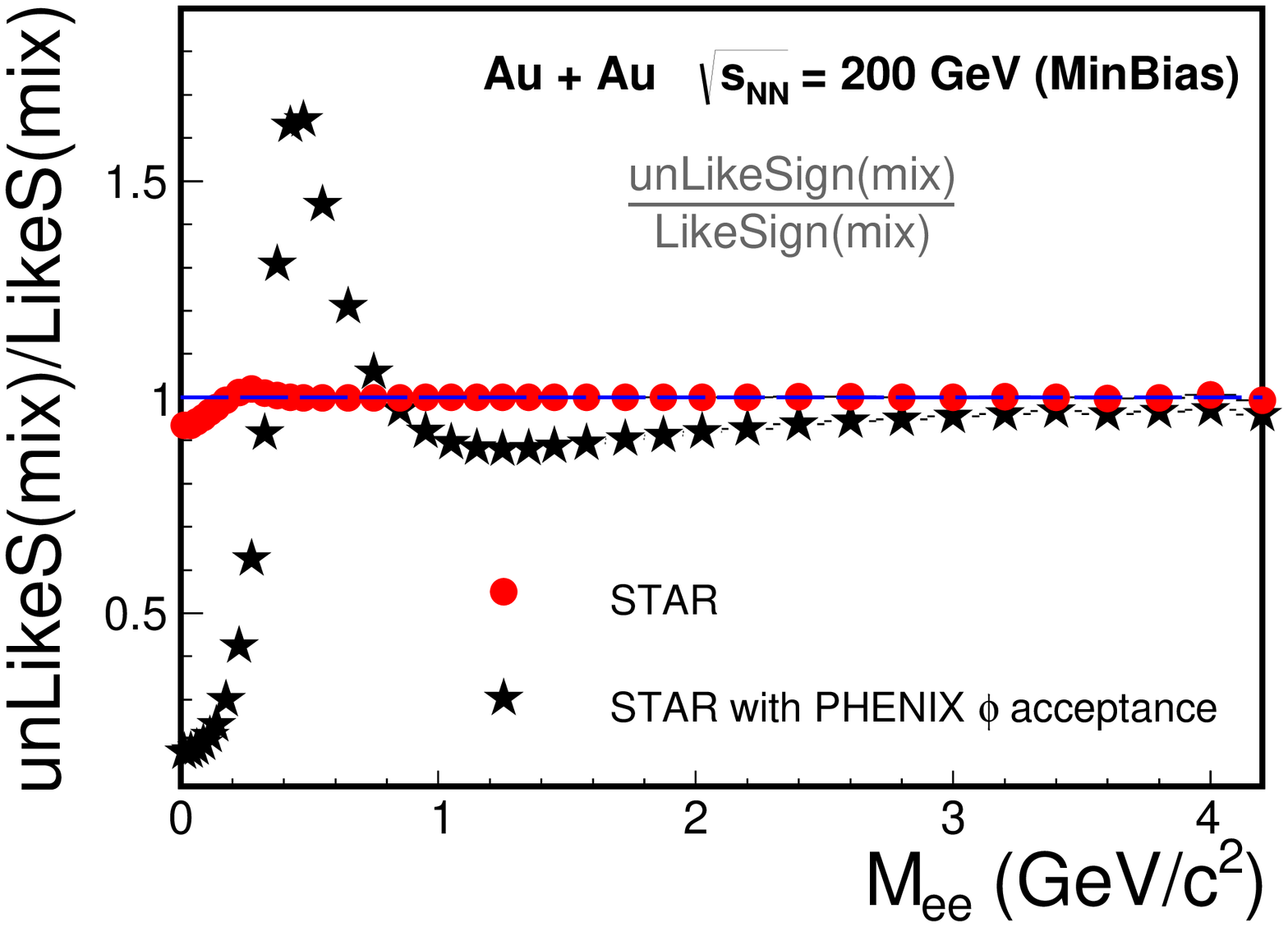}
\emn 
\centering
\bmn[b]{0.5\textwidth}
\includegraphics[width=1.0\textwidth]{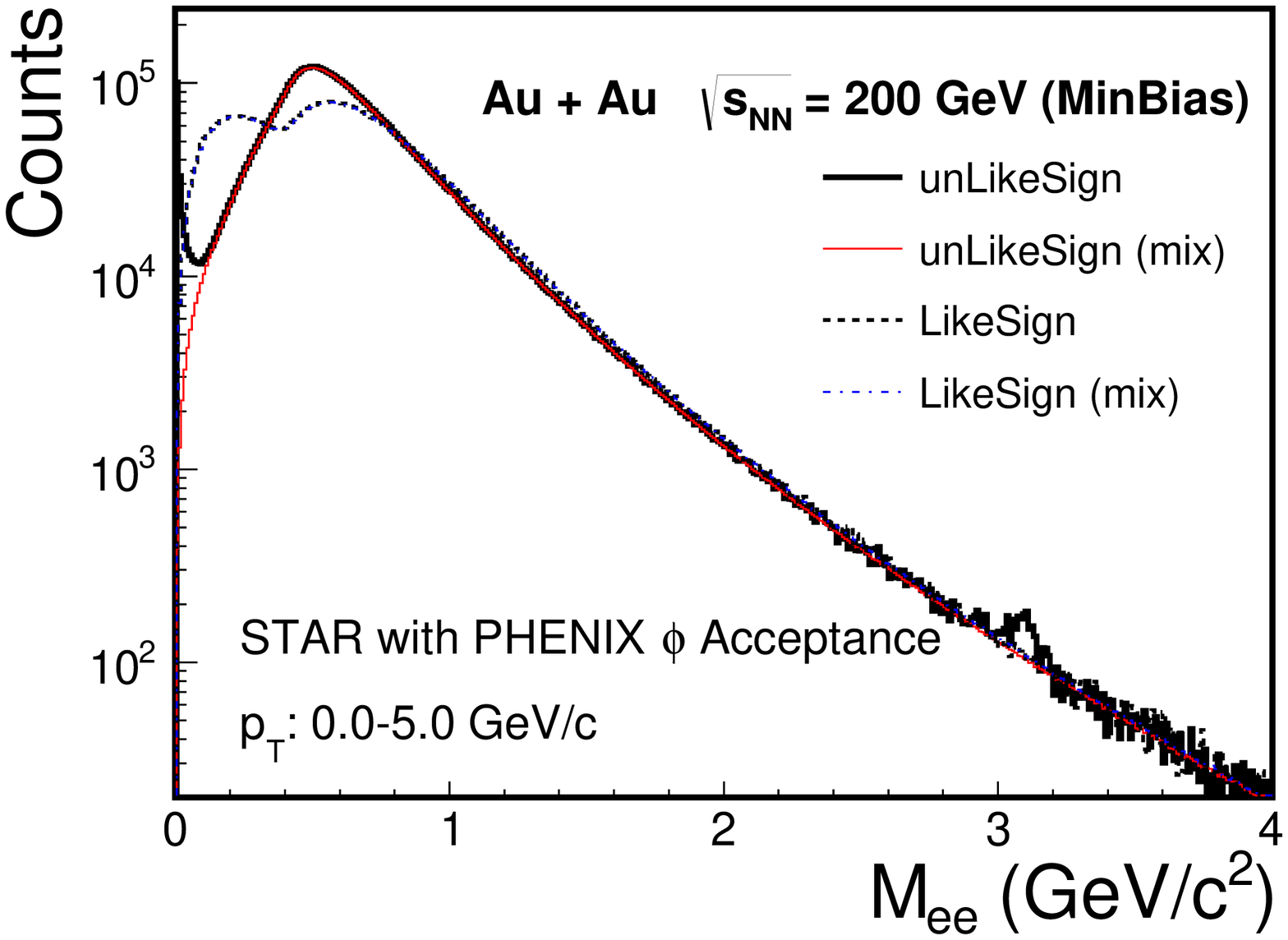}
\emn \\
\caption{(Color online) Left panel: unlike-sign/like-sign pair acceptance
  difference correction factor with the PHENIX $\phi$ acceptance (black stars)
  compared with the full acceptance (red circles). Right panel: dielectron pair
  mass distributions of the STAR data within the PHENIX $\phi$ acceptance; the
  inclusive unlike-sign distribution (black), the same event like-sign
  distribution (blue), and the mixed-event unlike-sign (red), like-sign
  (magenta) distributions.}
\label{acc_phenix_star}
\end{figure*}

The combinatorial background is subtracted from the inclusive unlike-sign pair distribution to obtain the raw signal, 
then the raw signal distribution is corrected for the detector efficiency.
Finally we obtained the signal dielectron invariant mass
spectrum from 200\,GeV minimum-bias Au$+$Au collisions and compared it to hadronic
cocktail simulations, shown in the left panel in Fig.~\ref{phenix_data1}.
In the low-mass region of 0.30-0.76 GeV/$c^2$, we observed an enhancement of 
a factor of 2.4$\pm$0.37(stat.)$\pm$0.38(sys.)$\pm$0.29(cocktail) 
when comparing the measured yield to the hadronic cocktail.
Selecting our data within the PHENIX $\phi$ acceptance 
does not appear to reproduce the large enhancement factor in the low-mass region
observed by the PHENIX collaboration~\cite{PHENIX}.

We added in the medium dielectron contributions from theoretical model
calculations. The right plot of Fig.~\ref{phenix_data1} shows the data compared
to the cocktail plus the broadened $\rho$ spectrum in the hadronic medium and
QGP thermal radiation. The particular calculation that is included here is only valid
for $M<$ 1.5\,GeV/$c^2$. The medium contribution from this model
(hadronic $\rho$ and QGP radiation) describes the observed low-mass excess very
well. 



\renewcommand{\floatpagefraction}{0.75}
\begin{figure*}
\centering
\bmn[b]{0.5\textwidth}
\includegraphics[width=0.9\textwidth]{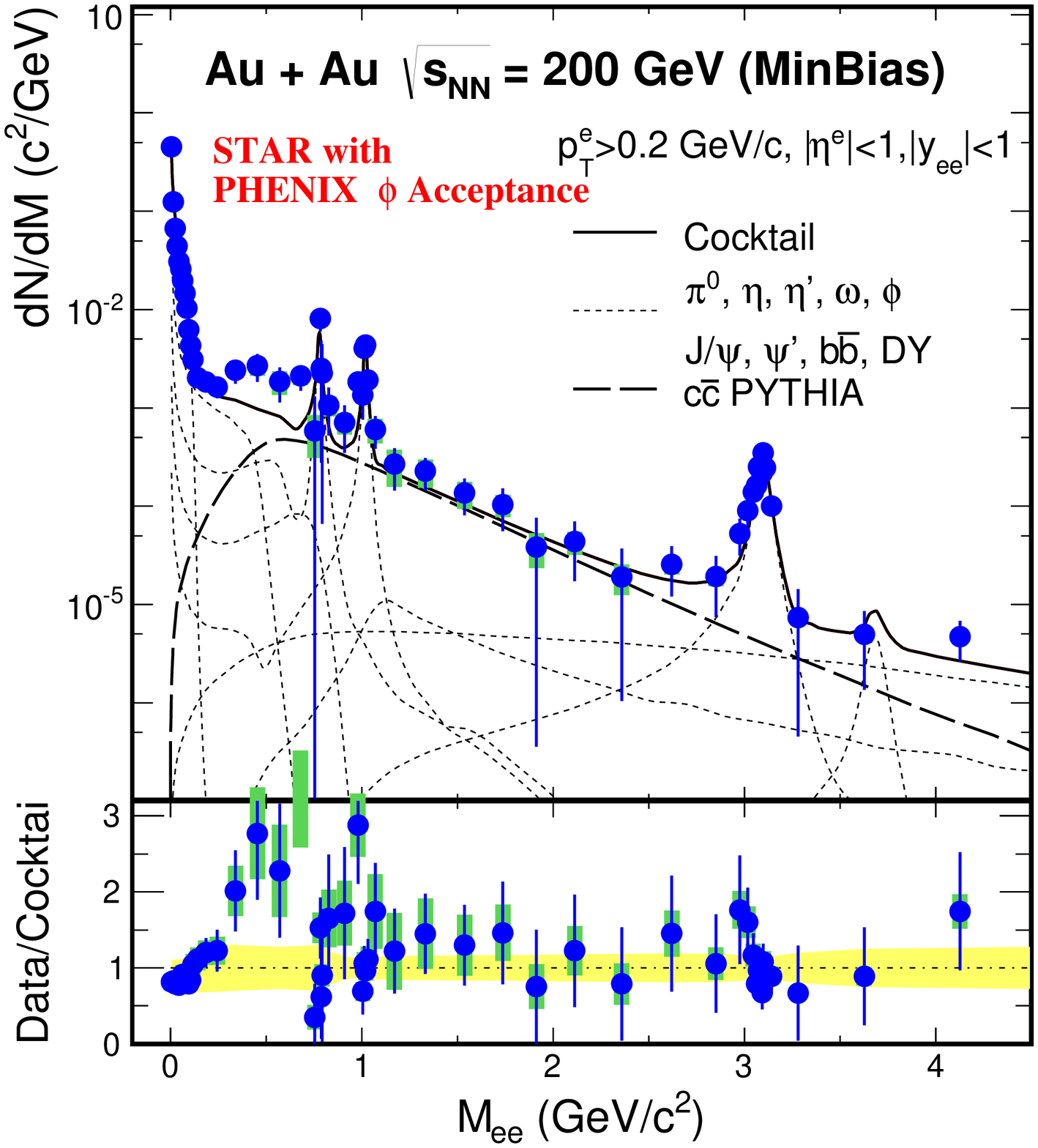}
\emn 
\centering
\bmn[b]{0.5\textwidth}
\includegraphics[width=0.9\textwidth]{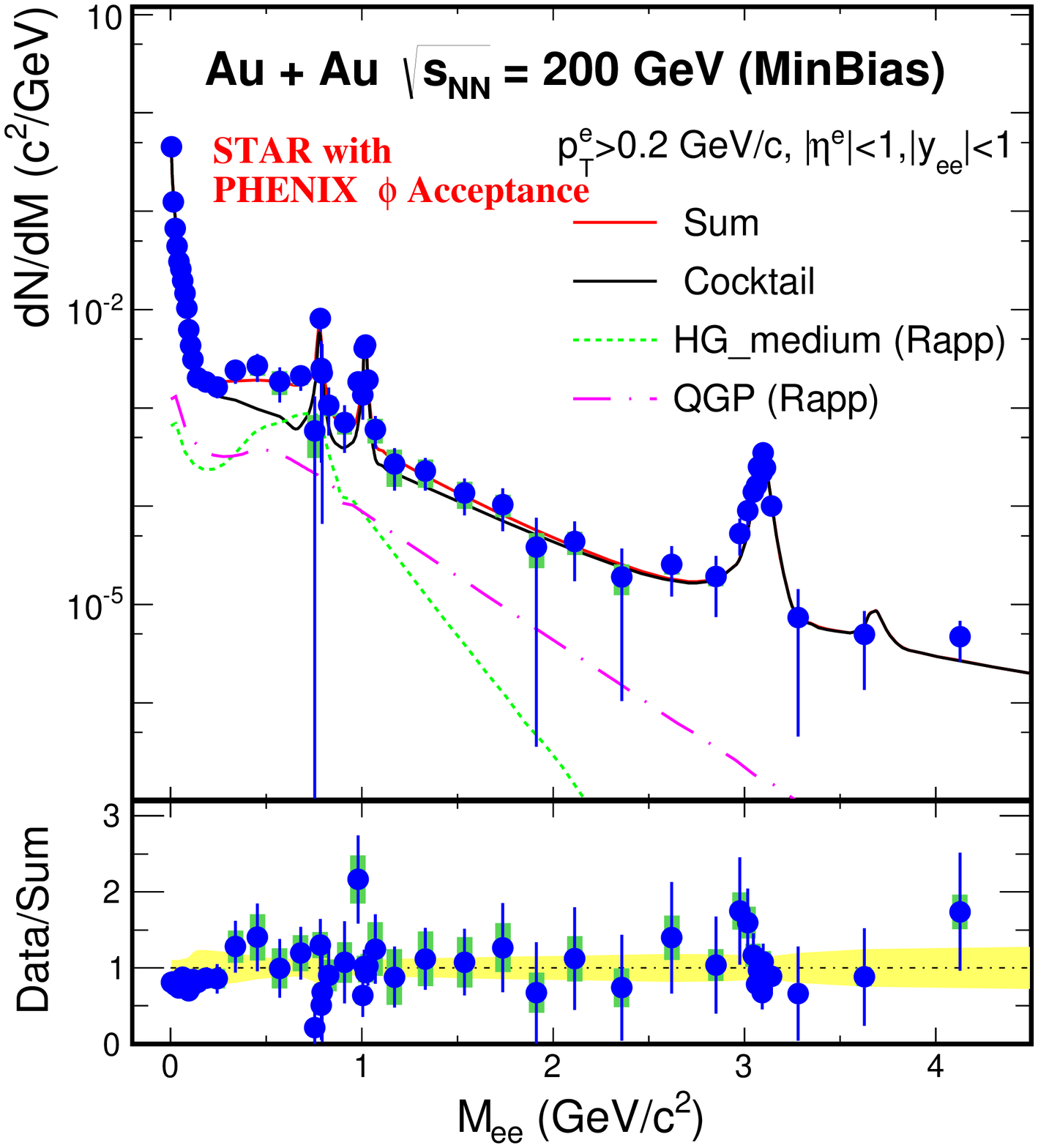}
\emn \\
\caption{(Color online) Left panels: efficiency corrected invariant mass spectra (blue solid dots) calculated using the STAR data filtered with the PHENIX azimuthal angle acceptance. The data points are compared to cocktail simulations shown as curves and the lower left panel shows the ratio of data to the cocktail sum. Right panels: the same data points compared to theoretical model calculations of medium vector meson and QGP contributions from Ref.~\cite{RappPriv}. The bottom right panel shows the ratio of data to the sum of the cocktail and the theory calculations of medium contributions.}
\label{phenix_data1}
\end{figure*}

\section{Cocktail comparison between STAR and PHENIX}
\label{Appendix:PHENIXvSTARcocktails}

In this appendix, we compare the cocktail simulation results between PHENIX and STAR. The details of the light hadron decays and Dalitz decays into dielectrons are described in~\cite{STARpp}. We used the $p$$+$$p$ input yields for cocktail calculations in this comparison and we folded in the PHENIX acceptance filter, described in Eq.~\ref{EQphenix}. Next, we compared the output to the PHENIX $p$$+$$p$ cocktail calculations.

The comparison for the total cocktail summed yield is shown in Fig.~\ref{comp_pp_allphxFilter}. The comparisons for each cocktail component are shown in Fig.~\ref{comp_pp_partphxFilter}.
We see that the cocktail calculations from both experiments agree reasonably well. There are some small differences in the $\eta$, $\omega$, $\phi$ Dalitz decay distributions which can be attributed to different choices of decay form factors in these Dalitz decays.

\begin{figure}
\centering{
\includegraphics[width=0.5\textwidth]{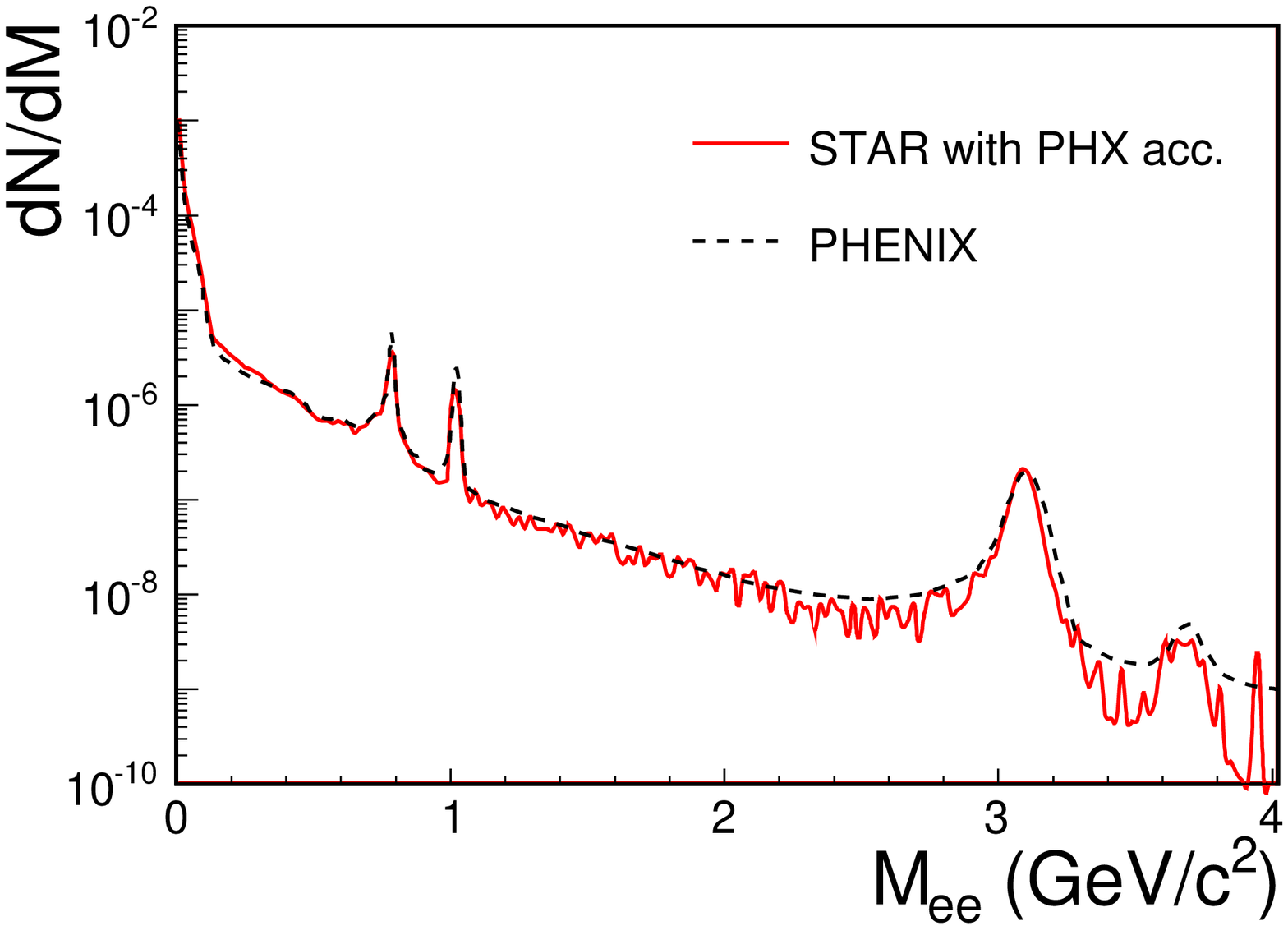}}
\caption[]{(Color online) Comparison of the total cocktail sum within the PHENIX azimuthal angle acceptance calculated by STAR (red solid) and PHENIX (black dashed) for $p$$+$$p$ 200\,GeV collisions.}
\label{comp_pp_allphxFilter}
\end{figure}

\begin{figure}
\centering{
\includegraphics[width=0.5\textwidth]{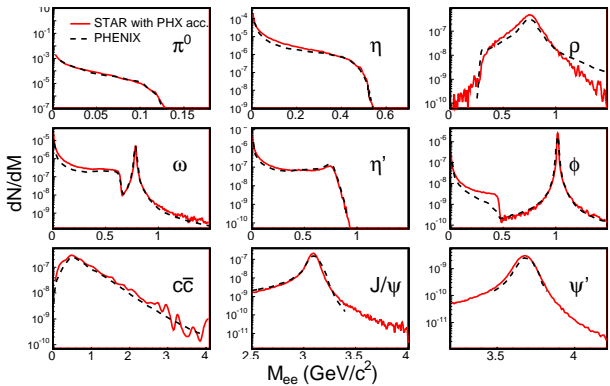}}
\caption[]{(Color online) Comparison of each cocktail component within the PHENIX azimuthal angle acceptance calculated by STAR (red solid) and PHENIX (black dashed) for $p$$+$$p$ 200\,GeV collisions.}
\label{comp_pp_partphxFilter}
\end{figure}

} 



\end{document}